\documentclass[a4paper,11pt]{article}

\usepackage{jheppub} 

\usepackage[T1]{fontenc} 
\usepackage{CJK}

\usepackage{hyperref}
\hypersetup{
	pdfstartview=FitH,
	bookmarksnumbered=true,
	colorlinks,
	breaklinks,
	urlcolor=RoyalBlue,
	linkcolor=RoyalBlue,
	citecolor=NavyBlue
}
\usepackage{graphicx}
\DeclareGraphicsExtensions{.pdf,.png,.jpg,.mps}
\usepackage{booktabs}
\usepackage[dvipsnames]{xcolor}
\usepackage{tikz}
\usetikzlibrary{calc,matrix,decorations.pathmorphing,decorations.markings,arrows.meta,positioning,intersections,mindmap,backgrounds}
\usepackage{cancel}

\usepackage{amsmath,amsthm,amsfonts,amssymb,amscd,mathtools
}
\usepackage{pifont}
\usepackage{ifsym}

\newcommand{\brief}[1]{{}}

\title{\boldmath Symbols from Bi-Projections}

\author[a,b,c]{Jianyu Gong (\begin{CJK}{UTF8}{gbsn}宫健宇\end{CJK}),}
\author[a]{You Wang (\begin{CJK}{UTF8}{gbsn}王悠\end{CJK}),}
\author[a]{Ellis Ye Yuan (\begin{CJK}{UTF8}{gbsn}袁野\end{CJK})}
\affiliation[a]{Zhejiang Institute of Modern Physics, Department of Physics, Zhejiang University,\\866 Yuhangtang Road, Hangzhou, Zhejiang 310058, China}

\affiliation[b]{State Key Laboratory of Dark Matter Physics, Shanghai Key Laboratory for Particle Physics and Cosmology, Key Laboratory for Particle Astrophysics and Cosmology (MOE), School of Physics and Astronomy, Shanghai Jiao Tong University, Shanghai 200240, China}
\affiliation[c]{Institute of Nuclear and Particle Physics (INPAC), Shanghai Jiao Tong University, Shanghai 200240, China}

\emailAdd{jianyu\_gong@sjtu.edu.cn}
\emailAdd{wangyou@zju.edu.cn}
\emailAdd{eyyuan@zju.edu.cn}

\abstract{We initiate a systematic framework for the analysis of analytic properties of finite Feynman integrals that are multiple polylogarithms. Based on the Feynman parameter representation in complex projective space, we make a complete classification of logarithmic singularities of the integral on its principal branch, by what we call touching configurations -- a geometric relationship between the integrand singularity and linear subspaces tied to boundary elements of the integral contour. These on the one hand indicate first entries of the symbol of the integral, and on the other hand induce a special set of new integrals that we call elementary discontinuities. These elementary discontinuities are derived through an operation called bi-projection, and actual discontinuities of the integral across logarithmic branch cuts are their linear combinations. By recursively applying the same analysis to the induced integrals one can fully construct the symbol of the original integral. We explicitly show how this analysis works at one loop in a massless hexagon and a box with two massive and two massless loop propagators. This framework may naturally extend to higher-loop integrals.}

\begin{document}

\maketitle
\flushbottom

\section{Introduction and summary}

\brief{Briefly review explorations related to symbol.}

Symbol has been an important tool ever since its introduction to the analysis of scattering amplitudes \cite{Goncharov:2009lql,Goncharov:2010jf}. It provides a powerful way to simplify expressions for loop-level amplitudes, and helps to reveal novel and interesting structures at loop level \cite{Abreu:2021vhb,Abreu:2017enx,Abreu:2017mtm}. In recent years, it also plays a central role in bootstrapping the frontier analytic results of loop-level amplitudes \cite{Dixon:2011pw,Dixon:2013eka,Dixon:2014iba,Drummond:2014ffa,Chicherin:2017dob,Caron-Huot:2019vjl,Caron-Huot:2020bkp,Dixon:2023kop,Hannesdottir:2024hke}.

Because the symbol is naturally defined recursively in terms of differentiation \cite{Duhr:2014woa,Weinzierl:2022eaz,Abreu:2022mfk}
\begin{align}\label{eq:symbolDefDiff}
	\mathrm{d}F=\sum_iF_i\times\mathrm{d}\log(R_i)\quad\Rightarrow\quad\mathcal{S}[F]=\sum_i\mathcal{S}[F_i]\otimes R_i,
\end{align}
once a system of differential equations is known for a set of integrals, the symbol for each function in the set can be read off easily. Nevertheless, in practice for a given Feynman integral one may not always prefer to solve the symbol in this way, as it requires the identification of master integrals and the use of IBPs (e.g., \cite{CHETYRKIN1981159,Kotikov:1990kg,Henn_2013}). This is particularly the case when the integral under study contains multiple parameters/scales. And so it should be also interesting to see if the symbol can be computed in a more straightforward manner from the definition of a Feynman integral. Recent explorations in this regard involve the bootstrap idea, where various consistency constraints are imposed on the relations between different entries in a symbol \cite{Caron-Huot:2018dsv,Caron-Huot:2019bsq,Drummond:2017ssj,Drummond:2018dfd}. And as an important step in this program, much attention has been paid on determining the symbol alphabet (i.e., the set of letters in all entries of the symbol), with the help of unitarity methods \cite{Abreu:2014cla,Abreu:2015zaa,Bourjaily:2017wjl,Bourjaily:2019iqr,Britto:2023rig,Britto:2024mna}, Landau equations \cite{Eden:1966dnq,Dennen:2015bet,Dennen:2016mdk,Prlina:2018ukf,Mizera:2021icv,Hannesdottir:2021kpd,Lippstreu:2022bib,Lippstreu:2023oio,Dlapa:2023cvx,Fevola:2023kaw,Fevola:2023fzn,Helmer:2024wax,Caron-Huot:2024brh}, Schubert analysis \cite{Yang:2022gko,He:2022ctv,Morales:2022csr,He:2022tph,He:2023umf,He:2024fij}, etc.

\brief{Symbol in view of discontinuities.}

Apart from the above developments, a different view angle can also inspire an alternative strategy in the direct derivation of symbol. Note that differentials reveal the last entries of a symbol, as shown in \eqref{eq:symbolDefDiff}. In reverse, the first entries of a symbol encode data on the logarithmic singularities of the corresponding function that are seen on its principal branch. Specifically, suppose the symbol of a function $G$ has the form
\begin{align}\label{eq:SGLDisc}
	\mathcal{S}[G]=\sum_iL_i\otimes\mathcal{S}[G_i].
\end{align}
Then on the one hand, $G$ has logarithmic branch points at locations where $L_i=0$ or $L_i=\infty$. On the other hand, its discontinuity across the branch cut tied to such branch point is $2\pi i\,G_i$ (for simplicity we will always omit the factor $2\pi i$ in later discussions). When these discontinuities are worked out, each $G_i$ can be treated as an independent function, which means we can freely consider its own analytic continuation without worrying how it arises from $G$. Being a function of one lower transcendental weight, the symbol of $G_i$ expects to share a very similar structure as \eqref{eq:SGLDisc}
\begin{align}
    \mathcal{S}[G_i]=\sum_jL_{ij}\otimes\mathcal{S}[G_{ij}],
\end{align}
such that its logarithmic branch points on the principal branch are indicated by $L_{ij}$'s, while $G_{ij}$ denotes the corresponding discontinuity. Such structures can be iterated until we get some function $G_{i_1i_2\cdots i_{w-1}}$ at the $w$-th step, which has the schematic form
\begin{align}
    G_{i_1i_2\cdots i_{w-1}}=R_{i_1i_2\cdots i_{w-1}}\,\log(T_{i_1i_2\cdots i_{w-1}}),
\end{align}
where both $R_{i_1i_2\cdots i_{w-1}}$ and $T_{i_1i_2\cdots i_{w-1}}$ are some algebraic functions. In this case the symbol can be simply read off from the result. If all the $L$'s can be identified during the above recursive procedure, then one can start with $\mathcal{S}[G_{i_1i_2\cdots i_{w-1}}]$ and inverse the analysis, so as to build up symbols of discontinuities in each layer, till $\mathcal{S}[G]$ is fully constructed at the end.
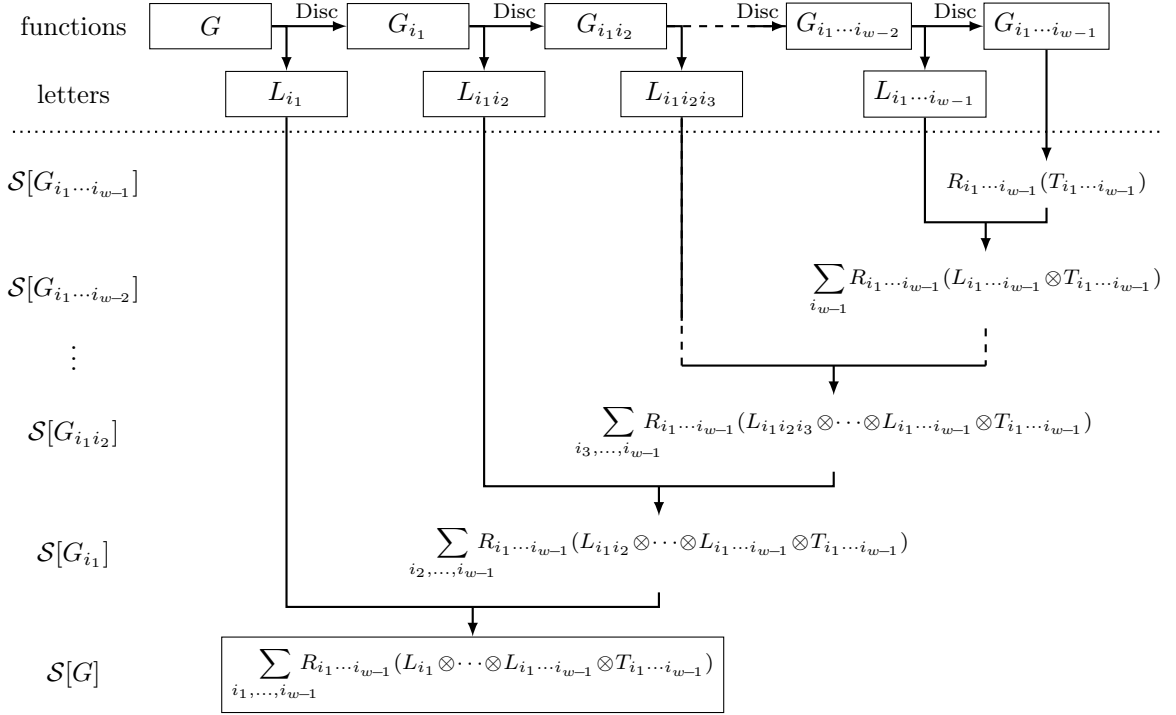
\begin{figure}[ht]
	\centering
	\begin{tikzpicture}[every node/.style={draw=black,minimum width=1.6cm,minimum height=.6cm}]
	    \node [draw=white] at (-1.8,0) {\small functions};
	    \node [draw=white] at (-1.8,-.9) {\small letters};
		\node (g0) at (0,0) {\small $G$};
		\node (g1) at (2.6,0) {\small $G_{i_1}$};
		\node (g2) at (5.2,0) {\small $G_{i_1i_2}$};
		\node (g3) at (8.4,0) {\small $G_{i_1\cdots i_{w-2}}$};
		\node (g4) at (11,0) {\small $G_{i_1\cdots i_{w-1}}$};
		\node (l1) at (1,-.9) {\small $L_{i_1}$};
		\node (l2) at (3.6,-.9) {\small $L_{i_1i_2}$};
		\node (l3) at (6.2,-.9) {\small $L_{i_1i_2i_3}$};
		\node (l4) at (9.4,-.9) {\small $L_{i_1\cdots i_{w-1}}$};
		\draw [thick,dotted] (-2.6,-1.4) -- +(15.1,0);
		\node [draw=white] at (-1.8,-2.1) {\small $\mathcal{S}[G_{i_1\cdots i_{w\!-\!1}}]$};
		\node [draw=white] (d4) at (11,-2.1) {\scriptsize $R_{i_1\cdots i_{w\!-\!1}}(T_{i_1\cdots i_{w\!-\!1}})$};
		\node [draw=white] at (-1.8,-3.5) {\small $\mathcal{S}[G_{i_1\cdots i_{w\!-\!2}}]$};
		\node [draw=white] (d3) at (10.2,-3.5) {\scriptsize $\displaystyle \sum_{i_{w\!-\!1}}\!R_{i_1\cdots i_{w\!-\!1}}(L_{i_1\cdots i_{w\!-\!1}}\!\otimes\! T_{i_1\cdots i_{w\!-\!1}})$};
		\node [draw=white] at (-1.8,-5.4) {\small $\mathcal{S}[G_{i_1i_2}]$};
		\node [draw=white] (d2) at (8.2,-5.4) {\scriptsize $\displaystyle \sum_{i_{3},\ldots,i_{w\!-\!1}}\hspace{-1em}R_{i_1\cdots i_{w\!-\!1}}(L_{i_1i_2i_3}\!\otimes\!\cdots\!\otimes\!L_{i_1\cdots i_{w\!-\!1}}\!\otimes\! T_{i_1\cdots i_{w\!-\!1}})$};
		\node [draw=white] at (-1.8,-7) {\small $\mathcal{S}[G_{i_1}]$};
		\node [draw=white] (d1) at (5.9,-7) {\scriptsize $\displaystyle \sum_{i_{2},\ldots,i_{w\!-\!1}}\hspace{-1em}R_{i_1\cdots i_{w\!-\!1}}(L_{i_1i_2}\!\otimes\!\cdots\!\otimes\!L_{i_1\cdots i_{w\!-\!1}}\!\otimes\! T_{i_1\cdots i_{w\!-\!1}})$};
		\node [draw=white] at (-1.8,-8.6) {\small $\mathcal{S}[G]$};
		\node (d0) at (3.45,-8.6) {\scriptsize $\displaystyle \sum_{i_{1},\ldots,i_{w\!-\!1}}\hspace{-1em}R_{i_1\cdots i_{w\!-\!1}}(L_{i_1}\!\otimes\!\cdots\!\otimes\!L_{i_1\cdots i_{w\!-\!1}}\!\otimes\! T_{i_1\cdots i_{w\!-\!1}})$};
		\draw [thick,-latex] (g0) -| (l1);
		\draw [thick,-latex] (g0) -- (g1);
		\draw [thick,-latex] (g1) -| (l2);
		\draw [thick,-latex] (g1) -- (g2);
		\draw [thick,dashed] (g2) -- (g3);
		\draw [thick,-latex] (g2) -| (l3);
		\draw [thick,-latex] ($(g2.east)!.7!(g3.west)$) -- (g3.west);
		\draw [thick,-latex] (g3) -| (l4);
		\draw [thick,-latex] (g3) -- (g4);
		\draw [thick,-latex] (g4) -- (d4);
		\draw [thick] (d4) -- (11,-2.6) -- (9.4,-2.6) -- (l4);
		\draw [thick,-latex] (10.2,-2.6) -- (d3);
		\draw [thick,dashed] (d3) -- (10.2,-4.5);
		\draw [thick] (10.2,-4.5) -- (6.2,-4.5);
		\draw [thick,dashed] (6.2,-4.5) -- (l3);
		\node [draw=white] at (-1.8,-4.3) {\small $\vdots$};
		\draw [thick] (l3.south) -- +(0,-2.7);
		\draw [thick,-latex] (8.2,-4.5) -- (d2);
		\draw [thick] (d2) -- (8.2,-6.1) -- (3.6,-6.1) -- (l2);
		\draw [thick,-latex] (5.9,-6.1) -- (d1);
		\draw [thick] (d1) -- (5.9,-7.7) -- (1,-7.7) -- (l1);
		\draw [thick,-latex] (3.45,-7.7) -- (d0);
		\coordinate [label=90:{\scriptsize Disc}] (p1) at (1.4,-.1);
		\coordinate [label=90:{\scriptsize Disc}] (p1) at (4,-.1);
		\coordinate [label=90:{\scriptsize Disc}] (p1) at (7.2,-.1);
		\coordinate [label=90:{\scriptsize Disc}] (p1) at (9.8,-.1);
	\end{tikzpicture}		
    \caption{Generic strategy of symbol construction by recursive study of discontinuities.}
	\label{fig:strategy}
\end{figure}

The above generic strategy is illustrated in Figure \ref{fig:strategy}. Here we see the workflow typically decomposes into three relatively distinct problems:
\begin{enumerate}
    \item Identification of the first entry letters.
    \item Derivation of all independent discontinuities.
    \item Construction of the symbol from the data collected in the first two problems.
\end{enumerate}
Because we target on a direct algorithm for the construction of symbols that avoids explicit computation of functions, in the first problem we need to find a way to extract first entry letters directly from an integral representation of the function. Correspondingly, in the second problem it already suffices if we manage to identify an integral representation for each independent discontinuity, without really performing the integration. In order to successfully build the full symbol in the third problem, we need to ensure that the letters and discontinuities collected in the first two problems are complete. While in Figure \ref{fig:strategy} we show schematic structures of the symbols to be built in this workflow, the actual analysis may be a bit more involved. Because there may exist algebraic relations among letters and linear relations among discontinuities, depending on the specific data that are available, each step in the symbol construction above may involve solving linear combinations among different symbol terms.

Based on the common strategy illustrated above, different views towards the first entry letters and discontinuities can in fact leads to different implementations of the strategy. Targeting on one-loop Feynman integrals, there exists an algorithm using the so-called spherical contours \cite{Arkani-Hamed:2017ahv} and another algorithm that utilizes a special type of contours which are interpreted as point projections \cite{Gong:2022erh}. While the former is very efficient for one-loop integrals, it heavily depends on special features at one loop and can barely generalize to higher-loop integrals. The latter views discontinuities in a more elementary manner and may potentially receive a higher-loop generalization, but it requires extra regularization procedure even at one loop when massless propagators are present, thus lacking efficiency. One aim of this paper is to solve these difficulties simultaneously. In fact, with several new observations we can do much better in dealing with the first two problems, so that the analysis in the third problem becomes simple.

For the first problem, we notice that the notion of certain \emph{ambient space} for faces of the simplex contour in Feynman parameter representations of integrals plays an essential role in the analysis of logarithmic singularities related to first entries of the symbol. When properly treated, it in fact offers a complete classification of such singularities in terms of geometric relations which we call \emph{touching configurations}. In particular, this classification universally applies to any Feynman integrals, regardless of loop numbers. For the case of one-loop Feynman integrals, we figure out a well-defined map from different singularities in this classification to the corresponding first entry letters.

For the second problem, based on the geometric classification of singularities above, we identify a special set of discontinuities which we call \emph{elementary discontinuities}. They are elementary in the sense that, on the one hand they are closely tied to each face element of the simplex contour (up to certain dimension), and on the other hand they offer a complete basis such that the actual discontinuities encountered in a given integral can always linearly decompose onto them. By a carefully designed way to ``fibrate'' the complex projective space of Feynman parameters, we introduce a natural parametrization for integrals such that an integral representation of each elementary discontinuity can be efficiently derived from the original integral. Such ``fibration'' is induced by projections through a pair of reference subspaces of the complex projective space, which generalizes the point projetions. In this sense we call such ``fibration'' and the corresponding derivation of discontinuities as \emph{bi-projections}.

With the above new ingredients we show how the symbol of one-loop integrals can be successfully determined in an example of a massless hexagon integral in 6d, and a box integral in 4d with two massive and two massless loop propagators. These two examples cover most of the features in this new analysis. It is worth to point out that in this paper we restrict our scope to integrals that expect to have uniform transcendental weight. For integrals with mixed weight, the analysis typically calls for a careful treatment of numerators in the integral, which is itself an independent and interesting question (see \cite{Arkani-Hamed:2017ahv} for a treatment in the context of spherical contours). We leave the analysis in this new setup for future explorations.

\brief{Outline.}

The remaining of this paper is planned as follows. In Section \ref{sec:preliminaries} we review necessary backgrounds on integrals in complex projective space and basic properties of Feynman parameter integrals, to set up context for later discussions. Section \ref{sec:touchingConfiguration} mainly deals with the first problem. Here we introduce the notion of ambient space and touching configurations. With these we draw a classification of singularities of integrals, and further illustrate methods to extract first entries for one-loop integrals. 
At the end of this section we motivate the notion of elementary discontinuities. Section \ref{sec:fibration} and \ref{sec:discHex} mainly deal with the second problem. As a preparation for properly computing these discontinuities, we discuss in detail a class of novel parametrizations for simplex contours based on bi-projections in Section \ref{sec:fibration}, and show their relations to the touching configurations. Then in Section \ref{sec:discHex} we use the hexagon example to explain how various discontinuities are computed in this new method. Section 6 mainly deals with the third problem. Here we collect the data computed in the previous sections and illustrate the reconstruction of symbol in the hexagon integral. We further provide the analysis of the box integrals in order to illustrate extra features that one may encounter when working on more general one-loop integrals. Along with this paper we also provide a self-explained \texttt{Mathematica} notebook as an ancillary file for the arXiv submission, which contains the complete analysis on the examples discussed in the paper. Finally, in Section \ref{sec:outlook} we briefly draw some related questions to be explored in future.

\section{Preliminaries}\label{sec:preliminaries}

The class of functions that we study in general is specified by integrals with rational integrands. To understand their analytic behavior, it is crucial to turn their integral definition to integrals in some complex projective space $\mathbb{CP}^d$. Here the complexification arises from the need of studying analytic continuations. The use of projective space ensures that the integral is defined on a compact domain, so as to avoid issues regarding singularities of the integrand at ``infinity'' (this can frequently occur for integrals defined on an affine space such as $\mathbb{R}^d$ or $\mathbb{C}^d$).  Such functions can appear in a wide variety of physical observables, ranging from scattering amplitudes, to energy correlators, and to cosmological correlators, etc.  In this paper, we primarily focus on the context of perturbative scattering amplitudes, especially at one loop, where such functions arise from the Feynman parameter representation of loop integrals.

\subsection{Integrals in complex projective space}\label{sec:intincp}

\brief{Properties of $\mathbb{CP}^n$}

Let us begin with a quick review of the complex projective space and integrals on it, to set up the context and conventions for later discussions. The complex projective space $\mathbb{CP}^{d}$ can be defined from $\mathbb{C}^{d+1}_*=\mathbb{C}^{d+1}\backslash\{(0,0,\ldots,0)\}$ by modding out the overall scale of its coordinates. Hence a point $X\in\mathbb{CP}^d$ can be represented by an $(d+1)$-tuple
\begin{align}
	X=[x_0:x_1:\cdots:x_d]\in\mathbb{C}^{d+1}_*,
\end{align}
under the equivalence relation
\begin{align}\label{eq:CPEquivalence}
	\left[x_0:x_1:\cdots:x_{d}\right]\sim\left[\lambda x_0:\lambda x_1:\cdots:\lambda x_{d}\right],\qquad\forall \lambda\neq0.
\end{align}
These are called homogeneous coordinates. When a coordinate of $X$ is non-zero, say $x_0$, we can use the above relation to set it to $x_0=1$. Then the remaining coordinates parameterize an affine complex space $\mathbb{C}^d$
\begin{align}\label{eq:AffinePatch}
	[1:x_1:x_2:\cdots:x_d]\in\mathbb{C}^d.
\end{align}
This is a genuine subspace of $\mathbb{CP}^d$ and is usually called its affine patch. The remaining points in $\mathbb{CP}^d$ necessarily have $x_0=0$. In this case, the other coordinates $\{x_1,x_2,\ldots,x_n\}$ cannot be simultaneously zero and still obey the equivalence relation \eqref{eq:CPEquivalence}. Therefore, they make up a complex projective space of one lower dimension
\begin{align}\label{eq:coordinatesAtInfinity}
	[0:x_1:x_2:\cdots:x_d]\in\mathbb{CP}^{d-1},
\end{align}
and the full $\mathbb{CP}^{d}$ is the disjoint union of these two subspaces
\begin{align}
	\mathbb{CP}^{d}=\mathbb{C}^d\sqcup\mathbb{CP}^{d-1},
\end{align}
where the latter can be viewed as the ``infinity'' in relation to the former. In fact, in many applications when we rewrite an integral in $\mathbb{C}^d$ into its counterpart in $\mathbb{CP}^{d}$, the specific operation in need is exactly to add in this infinity region $\mathbb{CP}^{d-1}$.
	
\brief{Integral on $\mathbb{CP}^d$}

An integral defined on $\mathbb{CP}^d$ can in general be expressed as
\begin{align}\label{eq:IntegralGeneral}
	\int_{C}\,\langle X\mathrm{d}X^d\rangle{\mathcal{F}(X)\over \mathcal{G}(X)},
\end{align}
where $C$ denotes the integral contour. $\langle X\mathrm{d}X^d\rangle$ is the standard volume element on $\mathbb{CP}^d$
\begin{align}
	\langle X \mathrm{d}X^d\rangle\equiv{1\over d!}\epsilon_{I_0I_1\cdots I_d}X^{I_0}\mathrm{d}X^{I_1}\wedge\mathrm{d}X^{I_2}\wedge\cdots\wedge \mathrm{d}X^{I_d},
\end{align}
where $\epsilon$ is the Levi-Civita symbol. $\mathcal{F}$ and $\mathcal{G}$ are homogeneous polynomials in $X$ with degree $f$ and $g$ respectively, satisfying the relation $g=f+d+1$. This relation originates from the invariance under the equivalence transformation $X\rightarrow \lambda X$.

\subsection{Projective automorphism of $\mathbb{CP}^d$ and the invariance of integrals}\label{sec:Projauto}

\brief{Automorphism of $\mathbb{CP}^n$}

$\mathbb{CP}^d$ is known to enjoy a $\mathrm{PGL}(d+1)$ projective automorphism, under which any integral defined on $\mathbb{CP}^{d}$ stays invariant. To understand this automorphism, one can choose an arbitrary set of $d+1$ reference points $V_i$ ($i=0,1,\ldots,d$), which are not all co-planar. Let us assume their coordinates to be
\begin{align}
	V_i=[v_{i0}:v_{i1}:\cdots:v_{id}],\qquad i=0,1,\ldots,d.
\end{align}
Then any point $V=[x_0:x_1:\cdots:x_d]\in\mathbb{CP}^d$ can be expressed as a linear combination
\begin{align}\label{eq:VasSumVi}
	V=\sum_{i=0}^dy_iV_i,
\end{align}
since $\{V_i\}$ naturally form a basis for the homogeneous coordinates. Note that when we simultaneously rescale all reference points $V_i\mapsto\lambda V_i$, RHS of \eqref{eq:VasSumVi} also undergoes the same rescaling, which is consistent with \eqref{eq:CPEquivalence}. Such rescaling can alternatively be viewed as acting on the coefficients
\begin{align}\label{eq:VRescaling}
	V\mapsto\lambda V=\sum_{i=0}^dy_i(\lambda V_i)=\sum_{i=0}^d(\lambda y_i)V_i.
\end{align}
This means that $[y_0:y_1:\cdots:y_d]$ form a new set of homogeneous coordinates for the same $\mathbb{CP}^d$. The original and the new coordinate systems are related by the general linear transformation
\begin{align}\label{eq:x2yv}
	x_i=\sum_{j=0}^dy_jv_{ji},\qquad i=0,1,\ldots,d.
\end{align}
The non-planarity of $V_i's$ guarantees that the above transformation is one-to-one. And the equivalence rescaling of $v_{ji}$ \eqref{eq:VRescaling} explains the projective nature of $\mathrm{PGL}(d+1)$ \footnote{The coordinate transformation \eqref{eq:x2yv} forms a $\mathrm{GL}(d+1)$. However, consider the transformation matrix $\lambda I$, where $I$ is the unit, the integral is totally the same as before because of the equivalent relation. Therefore, the actual transformation group should be $\mathrm{PGL}(d+1)=\mathrm{GL}(d+1)/\mathrm{Z}(\mathrm{GL}(d+1))$. $\mathrm{Z}(\mathrm{GL}(d+1))=\{\lambda I_{d+1}\}$ is called the center of $\mathrm{GL}(d+1)$.}.  On the other hand, if we choose different representative homogeneous coordinates of only one point in the basis, say $V_0$, even though geometrically this point remains the same, the elements $v_{0i}$ are rescaled but not the full transformation, hence this leads to a different $\mathrm{PGL}(d+1)$ transformation. So we conclude that any $d+1$ reference points together with their representative homogeneous coordinates determine an automorphism on $\mathbb{CP}^d$.

\brief{Invariance of integrals under automorphisms}

When we consider the above transformation in an integral, both the contour and the integrand are subject to change in general. If we abbreviate \eqref{eq:x2yv} as the matrix multiplication $X=\mathbf{M}Y$, where $\mathbf{M}$ is the transformation matrix, $\mathbf{M}_{ij}=v_{ji}$, then the volume elements are related by
\begin{align}
	\langle X\mathrm{d}X^d\rangle=\det(\mathbf{M})\,\langle Y\mathrm{d}Y^d\rangle,
\end{align}
and the homogeneous polynomials change to
\begin{align}
	\mathcal{F}(X)=\mathcal{F}(\mathbf{M}Y)\equiv \widetilde{\mathcal{F}}(Y),
\end{align}
(and similarly for $\mathcal{G}(X)$). Furthermore, we also need to use \eqref{eq:x2yv} to map every point in the contour $C$ to its image, which together form some other contour $\widetilde{C}$. Then the invariance of the integral under such an automorphism explicitly means
\begin{align}\label{eq:IntegralPGLInvariance}
	\int_C\langle X\mathrm{d}X^d\rangle\,\frac{\mathcal{F}(X)}{\mathcal{G}(X)}=\int_{\widetilde{C}}\langle Y\mathrm{d}Y^d\rangle\,\frac{\det(\mathbf{M})\,\widetilde{\mathcal{F}}(Y)}{\widetilde{\mathcal{G}}(Y)}.
\end{align}
Due to the above invariance, we treat integrals related by $\mathrm{PGL}(d+1)$ automorphisms as essentially the same integral, but they are expressed in different frames (of $\mathbb{CP}^d$).

\subsection{A simple example: dilogarithm}\label{sec:Li2}

It should be helpful to illustrate the above general discussions by a simple example. 

\subsubsection{Chen integral representation of $\mathrm{Li}_2(z)$}

Because the functions from Feynman integrals under our current scope all belong to the multiple polylogarithms, we can focus on the simplest yet non-trivial function of this kind, the dilogarithm $\mathrm{Li}_2(z)$. In the context of Chen's iterated integrals \cite{Chen:1977oja} it can be defined as
\begin{align}\label{eq:Li2ChenIntegral}
	\mathrm{Li}_2(z)&=-\int_0^z\frac{\mathrm{d}x_2}{x_2}\log(1-x_2)=\int_0^z\frac{\mathrm{d}x_2}{x_2}\int_0^{x_2}\frac{\mathrm{d}x_1}{1-x_1}.
\end{align}
For the time being, let us assume that both $z$ and the integration variables are real; then on $\mathbb{R}^2$ the contour in the above integral is clearly a triangle (in other words a $2$-simplex), whose three vertices locate at
\begin{align}
	v_0=(0,0),\qquad v_1=(z,z),\qquad v_2=(0,z).
\end{align}
This means any point $(x_1,x_2)$ inside the contour can be expressed as
\begin{align}\label{eq:barycentric}
	(x_1,x_2)=t_0v_0+t_1v_1+t_2v_2,\qquad t_i\geq0\text{ and }t_0+t_1+t_2=1.
\end{align}
On $\mathbb{R}^2$ the integrand is singular at two lines, specified by equations $x_2=0$ and $1-x_1=0$ respectively. So the integral is well-defined as long as $z\leq1$.\footnote{Note that $v_0=(0,0)$ is on the line $x_2=0$, which appears to make the integral ill-defined. However, considering the local coordinate around $v_0$ and performing the transformation $x_1=t y_1, x_2=t y_2$, the integrand scales like ${t \mathrm{d}y_1\mathrm{d}y_2}/{y_1(1-t y_2)}$, and it vanishes as $t\to0$. Hence this is not an issue.}

\begin{figure}[ht]
	\centering
	\begin{tikzpicture}
		\begin{scope}
			\draw [-Latex,Gray] (-.6,0) -- (3,0) node [right] {\small $x_1$};
			\draw [-Latex,Gray] (0,-.6) -- (0,2.6) node [above] {\small $x_2$};
			\coordinate [label=-45:{\small $v_0$}] (v0) at (0,0);
			\coordinate [label=-60:{\small $v_1$}] (v1) at (1,1);
			\coordinate [label=45:{\small $v_2$}] (v2) at (0,1);
			\fill [Gray!20,draw=black] (v0) -- (v1) -- (v2) -- cycle;
			\draw [thick,dashed] (-.4,0) -- (2.4,0) (1.5,-.4) -- (1.5,2.4);
			\foreach \i in {0,1,2} \fill [black] (v\i) circle [radius=1.5pt];
			\node [anchor=east] at (v2) {\small $z$};
			\node [anchor=north west] at (1.5,0) {\small $1$};
			\node [anchor=north] at (1,-.8) {\small (1) $z\leq 1$};
		\end{scope}
    	\begin{scope}[xshift=6cm]
			\draw [-Latex,Gray] (-.6,0) -- (3,0) node [right] {\small $x_1$};
			\draw [-Latex,Gray] (0,-.6) -- (0,2.6) node [above] {\small $x_2$};
			\coordinate [label=-45:{\small $v_0$}] (v0) at (0,0);
			\coordinate [label=-60:{\small $v_1$}] (v1) at (2,2);
			\coordinate [label=45:{\small $v_2$}] (v2) at (0,2);
			\fill [Gray!20,draw=black] (v0) -- (v1) -- (v2) -- cycle;
			\draw [thick,dashed] (-.4,0) -- (2.4,0) (1.5,-.4) -- (1.5,2.4);
			\foreach \i in {0,1,2} \fill [black] (v\i) circle [radius=1.5pt];
			\node [anchor=east] at (v2) {\small $z$};
			\node [anchor=north west] at (1.5,0) {\small $1$};
			\node [anchor=north] at (1,-.8) {\small (2) $z>1$};
		\end{scope}
	\end{tikzpicture}
	\caption{Iterated integral in $\mathbb{R}^2$ for $\mathrm{Li}_2(z)$. (1) The integral is well-defined when $z<1$; (2) The contour needs to deform in the imaginary direction in the neighborhood of $x_1=1$ when $z\geq 1$.}\label{fig:Li2}
\end{figure}
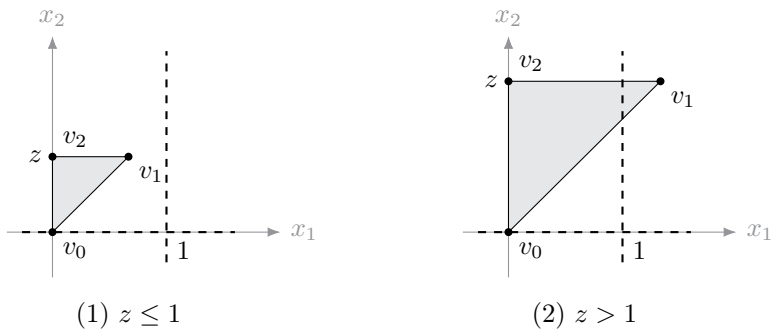

In contrast, as $z>1$ the contour starts to have overlap with the line $1-x_1=0$, see Figure \ref{fig:Li2}. In order to define the integral in such situation, one has to let the contour pass around the singularity by deforming in the imaginary direction. This requires us to extend our scope to the complex space $\mathbb{C}^2$. The two inequivalent ways of deformation give rise to two different integrals, which is responsible for the branch cut of $\mathrm{Li}_2(z)$ at $z>1$.

It is worth to note that as we analytically continue into complex space, the contour still maintains two real dimensional. As long as we do not encounter any singularities, an easy way to specify the contour is to first identify the location of its three vertices (even for complex values of $z$, e.g., $z=1+i$), and then specify the contour as the set of points following \eqref{eq:barycentric}, where the parameters $t_i$ are still kept \emph{real}. Of course, this is just one candidate in the equivalent class of contours related by continuous deformations. A more precise characterization of the contour will be discussed later in Section \ref{sec:simplexContour}.

\subsubsection{$\mathrm{Li}_2(z)$ as an integral on $\mathbb{CP}^2$}

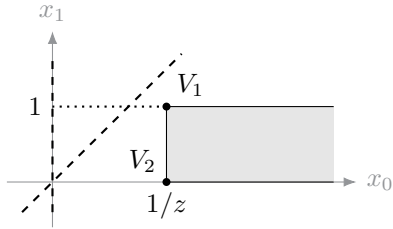
\begin{figure}[ht]
	\centering
	\begin{tikzpicture}
		\draw [-Latex,Gray] (-.6,0) -- (4,0) node [right] {\small $x_0$};
		\draw [-Latex,Gray] (0,-.6) -- (0,2) node [above] {\small $x_1$};
		\coordinate [label=20:{\small $V_1$}] (v1) at (1.5,1);
		\coordinate [label=150:{\small $V_2$}] (v2) at (1.5,0);
		\fill [gray!20] (v1) -- (v2) -- ++(2.2,0) -- ++(0,1) -- cycle;
		\draw [black] (3.7,1) -- (v1) -- (v2) -- ++(2.2,0);
		\draw [dotted,thick] (v1) -- ++(-1.5,0) node [left] {\small $1$};
		\draw [thick,dashed] (0,-.4,0) -- (0,1.7) (-.4,-.4) -- (1.7,1.7);
		\foreach \i in {1,2} \fill [black] (v\i) circle [radius=1.5pt];
		\node [anchor=north] at (v2) {\small $1/z$};
	\end{tikzpicture}
	\caption{Alternative integral for $\mathrm{Li}_2(z)$, induced by a different choice of affine patch.}
	\label{fig:Li2Alt}
\end{figure}

The branch cut of $\mathrm{Li}_2(z)$ is anchored at two branch points, $z=1$ and $z=\infty$, and the above analysis clearly explains the existence of the former. The branch point at $z=\infty$ also comes for a similar reason, but due to singularity of the integrand at infinity. To make this manifest we should put this integral in a projective space. Specifically, we first map the coordinates $(x_1,x_2)\mapsto [1:x_1:x_2]$, treating $\mathbb{C}^2$ as an affine patch of $\mathbb{CP}^2$, \eqref{eq:AffinePatch}. So the three vertices of the contour viewed in $\mathbb{CP}^2$ are
\begin{align}\label{eq:Li2V}
	V_0=[1:0:0],\qquad V_1=[1:z:z],\qquad V_2=[1:0:z].
\end{align}
To turn on the full homogeneous coordinates of the integration variables, we redefine $x_i\mapsto x_i/x_0$ and $\mathrm{d}x_i\mapsto \mathrm{d}x_i/x_0$ ($i=1,2$), where we temporarily treat $x_0$ as a non-zero constant, so that the coordinates become $[1:x_1:x_2]\mapsto[1:x_1/x_0:x_2/x_0]\sim[x_0:x_1:x_2]$. By replacing the volume element to the canonical one
\begin{align}
	\mathrm{d}x_1\wedge\mathrm{d}x_2\mapsto\frac{1}{x_0}\langle X\mathrm{d}X^2\rangle,
\end{align}
we obtain a representation of $\mathrm{Li}_2(z)$ as an integral genuinely defined on $\mathbb{CP}^2$
\begin{align}\label{eq:Li2ProjectiveIntegral}
	\mathrm{Li}_2(z)=\int_\triangle\frac{\langle X\mathrm{d}X^2\rangle}{x_0x_2(x_0-x_1)}.
\end{align}
Here $\triangle$ denotes the simplex contour determined by the vertices \eqref{eq:Li2V} via
\begin{align}
	X=t_0V_0+t_1V_1+t_2V_2,\qquad t_i\geq0.
\end{align}
Here we no longer impose the constraint $t_0+t_1+t_2=1$, due to the equivalence of homogeneous coordinates under rescaling. In the form of \eqref{eq:Li2ProjectiveIntegral}, we explicitly observe the singularities of the integrand at infinity, which is defined by $x_0=0$. Viewed in $\mathbb{CP}^2$, this infinity is nothing special compared to an ordinary hyperplane. In particular, when we perform the integration we can alternatively choose some other affine patch. For instance, we can choose the one specified by $x_2=1$, in which case the integral becomes
\begin{align}\label{eq:Li2ProjectiveIntegralAlt}
	\mathrm{Li}_2(z)=\int_{1/z}^{\infty}\mathrm{d}x_0\int_0^1\mathrm{d}x_1\,\frac{1}{x_0(x_0-x_1)},
\end{align}
and the infinity relative to this patch is defined by $x_2=0$ instead. From \eqref{eq:Li2V} we see that one of the contour vertices, $V_0$, is now located at the infinity. The integration region in \eqref{eq:Li2ProjectiveIntegralAlt} can be easily read off by keeping track of the vertices, as depicted in Figure \ref{fig:Li2Alt}. We see this new integral no longer has an iteration structure. 

In general, there exist an infinite number of integrals that lead to the same function, and the use of $\mathbb{CP}^d$ is a convenient way to unify a large class of them. This indicates that, despite the different appearance in various integrals, it is the \emph{geometry} behind the expressions that has an invariant meaning, and is ultimately responsible for the properties of the final function. 

\subsubsection{Integral in the canonical frame}\label{sec:CanonicalFrame}

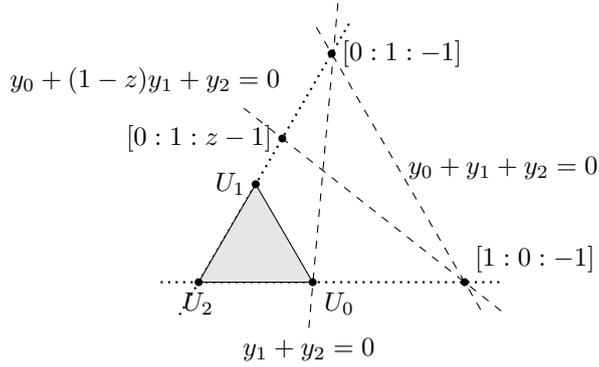
\begin{figure}[ht]
	\centering
	\begin{tikzpicture}
	    \def\R{1.5};
		\draw [dotted,thick] (-120:.5) -- (60:4) (180:.5) -- (0:4);
		\coordinate (u0) at (0:\R);
		\coordinate (u1) at (60:\R);
		\coordinate (u2) at (0,0);
		\draw [fill=gray!20] (u2) node [below] {\small $U_2$} -- (u0) node [below] {\small $\qquad U_0$} -- (u1) node [left] {\small $U_1$} -- cycle;
		\coordinate [label=60:{\small $[1:0:-1]$}] (p0) at (0:3.5);
		\coordinate [label=0:{\small $[0:1:-1]$}] (p1) at (60:3.5);
		\coordinate [label=180:{\small $[0:1:z-1]$}] (p2) at (60:2.2);
		\draw [dashed] ($(p0)!-.12!(p1)$) -- ($(p1)!-.12!(p0)$) ($(0:1.5)!-.2!(p1)$) node [below] {\small $y_1+y_2=0$} -- ($(p1)!-.2!(-30:1.5)$) ($(p0)!-.2!(p2)$) -- ($(p2)!-.24!(p0)$) node [above] {\small $y_0+(1-z)y_1+y_2=0\qquad\qquad\qquad\quad$};
		\node [anchor=west] at ($(p0)!.5!(p1)$) {\small $y_0+y_1+y_2=0$};
		\foreach \i in {0,1,2} \fill (p\i) circle [radius=1.5pt];
		\foreach \i in {0,1,2} \fill (u\i) circle [radius=1.5pt];
	\end{tikzpicture}
	\caption{Projective integral for $\mathrm{Li}_2(z)$ in the canonical frame. Dashed lines denote singularities of the integrand.}
	\label{fig:Li2CP2}
\end{figure}

Note that in \eqref{eq:Li2ProjectiveIntegral} the integrand singularities are completely fixed, while the shape of the contour depends on the variable $z$ via \eqref{eq:Li2V}. In practice we find it more convenient to switch to another frame such that the contour is fixed to be the \emph{canonical simplex}, whose vertices are specified by
\begin{align}
	U_i=[\underbrace{0:\cdots:0}_{i}:1:\underbrace{0:\cdots:0}_{d-i}],\qquad i=0,1,\dots,d.
\end{align}
This can always be achieved by a $\mathrm{PGL}(d+1)$ transformation such that $V_i\mapsto U_i$, following \eqref{eq:IntegralPGLInvariance}. In the explicit case of \eqref{eq:Li2ProjectiveIntegral}, we use the set of vertices of the original contour, \eqref{eq:Li2V}, as a basis to expand the homogeneous coordinates of an arbitrary point
\begin{align}
	X=y_0V_0+y_1V_1+y_2V_2,\qquad Y\equiv[y_0:y_1:y_2]\in\mathbb{CP}^2.
\end{align}
In this way the integral representation of $\mathrm{Li}_2(z)$ is transformed to
\begin{align}\label{eq:Li2Canonical}
	\mathrm{Li}_2(z)=\int_{\triangledown}\frac{z\,\langle Y\mathrm{d}Y^2\rangle}{(y_0+y_1+y_2)(y_1+y_2)(y_0+(1-z)y_1+y_2)}.
\end{align}
Throughout this paper we use $\triangledown$ to indicate the canonical simplex in $\mathbb{CP}^d$. The contour and the integrand singularities of this integral is illustrated in Figure \ref{fig:Li2CP2}. Whenever an integral has its contour being $\triangledown$, we say this integral is expressed in the \emph{canonical frame}. Such representation will be a convenient starting point for our later analysis.

\subsection{The Feynman parameter representation in complex projective space}

With the general setup in the previous discussions, we can readily write the Feynman parameter representation of any diagram as integrals on $\mathbb{CP}^{d}$, where $d+1$ corresponds to the number of loop propagators. Coordinates for $\mathbb{CP}^{d}$ is exactly made up by the $d+1$ Feynman parameters, i.e., $X=[x_0:x_1:\cdots:x_{d}]$. For scalar diagrams this representation is then expressed as the following integral in the canonical frame
\begin{equation}\label{eq:FeynmanParameterGeneric}
	\int_{\triangledown}\langle X\mathrm{d}X^{d}\rangle{\mathcal{U}^{a+(L+1)D/2}\prod_{i}x_i^{a_i-1}\over{(-\mathcal{V}+\mathcal{U}\sum_im_i^2x_i)^{a-LD/2}}}.
\end{equation}
Here $a_i$ denotes the power of the $i$-th propagator, $L$ the number of loops, and $D$ the spacetime dimensions. $\mathcal{U}$ and $\mathcal{V}$ are the standard Symanzik polynomials \footnote{Here $T^1$ is the set of all possible trees obtained by cutting internal legs and $T^2$ the set of all possible pairs of disjoint trees obtained by cutting the same diagram. $q^T$ denotes the total momentum flowing from one side to the other side of the disjoint diagram.}
\begin{equation}
	\mathcal{U}(x)=\sum_{T\in T^1}\prod_{i\notin T}x_i,\qquad \mathcal{V}(x)=\sum_{T\in T^2}(q^T)^2\prod_{i\notin T}x_i.
\end{equation}
Both polynomials are homogeneous, and it is easy to verify the weight of $X$ between the numerator and the denominator is balanced. 

The derivation of \eqref{eq:FeynmanParameterGeneric} starts with the more familiar form of Feynman parameter representation in real space
\begin{equation}\label{eq:feynparaint}
	\int_{0}^\infty\mathrm{d}x_0\cdots\int_{0}^\infty\mathrm{d}x_{d}\delta\left(1-\sum_{i=0}^{d}x_i\right)\prod_{i=0}^{d}x_i^{a_i-1}{\mathcal{U}^{a+(L+1)d/2}\over{(-\mathcal{V}+\mathcal{U}\sum_im_i^2x_i)^{a-Ld/2}}}.
\end{equation}
According to the Cheng-Wu theorem \cite{Cheng:1987}, the integral remains the same if the $\delta$ is replaced by $\delta\left(1-\sum_S x_i\right)$, where $S$ is a non-empty subset of $x_i$. Consequently, the integral should be invariant if one sets any of $x_i=1$, which is definitely the property of integrals in the complex projective space, so that \eqref{eq:FeynmanParameterGeneric} is directly obtained by replacing the measure and the delta function by the standard $\mathbb{CP}^{d}$ volume element
\begin{equation}
	\mathrm{d}x_0\cdots\mathrm{d}x_{d}\delta\left(1-\sum_{i=0}^dx_i\right)\longrightarrow\langle X\mathrm{d}X^{d}\rangle.
\end{equation}

In this paper, we mainly focus on one-loop integrals, which means the degree of $\mathcal{U}$ and $\mathcal{V}$ are separately $1$ and $2$. In this case the integrals take the schematic form
\begin{equation}\label{eq:OneLoopIntegralsGeneric}
	\int_{\triangledown}{\langle X\mathrm{d}X^{d}\rangle N[X^k]\over(XQX)^{{d+k+1}\over2}}
\end{equation}
where $N[X^k]$ denotes some homogeneous polynomial of degree $k$. Hence the integrand singularities always form a quadric. Because our method requires the integrand to be a rational function, we restrict out scope to Feynman integrals that are finite, hence no dimensional regularization is introduced.

\section{Touching configurations and emergence of integral singularities}\label{sec:touchingConfiguration}

In the discussion of $\mathrm{Li}_2(z)$ in Section \ref{sec:Li2} we observe that a logarithmic singularity of the integral emerges as a $0$-face of the contour simplex touches the integrand singularity. This is a natural generalization of one-dimensional integrals that computes $\log(z)$, e.g.,
\begin{align}
    \log(z)=\int_1^z\frac{\mathrm{d}x_1}{x_1}\equiv\int_{[1:1]}^{[1:z]}\frac{\langle X\mathrm{d}X\rangle}{x_0x_1},
\end{align}
where singularity at $z=0$ or $z=\infty$ develops as the $0$-face $[1:z]$ touches the line $x_1=0$ or $x_0=0$, respectively. Hence one may naturally expect that such configuration of $0$-faces of the contour simplex will continue to serve as a basic mechanism for the emergence of logarithmic singularities even for integrals in $\mathbb{CP}^{d>2}$. 

However, apparently this cannot be sufficient. Already in the integral \eqref{eq:Li2ProjectiveIntegral} we see that the $0$-face $V_0$ is always on top of an integrand singularity while the integral stays finite. As a more extreme example, we can consider a massless box with four massive corners (to simplify, let us assume that each corner has a massive external line with the same mass), whose Feynman parametrization is
\begin{align}\label{eq:masslessbox}
	I_{\text{massless box}}=\int_\triangledown\frac{\langle X\mathrm{d}X^3\rangle}{(s\,x_0x_2+t\,x_1x_3+m^2(x_0x_1+x_1x_2+x_2x_3+x_3x_0))^2}.
\end{align}
Here all the four $0$-faces of the contour reside on the quadric singularity, since the quadratic polynomial in the denominator is linear in any of the Feynman parameters (note that the contour is the canonical simplex). Therefore this integral cannot have singularities emerged in the above manner. On the other hand, it is known that the letters in the first entries of its symbol are $s$, $t$ and $m^2$, such that a logarithmic singularity is encountered as we set either of these variables to zero. Upon such a singularity, e.g., when $s=0$, the quadric in the denominator of the integrand reduces to
\begin{align}
	t\,x_1x_3+m^2(x_0x_1+x_1x_2+x_2x_3+x_3x_0),
\end{align}
which contains none of the monomials $x_0^2$, $x_0x_2$ and $x_2^2$. Hence in this case the quadric contains not only the 0-faces $U_0$ and $U_2$, but also even the entire 1-face joining these two $0$-faces.

The incidence relations observed in the above examples, both for $0$-faces and $1$-faces of the contour, belong to a broader class of configurations which we would like to call \emph{touching configurations}. Viewing in the canonical frame, the occurrence of a touching configuration typically requires certain constraints to be imposed on the parameters for the singularity hypersurface of the integrand, such as $s=0$ in the above example. In general, we can classify touching configurations according to the number of these constraints. A touching configuration is more singular when it imposes more constraints. An integral with a touching configuration does not necessarily mean the integral itself is ill-defined, such as the \eqref{eq:Li2ProjectiveIntegral} for $\mathrm{Li}_2(z)$. But it will become singular when a more singular touching configuration is reached by setting exactly one extra condition on the parameters of the integral. As already indicated in the above example, this one condition is responsible for the first entries of the symbol.

We propose that this notion of touching configuration is the key to a unified description for the emergence of all singularities of Feynman integrals. In this section we will make precise definition of the touching configurations, and classify their types with a focus on one-loop integrals.

\subsection{The contour simplex: its faces and their ambient spaces}\label{sec:simplexContour}

In order to draw a concrete definition of touching configurations and analyze their relations, we need to first clarify some structural properties of the contour simplex.

\begin{figure}[ht]
	\centering
	\begin{tikzpicture}
		\begin{scope}
			\draw [dotted,thick] (-2,0) -- (2,0) node [right] {\small $\mathbb{CP}^1$};
			\coordinate [label=90:{\small $V_0$}] (p0) at  (-1.2,0);
			\coordinate [label=90:{\small $V_1$}] (p1) at (1.2,0);
			\draw [thick] (p0) -- (p1);
			\foreach \i in {0,1} \fill (p\i) circle [radius=2pt];
			\node [anchor=north] at (0,-.6) {\small (1)};
		\end{scope}
		\begin{scope}[xshift=6cm]
            \coordinate [label=90:{\small $V_0$}] (p0) at  (-1.2,0);
			\coordinate [label=90:{\small $V_1$}] (p1) at (1.2,0);
			\draw [thick] (p0) .. controls +(30:.8) and +(-150:.8) .. (p1);
			\foreach \i in {0,1} \fill (p\i) circle [radius=2pt];
			\node [anchor=south west] at (2.2,.4) {\small $\mathbb{CP}^1$};
			\draw (2.2,.9) |- (3,.4);
			\node [anchor=north] at (0,-.6) {\small (2)};
		\end{scope}
	\end{tikzpicture}
	\caption{Two ways of illustrating the $1$-simplex contour and its $\mathbb{CP}^1$ ambient space on a piece of paper: (1) Treat the paper as an affine patch of $\mathbb{CP}^2$, so that $\mathbb{CP}^1$ looks like a line. We use solid line to represent the contour and dotted line to represent the ambient space. (2) Treat the paper as an affine pathch of the $\mathbb{CP}^1$ ambient space, so that deformation of the contour is manifest.}
	\label{fig:illustrateCP1}
\end{figure}
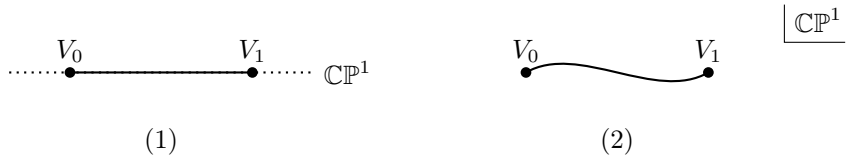

As already stated before, while we consider an integral in some $\mathbb{CP}^d$, its contour is a simplex of \emph{real} dimension $d$. The simplest non-trivial example is the case of $\mathbb{CP}^1$, as depicted in Figure \ref{fig:illustrateCP1}(2). 
Here the contour is a real curve joining its two end points $V_0$ and $V_1$. When these end points are fixed, the contour itself can still be continuously deformed as long as it does not hit any singularity of the integrand. Due to this, it is typically not sufficient to merely consider one specific contour. Instead one should consider the entire class of contours related by continuous deformations. Note that $V_0$ and $V_1$ can span the entire $\mathbb{CP}^1$ by taking linear combinations of their homogeneous coordinates (with complex-valued coefficients), and the contour can be deformed within this $\mathbb{CP}^1$. For this reason we denote this $\mathbb{CP}^1$ by $\overline{V_0V_1}$, and call it the \emph{ambient space} of the contour. We also denote the contour by $\underline{V_0V_1}$, in order to distinguish it from its ambient space. Both $\underline{V_0V_1}$ and $\overline{V_0V_1}$ have the same dimensionality, but the former is real while the latter complex. Sometimes when we illustrate both object in a higher-dimensional space, we may draw them on a same line like in Figure \ref{fig:illustrateCP1}(1), but it is important to keep in mind the difference between these two objects. Note that each of the end points can be viewed as a zero-dimensional simplex contour on its own, and in this case its ambient space $\overline{V_i}$ is nothing but the contour $\underline{V_i}$ itself, which is totally fixed. In this case it is fine that we do not distinguish them and denote both as $V_i$.

Things become more interesting as we move to higher dimensions. Take $\mathbb{CP}^2$ for instance. Here we can specify the contour 2-simplex $\underline{V_0V_1V_2}$ by fixing its three $0$-faces. The span $\overline{V_0V_1V_2}$ is the entire $\mathbb{CP}^2$ and provides the ambient space for $\underline{V_0V_1V_2}$. Moreover, the contour simplex $\underline{V_0V_1V_2}$ contains three 1-faces, $\underline{V_0V_1}$, $\underline{V_0V_2}$ and $\underline{V_1V_2}$, each of which is a 1-dimensional contour on its own. As we continuously deform $\underline{V_0V_1V_2}$, we need to make sure that these 1-faces are deformed within their own ambient spaces $\overline{V_0V_1}$, $\overline{V_0V_2}$ and $\overline{V_1V_2}$, respectively, which are $\mathbb{CP}^1$ subspaces of the $\mathbb{CP}^2$.

Generalizing this picture to an arbitrary $\mathbb{CP}^d$, in this space we can define a contour $d$-simplex $\underline{V_0V_1\cdots V_d}$ by fixing the $d+1$ $0$-faces. For any subset of labels $\{i_0,i_1,\ldots,i_k\}\subset\{0,1,\ldots,d\}$, the corresponding $0$-faces specify a $k$-face $\underline{V_{i_0}V_{i_1}\cdots V_{i_k}}$ of this contour ($k=0,1,\ldots,d$), whose ambient space is their span $\overline{V_{i_0}V_{i_1}\cdots V_{i_k}}$, which is a $\mathbb{CP}^k$ subspace of $\mathbb{CP}^d$. When we continuously deform the contour, every $k$-face can only deform within its own corresponding ambient space
\begin{align}
    \underbrace{\underline{V_{i_0}V_{i_1}\cdots V_{i_k}}}_{k\text{-simplex}}\subset\underbrace{\overline{V_{i_0}V_{i_1}\cdots V_{i_k}}}_{\mathbb{CP}^k}\subset\mathbb{CP}^d.
\end{align}
Note that among all the various ambient spaces, those with codimension 1 are very special, as all the other ambient spaces with lower dimensions can be obtained by their intersections. Each of such codim-1 ambient spaces is spanned by all but one of the $0$-faces, say $V_i$, and so for simplicity we can denote it by $B_i$. Then the above characterization of the contour $d$-simplex can be stated in a more mathematically rigorous way in terms of the relative homology $\mathrm{H}_d(\mathbb{CP}^d,\cup_iB_i)$.

When we write an integral in the canonical frame, the contour $d$-simplex is $\underline{U_0U_1\cdots U_d}$. In this case the ambient space $B_i$ is simply the hyperplane defined by $x_i=0$. Then following \eqref{eq:coordinatesAtInfinity} one can easily see that the original homogeneous coordinates naturally induce a coordinate frame $[x_0:\cdots:x_{i-1}:0:x_{i+1}:\cdots:x_d]$ upon this $\mathbb{CP}^{d-1}$ subspace, in which the corresponding $(d-1)$-face $\underline{U_0\cdots U_{i-1}U_{i+1}\cdots U_d}$ is again a canonical simplex. Such relation recursively holds for all lower-dimensional faces as well. Due to this simplicity, we will stick to the canonical frame as we discuss touching configurations in the following.

\subsection{Touching configurations}

Since a contour as well as its faces (except for 0-faces) can in general be continuously deformed, when analyzing potential singularities of an integral, it does not make much sense to look for configurations where a specific contour intersects the integrand singularity at some location. Alternatively, from the algebraic point of view, a singularity is typically reached by imposing a certain algebraic equation on the variables, such as setting $z=0$ in $\log(z)$. For a Feynman integral such equation should in principle describe relationships between the contour and the integrand, but it is very weird to mix things with real dimensions (from the contour) and things with complex dimensions (from the integrand) in the same equation. These are the basic motivations for introducing the ambient spaces for various faces of the contour simplex in the previous subsection.

	
With the notion of ambient spaces, we can define and classify these configurations in a quantitative manner.

{\bf Definition:} In $\mathbb{CP}^d$, a \emph{touching configuration} of a projective integral (with a simplex contour) is the configuration when the ambient spaces of one or more genuine faces of the contour simplex are contained in the singularity hypersurface of the integrand.

Without loss of generality, let us think about a $k$-face $\underline{V_0V_1\cdots V_k}$ ($k<d$), its ambient space $\overline{V_0V_1\cdots V_k}$ is a $\mathbb{CP}^k$ spanned by $V_0,V_1,\ldots,V_k$. Let us use $S$ to denote the integrand singularity hypersurface, i.e., the zero loci of $\mathcal{G}(X)$ in \eqref{eq:IntegralGeneral}, then the statement that $\overline{V_0V_1\cdots V_k}\subset S$ is equivalent to
\begin{align}
	\mathcal{G}(t_0V_0+t_1V_1+\cdots+t_kV_k)=0,\qquad\forall [t_0:t_1:\cdots:t_k]\in\mathbb{CP}^k.
\end{align}
In particular, when we work in the canonical frame, $V_i=U_i$, then the above statement simplifies to
\begin{align}\label{eq:touchingConditionCanonical}
	\mathcal{G}([x_0:x_1:\cdots:x_k:\underbrace{0:0:\cdots:0}_{d-k}])\equiv0,
\end{align}
or in other words, the polynomial $\mathcal{G}(X)$ is free of any monomials that are constructed solely with $\{x_0,x_1,\ldots,x_k\}$.

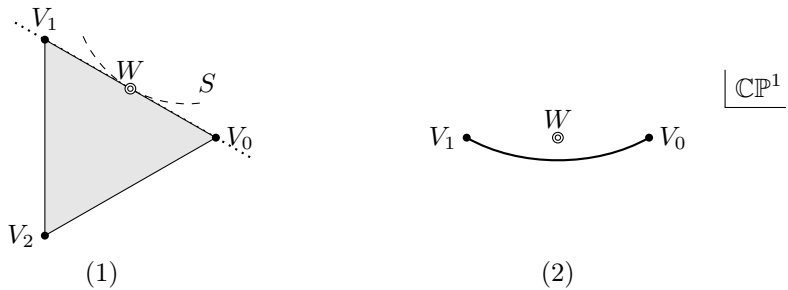
\begin{figure}[ht]\label{fig:tanconfig}
	\centering
	\begin{tikzpicture}
		\begin{scope}
			\coordinate [label=0:{\small $V_0$}] (v0) at (0:1.5);
			\coordinate [label=90:{\small $V_1$}] (v1) at (120:1.5);
			\coordinate [label=180:{\small $V_2$}] (v2) at (-120:1.5);
			\coordinate [label=90:{\small $W$}] (m) at ($(v0)!.5!(v1)$);
			\draw [dotted,thick] ($(v0)!-.2!(v1)$) -- ($(v1)!-.2!(v0)$);
			\fill [gray!20,draw=black] (v0) -- (v1) -- (v2) -- cycle;
			\draw [dashed] (m) arc [start angle=-120,end angle=-160,radius=1.5] (m) arc [start angle=-120,end angle=-80,radius=1.5] node [anchor=south] {\small $S$};
			\draw [fill=white] (m) circle [radius=2pt];
			\draw [fill=white] (m) circle [radius=1pt];
			\foreach \i in {0,1,2} \fill (v\i) circle [radius=1.5pt];
			\node [anchor=north] at (0,-1.5) {\small (1)};
		\end{scope}
		\begin{scope}[xshift=6cm]
			\coordinate [label=180:{\small $V_1$}] (p1) at  (-1.2,0);
			\coordinate [label=0:{\small $V_0$}] (p2) at (1.2,0);
			\draw [thick] (p1) .. controls +(-30:.8) and +(-150:.8) .. (p2);
			\foreach \i in {1,2} \fill (p\i) circle [radius=1.5pt];
			\foreach \r in {2,1} \draw [fill=white] (0,0) circle [radius=\r pt];
			\node [anchor=south] at (0,0) {\small $W$};
			\node [anchor=south west] at (2.2,.4) {\small $\mathbb{CP}^1$};
			\draw (2.2,.9) |- (3,.4);
			\node [anchor=north] at (0,-1.5) {\small (2)};
		\end{scope}
	\end{tikzpicture}
	\caption{(1) A configuration in $\mathbb{CP}^2$ that appears but does not yield any singularity. (2) Viewing inside $\overline{V_0V_1}$, $S$ only intersects it at a point $W$, and the contour can deform away from it.}
	\label{fig:tangent}
\end{figure}

Since $\underline{V_0V_1\cdots V_k}$ is fully restricted within $\overline{V_0V_1\cdots V_k}$, when $\overline{V_0V_1\cdots V_k}\subset S$ in the touching configuration, the $k$-face $\underline{V_0V_1\cdots V_k}$ has to live entirely on $S$ as well, no matter how it deforms. When we use figures on the real slice to illustrate geometries in $\mathbb{CP}^d$, sometimes it is very tempting to mistakenly think that $\underline{V_0V_1\cdots V_k}$ may touch $S$ by being ``tangent'' to it, such as the one presented in Figure \ref{fig:tangent}. However, it only makes sense to talk about tangency between $S$ and $\overline{V_0V_1\cdots V_k}$, since such relation should be characterized by some algebraic equation. In Figure \ref{fig:tangent}, $S$ only carves out a point in $\overline{V_0V_1}$, and the $1$-face $\underline{V_0V_1}$ can easily deform away from it. This is another evidence for the necessity of the notion of ambient space.

The only case where one can study the emergence of singularities in terms of the faces themselves is $k=0$, since a 0-face is its own ambient space. This case is the direct analogue of the Landau analysis in $\mathbb{CP}^1$, and is the source for the method of point projections introduced previously in \cite{Gong:2022erh}.

In the analysis of integrals, a crucial step is the quest for conditions that lead to touching configurations. First of all, it is useful to note that such conditions are not all independent of each other. For simplicity and concreteness, let us work in the canonical frame and assume that $S$ is irreducible. The degree $g$ of $S$ is determined by the degree of its defining polynomial $\mathcal{G}(X)$. For a $k$-face, e.g., $\underline{U_0U_1\cdots U_k}$, we would like to consider the extreme case when the ambient spaces of all its own genuine faces (excluding $\underline{U_0U_1\cdots U_k}$ itself) are already contained inside $S$. By \eqref{eq:touchingConditionCanonical} we learn that any monomial in $\mathcal{G}(X\in\overline{U_0U_1\ldots U_k})$ that does not contain the full set of variables $\{x_0,x_1,\ldots,x_k\}$ has to be absent. If there exists any such monomial, e.g., $x_0^{p_0}x_1^{p_1}\cdots x_l^{p_l}$, with $l<k$ and $\sum_{i=0}^lp_i=g$, then we have $\overline{U_0U_1\ldots U_l}\not\subset S$. This violates our assumptions. Since there are altogether $k+1$ variables related to this $k$-face, the situation divides into two cases:
\begin{itemize}
	\item When $k\geq g$, it means that the polynomial $\mathcal{G}(X\in\overline{V_0V_1\cdots V_k})=0$. Therefore $\overline{V_0V_1\cdots V_k}\subset S$ as well, which requires no further conditions.
	\item When $k<g$, then the polynomial $\mathcal{G}(X\in\overline{V_0V_1\cdots V_k})$ is non-vanishing, so that extra condition needs to be imposed in order to force $\overline{V_0V_1\cdots V_k}\subset S$. In particular, in the special case of $k=g-1$, we uniquely have
	\begin{align}
		\mathcal{G}(X\in\overline{V_0V_1\cdots V_k})\propto x_0x_1\cdots x_k.
	\end{align}
	Hence based on the assumptions above, the condition $\overline{V_0V_1\cdots V_k}\subset S$ is controlled by the single parameter in front of this unique monomial.
\end{itemize}
In consequence, regardless of the dimension of $\mathbb{CP}^d$, when the degree of $S$ is $g$, there is no need to study $k$-faces of the contour with $k\geq g$ in the enumeration of touching configurations. Due to this, in a specific integral, the set of touching configurations is not only finite, but in fact quite controlled.
When $S$ is reducible, it suffices to study each irreducible component at a time. Therefore, we can classify the problem of identifying touching configurations by the degree of such irreducible components. Up to one-loop Feynman integrals, this degree is at most two.

\subsection{Stratification of touching configurations for hyperplanes}\label{sec:touchingHyperplane}

Now let us work on explicit examples to illustrate the ideas introduced in the previous subsection. In this section we fix $g=1$, and so the integrand singularity $S$ is merely a hyperplane $H$. Let us denote the polynomial for this singularity as
\begin{align}
	\mathcal{G}(X)=HX\equiv \sum_{i=0}^dh_ix_i
\end{align}
in the canonical frame.

Let us first consider a $0$-face, e.g., $U_0$. Because in this case its ambient space is the face itself, and it is zero dimensional, there are only two possible touching configurations. The first is trivial, i.e., $U_0$ is at a generic position, and is off $H$. In this case the coefficients $h_i$ are completely free. The second is the configuration $U_0\subset H$, which is reached by setting one condition $h_0=0$.

\begin{figure}[ht]
	\centering
	\begin{tikzpicture}
		\begin{scope}
			\draw [dashed] (-1.5,0) -- (1.5,0) node [right] {\small $H$};
			\fill (0,.5) circle [radius=1.5pt] node [above] {\small $U_0$};
			\node [anchor=east] at (-2.5,.4) {\small Stratum 0:};
		\end{scope}
		\draw [-Stealth,thick] (0,-.4) -- (0,-1.2);
		\node [anchor=west] at (.2,-.8) {\small $h_0=0$};
		\begin{scope}[yshift=-2cm]
			\draw [dashed] (-1.5,0) -- (1.5,0) node [right] {\small $H$};
			\fill (0,0) circle [radius=1.5pt] node [above] {\small $U_0$};
			\node [anchor=east] at (-2.5,.2) {\small Stratum 1:};
		\end{scope}
	\end{tikzpicture}
	\caption{Stratifying touching configurations of a $0$-face in relation to a hyperplane singularity $H$.}
	\label{fig:strata0faceH}
\end{figure}
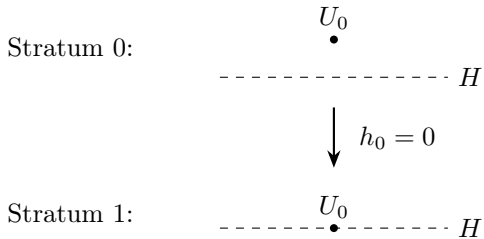

Here we observe that the touching configurations can be classified into different strata, according to the number of constraints imposed on the parameters in $\mathcal{G}(X)$. When no constraints are imposed, we always assume that all parameters are generically non-zero. In this case we have exactly two strata and one configuration type in each stratum, and the two strata are linked by exactly one constraint, as is illustrated in Figure \ref{fig:strata0faceH}.
Since we always stay in the canonical frame, analytic continuation of the integral only affects parameters in $\mathcal{G}(X)$, hence deforms $H$, while the contour is kept to be $\triangledown$. In this setup, the integral can potentially become singular when the configuration jumps from one stratum to another with one higher order. Figure \ref{fig:strata0faceH} provides the simplest scenario for this jump, and we will see more along with the discussion. Note that we do not consider any jumps between strata of order separation higher than 1, since that typically requires the simultaneous imposition of two or more conditions. In that case we can always resolve the situation by imposing the conditions one after another by deforming the way of analytic continuation.

\begin{figure}[ht]
	\centering
	\begin{tikzpicture}
		\begin{scope}
			\draw [dashed] (-1.5,0) -- (1.5,0) node [right] {\small $H$};
			\draw [dotted,thick] (-1,-.2) -- (1.5,.8) node [right] {\small $\overline{U_0U_1}$};
			\draw (0,.2) -- (1,.6);
			\fill (0,.2) circle [radius=1.5pt] node [above] {\small $U_0$};
			\fill (1,.6) circle [radius=1.5pt] node [above] {\small $U_1$};
			\node [anchor=east] at (-2.5,.5) {\small Stratum 0:};
		\end{scope}
		\draw [-Stealth,thick] (0,-.4) -- (0,-1.2);
		\node [anchor=west] at (.2,-.8) {\small $h_0=0$};
		\begin{scope}[yshift=-2.4cm]
			\draw [dashed] (-1.5,0) -- (1.5,0) node [right] {\small $H$};
			\draw [dotted,thick] (-1,-.2) -- (1.5,.8) node [right] {\small $\overline{U_0U_1}$};
			\draw (-.5,0) -- (.5,.4);
			\fill (-.5,0) circle [radius=1.5pt] node [above] {\small $U_0$};
			\fill (.5,.4) circle [radius=1.5pt] node [above] {\small $U_1$};
			\node [anchor=east] at (-2.5,.5) {\small Stratum 1:};
		\end{scope}
		\draw [-Stealth,thick] (0,-2.8) -- (0,-3.6);
		\node [anchor=west] at (.2,-3.2) {\small $h_1=0$};
		\begin{scope}[yshift=-4.4cm]
			\draw [dashed] (-1.5,0) -- (1.5,0) node [right] {\small $\overline{U_0U_1}\subset H$};
			\draw (-.5,0) -- (.5,0);
			\fill (-.5,0) circle [radius=1.5pt] node [above] {\small $U_0$};
			\fill (.5,0) circle [radius=1.5pt] node [above] {\small $U_1$};
			\node [anchor=east] at (-2.5,.2) {\small Stratum 2:};
		\end{scope}
	\end{tikzpicture}
	\caption{Stratifying touching configurations of a $1$-face in relation to a hyperplane singularity $H$.}
	\label{fig:strata1faceH}
\end{figure}
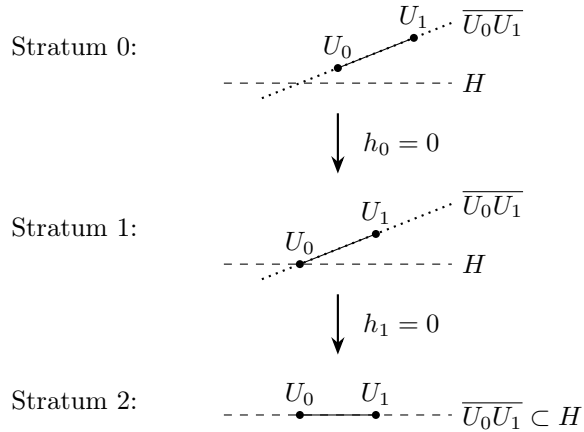

According to the previous subsection, when $g=1$ there is no need to consider touching configurations for $1$- and higher faces. It is also very simple to understand from pictures why this is true. Consider the $1$-face $\underline{U_0U_1}$. In Stratum 1 we can have $U_0\subset H$ by setting $h_0=0$, while leaving $U_1$ generic \footnote{Alternatively we can set $U_1\subset H$, but it is of the same configuration type, so we do not list it explicitly.}. There is also Stratum 2, where (based on Strata 1) we further let $U_1\subset H$. Being linear, in this situation we can view $H$ as the span of $U_0$, $U_1$ together with possibly other points, which means $\overline{U_0U_1}\subset H$, see Figure \ref{fig:strata1faceH}. Although there appears to be a new touching configuration, the jump from Stratum 1 to Stratum 2 is merely identical to that in the stratification problem of a $0$-face in Figure \ref{fig:strata0faceH}, but with the $0$-face being $U_1$ instead. Therefore all the jumps in this case are not new, given that Figure \ref{fig:strata0faceH} has been studied. This holds similarly for higher faces as well.
This example indicates that, for an integral whose integrand singularity $S$ only contains linear irreducible components, we only need stratifications of touching configurations for $0$-faces like in Figure \ref{fig:strata0faceH}, but this should be done for every $0$-face.

\subsection{Stratification of touching configurations for quadrics}\label{sec:touchingQuadric}

Next we move on to the case $g=2$, where the singularity hypersurface $S$ is a quadric $Q$, and we denote its corresponding polynomial as
\begin{align}
	\mathcal{G}(X)=Q(X)\equiv\sum_{0\leq i\leq j\leq d}q_{ij}x_ix_j
\end{align}
in the canonical frame.

In this case we do not need to independently discuss the stratification of touching configurations for $0$-faces, but directly study $1$-faces. This is because the former is included in the latter as a subset of the stratification.

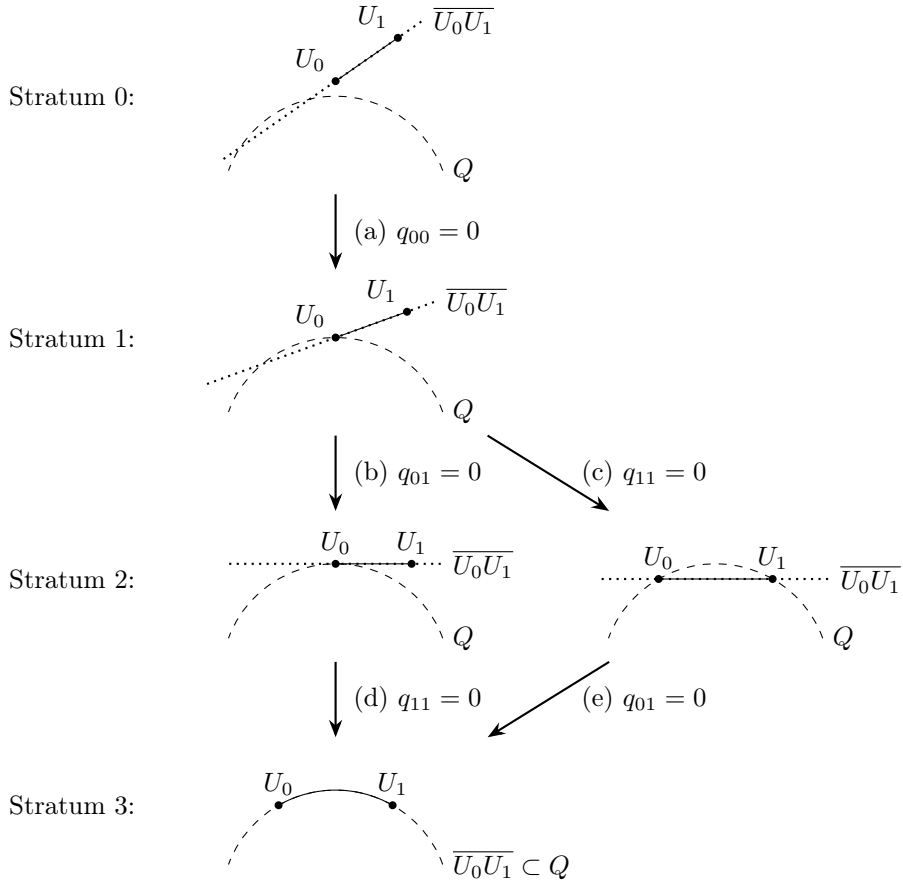
\begin{figure}[ht]
	\centering
	\begin{tikzpicture}
		\begin{scope}
		    \draw [dashed] (20:1.5) node [right] {\small $Q$} arc [start angle=20,end angle=160,radius=1.5];
			\draw [dotted,thick] ($(0,1.7)+(-145:1.8)$) -- ++(35:3.2) node [right] {\small $\overline{U_0U_1}$};
			\draw (0,1.7) -- +(35:1);
			\fill (0,1.7) circle [radius=1.5pt] node [anchor=south east] {\small $U_0$};
			\fill ($(0,1.7)+(35:1)$) circle [radius=1.5pt] node [anchor=south east] {\small $U_1$};
			\node [anchor=east] at (-2.5,1.5) {\small Stratum 0:};
		\end{scope}
		\draw [-Stealth,thick] (0,.2) -- (0,-.8);
		\node [anchor=west] at (.1,-.3) {\small (a) $q_{00}=0$};
		\begin{scope}[yshift=-3.2cm]
			\draw [dashed] (20:1.5) node [right] {\small $Q$} arc [start angle=20,end angle=160,radius=1.5];
	        \draw [dotted,thick] ($(0,1.5)+(-160:1.8)$) -- ++(20:3.2) node [right] {\small $\overline{U_0U_1}$};
			\draw (0,1.5) -- +(20:1);
			\fill (0,1.5) circle [radius=1.5pt] node [anchor=south east] {\small $U_0$};
			\fill ($(0,1.5)+(20:1)$) circle [radius=1.5pt] node [anchor=south east] {\small $U_1$};
			\node [anchor=east] at (-2.5,1.5) {\small Stratum 1:};
		\end{scope}
		\draw [-Stealth,thick] (0,-3) -- (0,-4);
		\node [anchor=west] at (.1,-3.5) {\small (b) $q_{01}=0$};
		\draw [-Stealth,thick] (2,-3) -- (3.6,-4);
		\node [anchor=west] at (3.1,-3.5) {\small (c) $q_{11}=0$};
		\begin{scope}[yshift=-6.2cm]
			\draw [dashed] (20:1.5) node [right] {\small $Q$} arc [start angle=20,end angle=160,radius=1.5];
			\draw [dotted,thick] (-1.4,1.5) -- (1.4,1.5) node [right] {\small $\overline{U_0U_1}$};
			\draw (0,1.5) -- +(0:1);
			\fill (0,1.5) circle [radius=1.5pt] node [above] {\small $U_0$};
			\fill (1,1.5) circle [radius=1.5pt] node [above] {\small $U_1$};
			\node [anchor=east] at (-2.5,1.3) {\small Stratum 2:};
		\end{scope}
		\draw [-Stealth,thick] (0,-6) -- (0,-7);
		\node [anchor=west] at (.1,-6.5) {\small (d) $q_{11}=0$};
		\draw [-Stealth,thick] (3.6,-6) -- (2,-7);
		\node [anchor=west] at (3.1,-6.5) {\small (e) $q_{01}=0$};
		\begin{scope}[xshift=5cm,yshift=-6.2cm]
			\draw [dashed] (20:1.5) node [right] {\small $Q$} arc [start angle=20,end angle=160,radius=1.5];
			\draw [dotted,thick] ($(120:1.5)!-.5!(60:1.5)$) -- ($(60:1.5)!-.5!(120:1.5)$) node [right] {\small $\overline{U_0U_1}$};
			\draw (120:1.5) -- (60:1.5);
			\fill (120:1.5) circle [radius=1.5pt] node [above] {\small $U_0$};
			\fill (60:1.5) circle [radius=1.5pt] node [above] {\small $U_1$};
		\end{scope}
		\begin{scope}[yshift=-9.2cm]
			\draw [dashed] (20:1.5) node [right] {\small $\overline{U_0U_1}\subset Q$} arc [start angle=20,end angle=160,radius=1.5];
			\draw (60:1.5) arc [start angle=60,end angle=120,radius=1.5];
			\fill (120:1.5) circle [radius=1.5pt] node [above] {\small $U_0$};
			\fill (60:1.5) circle [radius=1.5pt] node [above] {\small $U_1$};
			\node [anchor=east] at (-2.5,1.3) {\small Stratum 3:};
		\end{scope}
	\end{tikzpicture}
	\caption{Stratifying touching configurations of a $1$-face in relation to a quadric singularity $Q$.}
	\label{fig:strata1faceQ}
\end{figure}

First of all, note that when restricting to $\overline{U_0U_1}$ the quadratic polynomial generically reads
\begin{align}\label{eq:QU0U1}
	Q(X\in\overline{U_0U_1})=q_{00}x_0^2+q_{01}x_0x_1+q_{11}x_1^2.
\end{align}
There are in total three independent parameters, so one can expect that in this case the highest stratum has order 3, corresponding to all parameters being zero. The entire strata now have the structure shown in Figure \ref{fig:strata1faceQ}.
Here Stratum 0 still represents the trivial touching where all parameters are completely generic. Stratum 1 records touching configuration where one of the $0$-faces is contained in $Q$, e.g., $U_0\subset Q$ as depicted in Figure \ref{fig:strata1faceQ} (so correspondingly $q_{00}=0$). Based on Stratum 1 we can further build Stratum 2 by imposing one extra constraint. Here for the first time we observe two different types of touching configurations in a single stratum. In the former configuration we require $\overline{U_0U_1}$ to be tangent to $Q$ at $U_0$, which means $q_{00}=q_{01}=0$. This is still a case where only the zero-dimensional ambient space $U_0\subset Q$, but now it is contained ``twice'' as compared to that in Stratum 1, which is allowed by the degree of $Q$. In the latter configuration we require both $0$-faces to be contained in $Q$, and only once for each, so that $q_{00}=q_{11}=0$. This is exactly the case which we encounter in the previous massless box example \eqref{eq:masslessbox}. Lastly we also have Stratum 3, where all three parameters in \eqref{eq:QU0U1} are set to zero, so that we fully have $\overline{U_0U_1}\subset Q$.

Next we check the five possible jumps between different strata, which are labeled as in Figure \ref{fig:strata1faceQ}. Among them (a), (c) and (d) are in fact of the same type, because they are effectively equivalent to sending one $0$-face onto $Q$, i.e., the unique jump in the stratification for a $0$-face similar to the one in Figure \ref{fig:strata0faceH}. This can also be seen from the fact that the relevant parameters, $q_{00}$ or $q_{11}$, come from monomials of only one Feynman parameter. On the other hand, the jumps (b) and (e) are essentially tied to the behavior of the $1$-face $\underline{U_0U_1}$ and its ambient space.

\subsection{First symbol entries}\label{sec:firstEntries}

In the previous subsections we have investigated stratification of touching configurations and identified jumps among different strata. Now we move on to relate this picture to first entries of the symbol. Recall that a logarithmic singularity of the integral is produced whenever a first entry of its symbol is set to zero (or infinity). This is exactly one equation, which imposes one extra constraint on the geometric configuration of $S$. Due to this, it is very natural to associate the first entries with the jumps in the stratifications discussed previously.

Let us first check the case when $S$ is a hyperplane $H$, as described in Section \ref{sec:touchingHyperplane}. In this case the only relevant jump is depicted in Figure \ref{fig:strata0faceH}, and the condition for that is
\begin{align}
	H(U_0)\equiv h_0=0.
\end{align}
Correspondingly we take $h_0$ as one of the first entries. This has already been adopted in the first entry studies of Aomoto polylogarithms in \cite{aomoto_1982}.

When we move on to the quadric case discussed in Section \ref{sec:touchingQuadric}, all possible jumps are illustrated in Figure \ref{fig:strata1faceQ}. Here things become more interesting. Although all the three jumps (a), (c) and (d) are equivalent to sending a $0$-face onto the quadric, the way to read off first entries has to differ.
\begin{enumerate}
	\item[(a)] In the jump (a), because the condition $U_0\subset Q$ is tied to the algebraic equation $q_{00}=0$, it is very tempting to take $q_{00}$ as the corresponding first entry. This is in fact not quite correct. Note that in Stratum 0 $Q$ generically intersects $\overline{U_0U_1}$ at two distinguished points. When jumping to Stratum 1, in this ambient space $U_0$ has to coincide with either of these two points, and the two situations are related by analytic continuations, thus leading to a two-fold ambiguity. Due to this, in order to work out the first entries for this jump, we should restrict to $Q(X\in\overline{U_0U_1})$, \eqref{eq:QU0U1}, and work out its two zero loci
	\begin{align}
		[x_0:x_1]=[1:\frac{-q_{01}\pm\sqrt{q_{01}^2-4q_{00}q_{11}}}{2q_{11}}].
	\end{align}
	When the jump occurs, either of the two approaches $[x_0:x_1]=[1:0]$, and so from this jump we can in fact read off two first entries
	\begin{align}
		\frac{-q_{01}+\sqrt{q_{01}^2-4q_{00}q_{11}}}{2q_{11}}\quad\text{and}\quad\frac{-q_{01}-\sqrt{q_{01}^2-4q_{00}q_{11}}}{2q_{11}}.
	\end{align}
	As we will see in Section \ref{sec:symbolConstruction}, the factor $q_{11}$ in the denominator is not essential, since these two first entries always appear in terms of their ratio.
	\item[(c)(d)] In contrast to the jump (a), in these two cases the crucial difference is that we already have $U_0\subset Q$. Since $U_0$ and $U_1$ have to be distinct, this leaves no ambiguities as to how the jump occurs. This can also be explicitly seen in formulas. For the jump (c), in Stratum 1 we have
	\begin{align}\label{eq:QStratum1}
		Q(X\in\overline{U_0U_1})=x_1(q_{01}x_0+q_{11}x_1)
	\end{align}
	which factorizes. So this essentially reduces to the case of hyperplane singularities. For the jump (d), in Stratum 2 we have
	\begin{align}
		Q(X\in\overline{U_0U_1})=q_{11}x_1^2.
	\end{align}
	This can be viewed as a special case of \eqref{eq:QStratum1}, and we can only have $q_{11}=0$ for the jump to occur. Therefore, in both cases we directly identify $q_{11}$ as the corresponding first entry.
\end{enumerate}
In the above jumps it should be noted that, while we are effectively studying the stratification of touching configurations for $0$-faces, we are studying this in the context of a $1$-face $\underline{U_0U_1}$ and the $0$-face comes as its boundary. Consequently the first entries are read off in view of the ambient space $\overline{U_0U_1}$ by restricting $Q$ onto it. In $\mathbb{CP}^{d\geq 2}$ a $0$-face simultaneously belongs to several different $1$-faces. This means that when extracting the first entries for, say $U_0\subset Q$, we should study the above problem in relation to each $\overline{U_0U_i}$. In practice this may result in several different first entries, all of which effectively lead to the same geometric conditions and are needed for later analysis on the symbol.

Then for the two jumps essentially tied to $\overline{U_0U_1}$, the first entries are obtained as follows:
\begin{enumerate}
	\item[(b)] Because a quadric has degree 2, a point can not only belong to it, but also belong to it twice. The jump (b) describes the change in this incidence number. Since before the jump we already have $U_0\subset Q$, this change also occurs in a unique way, like the situation in jump (c). The configuration before the jump is again \eqref{eq:QStratum1}. $U_0\in Q$ twice after the jump requires that this polynomial is free of linear term in $x_0$, i.e., $q_{01}=0$. So for this jump we obtain a first entry $q_{01}$.
	\item [(e)] In this case both vertices of $\overline{U_0U_1}$ are already located on $Q$ and the configuration is
	\begin{align}
		Q(X\in\overline{U_0U_1})=q_{01}x_0x_1.
	\end{align}
	Like in (d), the configuration can become more singular only by requiring the above monomial to vanish entirely. So the corresponding first entry is $q_{01}$.
\end{enumerate}
In summary, we see that except for the jump (a) which contains a two-fold ambiguity, in all other situations the first entries are simply read off from relevant coefficients in the polynomial $Q(X\in\overline{U_0U_1})$.

\subsection{Elementary discontinuities}\label{sec:discontinuityFirstLook}

In the previous subsection we have provided a prescription to determine first entries of symbol. One may already ask the following questions: when a first entry $f$ is determined, why should it appear exactly as $f$, rather than the form $f^p$ with some other real number $p$? After all, as an algebraic equation the condition $f=0$ is the same as $f^p=0$ for any positive $p$, and as $f^p=\infty$ for any negative $p$.

The answer to this question has to do with the behavior of the integral under analytic continuation around $f=0$, hence tied to its discontinuity. In the simplest situation, imagine we plainly put $f^p$ inside a log, then we have the identity
\begin{align}\label{eq:logexample}
	\log(f^p)=p\,\log(f)=\int_0^\infty\frac{(f^p-1)\mathrm{d}t}{(t+1)(t+f^p)}.
\end{align}
Consider analytically continuing $f$ by setting $f=\epsilon\,e^{i\theta}$ for some fixed infinitesimal $\epsilon$ and increasing $\theta$ from $0$ to $2\pi$. On the one hand, the discontinuity of $\log(f^p)$ during the above continuation is $2\pi i\,p$, where $2\pi i$ is the discontinuity of $\log(f)$. On the other hand, in the above integral representation, as we continue $f$ the singularity point $t=-f^p$ moves around $t=0$, i.e., a $0$-face of the contour. Hence this singularity point pushes the contour along the way. In order that this singularity point comes back to its original position at the end of the analytic continuation, we see the power $p$ has to be an integer. This power also determines how many times the singularity point circles around $t=0$ during the process. Therefore, the discontinuity contour, i.e., the difference between the contours after and before the analytic continuation, is a closed contour wrapping around the integrand singularity $t=-f^p$, with a winding number $p$. This means that we can read off the power $p$ by inspecting the winding number of the discontinuity contour.

From the above example we see that once an integral representation for a function is given, the discontinuity of the function can be represented using the same integrand but with a different contour, and the new contour typically contains a part that merely computes residues of the integrand. This is a common theme of the integrals under our current study. Hence it is important to have a systematic way to construct these new contours for each existing discontinuity. While the rigorous procedure needs care and will be discussed in detail in the next two sections, it is helpful to first get some intuition on the possible patterns of the discontinuity contours.

Such intuition can be obtained by restricting the integral to the real slice of $\mathbb{CP}^d$. In doing this we also keep the parameters in the integrand real. In the example \eqref{eq:logexample} this means $f^p\in\mathbb{R}$. The resulting real integral is well-defined when $f^p>0$, so that $S$ (i.e., the poles) are outside of the integration region. To study the discontinuity around $f^p=0$, we continue this parameter along the real axis from $f^p>0$ to $f^p<0$. When this happens, the intersection between $S$ and the integral contour becomes non-empty, which is the pole $t=-f^p$. This creates an ambiguity, since the contour has to deform into imaginary direction whenever it hits the pole. The difference between the two choices of deformation exactly leads to the discontinuity. From this point of view, we see that a logarithmic singularity occurs whenever some part of $S$ starts to invade the inside of the real contour, and the corresponding discontinuity is tied to local properties in the neighborhood of their intersection.

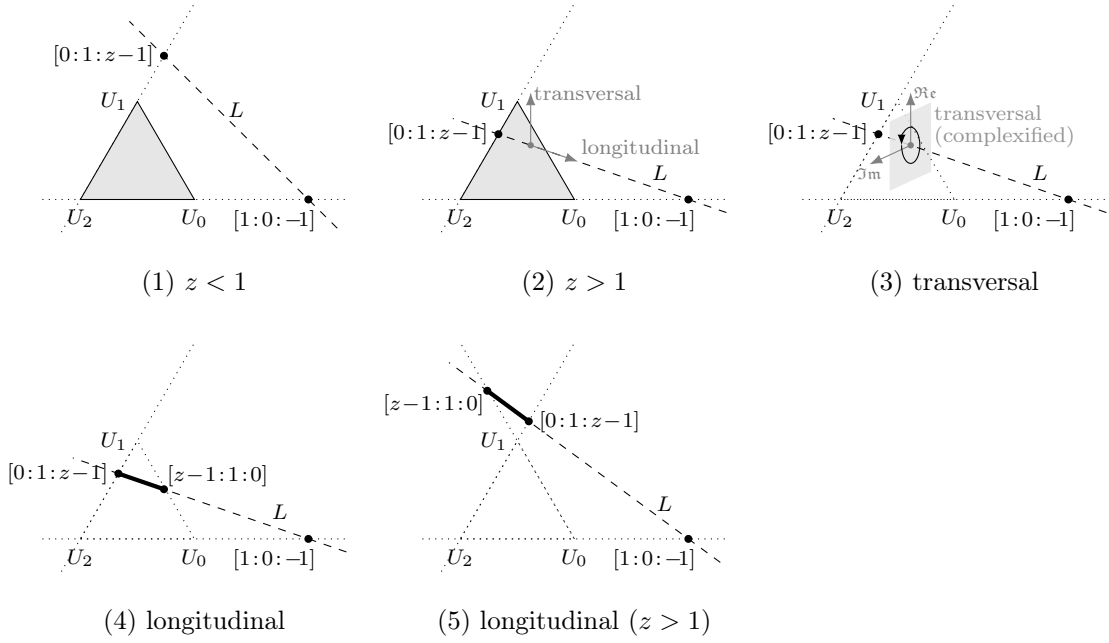
\begin{figure}[ht]
	\centering
	\begin{tikzpicture}
		\begin{scope}
			\draw [dotted] (-120:.5) -- (60:3) (180:.5) -- (0:3.5);
			\draw [fill=gray!20] (0,0) node [below] {\scriptsize $U_2$} -- (0:1.5) node [below] {\scriptsize $U_0$} -- (60:1.5) node [left] {\scriptsize $U_1$} -- cycle;
			\coordinate [label=-90:{\scriptsize $[1\!:\!0\!:\!-\!1]\qquad\quad$}] (p0) at (0:3);
			\coordinate [label=180:{\scriptsize $[0\!:\!1\!:\!z\!-\!1]$}] (p2) at (60:2.2);
			\draw [dashed] ($(p0)!-.2!(p2)$) -- ($(p2)!-.24!(p0)$);
			\foreach \i in {0,2} \fill (p\i) circle [radius=1.5pt];
			\node [anchor=south] at ($(p0)!.5!(p2)$) {\scriptsize $L$};
			\node [anchor=north] at (1.5,-.8) {\small (1) $z<1$};
		\end{scope}
		\begin{scope}[xshift=5cm]
			\draw [dotted] (-120:.5) -- (60:3) (180:.5) -- (0:3.5);
			\draw [fill=gray!20] (0,0) node [below] {\scriptsize $U_2$} -- (0:1.5) node [below] {\scriptsize $U_0$} -- (60:1.5) node [left] {\scriptsize $U_1$} -- cycle;
			\coordinate [label=-90:{\scriptsize $[1\!:\!0\!:\!-\!1]\qquad\quad$}] (p0) at (0:3);
			\coordinate [label=180:{\scriptsize $[0\!:\!1\!:\!z\!-\!1]$}] (p2) at (60:1);
			\draw [dashed] ($(p0)!-.2!(p2)$) -- ($(p2)!-.24!(p0)$);
			\coordinate (q) at ($(p0)!.83!(p2)$);
			\draw [gray,-Latex] (q) -- ($(p0)!.58!(p2)$) node [anchor=south west,inner sep=1pt] {\scriptsize longitudinal};
			\draw [gray,-Latex] (q) -- ++(0,.7) node [anchor=west,inner sep=1pt] {\scriptsize transversal};
			\fill [gray] (q) circle [radius=1.2pt];
			\foreach \i in {0,2} \fill (p\i) circle [radius=1.5pt];
			\node [anchor=south] at ($(p0)!.15!(p2)$) {\scriptsize $L$};
			\node [anchor=north] at (1.5,-.8) {\small (2) $z>1$};
		\end{scope}
		\begin{scope}[xshift=10cm]
			\draw [dotted] (-120:.5) -- (60:3) (180:.5) -- (0:3.5);
			\draw [dotted] (0:1.5) node [below] {\scriptsize $U_0$}  -- (0,0) node [below] {\scriptsize $U_2$} -- (60:1.5) node [left] {\scriptsize $U_1$} -- ($(60:1.5)!.25!(0:1.5)$);
			\coordinate [label=-90:{\scriptsize $[1\!:\!0\!:\!-\!1]\qquad\quad$}] (p0) at (0:3);
			\coordinate [label=180:{\scriptsize $[0\!:\!1\!:\!z\!-\!1]$}] (p2) at (60:1);
			\coordinate (q) at ($(p0)!.83!(p2)$);
			\draw [dashed] (q) -- ($(p2)!-.24!(p0)$);
			\fill [gray!20,opacity=.85] (30:.75) -- ++(90:.66) -- ++(25:.6) node [anchor=north west,inner sep=1pt,align=left] {\scriptsize\color{gray}transversal\\[-.6em]\scriptsize\color{gray}(complexified)} -- ++(-90:.92) -- ++(-155:.6) -- cycle;
			\fill [gray] (q) circle [radius=1.2pt];
			\draw [gray,-Latex] (q) -- ++(90:.7) node [anchor=west,inner sep=1pt] {\tiny $\mathfrak{Re}$};
			\draw [gray,-Latex] (q) -- ++(-155:.6) node [anchor=north,inner sep=1pt] {\tiny $\mathfrak{Im}$};
			\draw [-latex] ($(q)+(180:.12)$) arc [start angle=-180,end angle=180,x radius=.12cm,y radius=.24cm];
			\draw [dotted] ($(60:1.5)!.25!(0:1.5)$) -- (0:1.5);
			\draw [dashed] ($(p0)!-.2!(p2)$) -- (q);
			\foreach \i in {0,2} \fill (p\i) circle [radius=1.5pt];
			\node [anchor=south] at ($(p0)!.15!(p2)$) {\scriptsize $L$};
			\node [anchor=north] at (1.5,-.8) {\small (3) transversal};
		\end{scope}
		\begin{scope}[yshift=-4.5cm]
			\draw [dotted] (-120:.5) -- (60:3) (180:.5) -- (0:3.5);
			\draw [dotted] (0,0) node [below] {\scriptsize $U_2$} -- (0:1.5) node [below] {\scriptsize $U_0$} -- (60:1.5) node [left] {\scriptsize $U_1$} -- cycle;
			\coordinate [label=-90:{\scriptsize $[1\!:\!0\!:\!-\!1]\qquad\quad$}] (p0) at (0:3);
			\coordinate [label=180:{\scriptsize $[0\!:\!1\!:\!z\!-\!1]$}] (p2) at (60:1);
			\draw [dashed] ($(p0)!-.2!(p2)$) -- ($(p2)!-.24!(p0)$);
			\draw [ultra thick] (p2) -- ($(p0)!.76!(p2)$);
			\foreach \i in {0,2} \fill (p\i) circle [radius=1.5pt];
			\fill ($(p0)!.76!(p2)$) circle [radius=1.5pt];
			\node [anchor=south west,inner sep=1pt] at ($(p0)!.76!(p2)$) {\scriptsize $[z\!-\!1\!:\!1\!:\!0]$};
			\node [anchor=south] at ($(p0)!.15!(p2)$) {\scriptsize $L$};
			\node [anchor=north] at (1.5,-.8) {\small (4) longitudinal};
		\end{scope}
		\begin{scope}[xshift=5cm,yshift=-4.5cm]
			\draw [dotted] (-120:.5) -- (60:3) (180:.5) -- (0:3.5) (0:1.5) -- ($(60:1.5)!-1!(0:1.5)$);
			\draw [dotted] (0,0) node [below] {\scriptsize $U_2$} -- (0:1.5) node [below] {\scriptsize $U_0$} -- (60:1.5) node [left] {\scriptsize $U_1$} -- cycle;
			\coordinate [label=-90:{\scriptsize $[1\!:\!0\!:\!-\!1]\qquad\quad$}] (p0) at (0:3);
			\coordinate [label=0:{\scriptsize $[0\!:\!1\!:\!z\!-\!1]$}] (p2) at (60:1.8);
			\draw [dashed] ($(p0)!-.2!(p2)$) -- ($(p2)!-.5!(p0)$);
			\draw [ultra thick] (p2) -- ($(p2)!-.26!(p0)$);
			\foreach \i in {0,2} \fill (p\i) circle [radius=1.5pt];
			\fill ($(p2)!-.26!(p0)$) circle [radius=1.5pt];
			\node [anchor=north east,inner sep=1pt] at ($(p2)!-.26!(p0)$) {\scriptsize $[z\!-\!1\!:\!1\!:\!0]$};
			\node [anchor=south] at ($(p0)!.15!(p2)$) {\scriptsize $L$};
			\node [anchor=north] at (1.5,-.8) {\small (5) longitudinal ($z>1$)};
		\end{scope}
	\end{tikzpicture}
	\caption{Discontinuity contour for $\mathrm{Li}_2(z)$. (1) Original integral when $z>1$. (2) Original integral when $z<1$; in the neighborhood of $L$ the contour can decompose into a transversal part and a longitudinal part. (3) Circle-shaped transversal part of the discontinuity contour. (4) Longitudinal part of the discontinuity contour (thick solide line segment). (5) Longitudinal part of the discontinuity contour when $z$ is analytically continued to $(1,+\infty)$.}
	\label{fig:Li2Disc}
\end{figure}
To observe the above view point more clearly, let us further check the discontinuity of $\mathrm{Li}_2(z)$ around $z=1$. Based on the representation in the canonical frame \eqref{eq:Li2Canonical}, we see that only one irreducible component of $S$ can deform as we analytically continue $z$, which is defined by the equation $y_0+(1-z)y_1+y_2=0$. Let us call it $L$. Restricted to the real slice, this integral is well-defined when $z<1$, as shown in Figure \ref{fig:Li2Disc}(1). When we continue the parameter to the region $z>1$, the canonical contour $\triangledown$ starts to have non-empty intersection with $L$, such as in Figure \ref{fig:Li2Disc}(2). Obviously, in this case singularities arise in the neighborhood of this intersection during integration. It is useful to note that in this neighborhood the integral consists of a longitudinal part, which integrates along $L$, and a one-dimensional transversal part. The transversal part locally integrates along a direction transverse to $L$ for every point in the intersection, although the specific direction depends on the choice of parameterization of the contour \footnote{We will see different choices of this transversal direction explicitly in Section \ref{sec:discHex}}. The intersection looks like a pole for the transversal part of the integral, yielding an ambiguity to this integral in the same way as what happens in the previous example of a logarithmic function. Therefore, is shown in Figure \ref{fig:Li2Disc}(3), to compute the discontinuity one should modify the transversal part of the contour into a circle that wraps around the intersection (in the complex plane obtained by complexifying the transversal direction). The direction and winding number of this circle depends on the details of analytic continuation. Meanwhile, the longitudinal part of the integral has nothing to do with the emergence of the singularity, and so the contour for this part stays as $\triangledown\cap L$ (Figure \ref{fig:Li2Disc}(4)). Therefore, we expect the discontinuity contour to have the shape of a tube, which is a product of a circle-shaped transversal part and a longitudinal part $\triangledown\cap L$.

Once the discontinuity is obtained, it itself stands as an independent function. In other words, one can consider its analytic continuation without worrying about the branch cut of the original function, from which the discontinuity arises. In the case of $\mathrm{Li}_2(z)$, this manifests in the fact that $\log(z)$ is analytic in $\mathfrak{Re}(z)>0$ even though it comes from the branch cut of the former in $z>1$. At the level of integral representations, this means that once we learn how to define a new contour for the discontinuity, the definition itself does not really rely on whether $\triangledown\cap L$ is non-empty or not. This can be seen explicitly in the previous example. The definition for the circle-shaped transversal part has nothing to do with the value of $z$. As for the longitudinal part, in the case of $z>1$ we learn that $\triangledown\cap L$ is a line segment (a $1$-simplex) along $L$, with two $0$-faces at $[0:1:z-1]$ and $[z-1:1:0]$ respectively. So we can take the characterization of $0$-faces as the defining property of the longitudinal part, and safely consider analytic continuation even to the region $z>1$ (where $\triangledown\cap L=\varnothing$), as shown in Figure \ref{fig:Li2Disc}(5). In particular, as we reach $z=1$ this $1$-simplex shrinks to a point, which make the integral vanish, and correspondingly we have $\log(1)=0$.

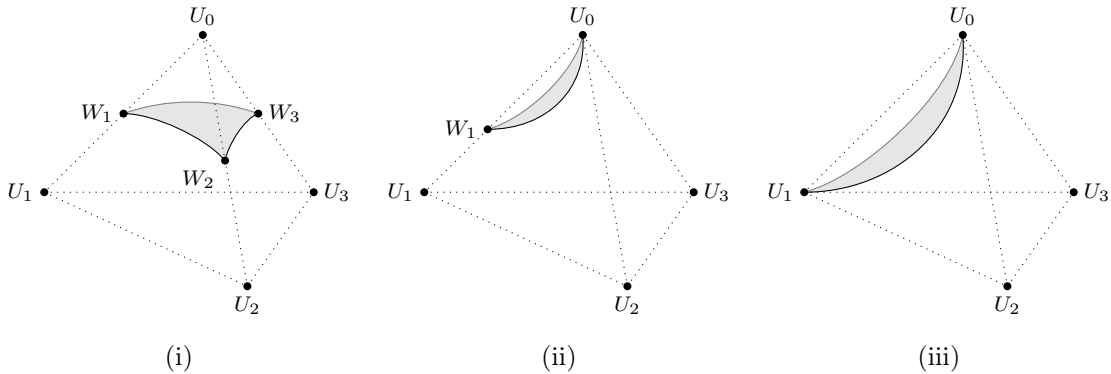
\begin{figure}[ht]
	\centering
	\begin{tikzpicture}
		\def\R{1.8}
		\begin{scope}
			\coordinate [label=90:{\scriptsize $U_0$}] (v0) at (80:\R);
			\coordinate [label=180:{\scriptsize $U_1$}] (v1) at (190:\R);
			\coordinate [label=270:{\scriptsize $U_2$}] (v2) at (-60:\R);
			\coordinate [label=0:{\scriptsize $U_3$}] (v3) at (-10:\R);
			\coordinate [label=180:{\scriptsize $W_1$}] (i1) at ($(v0)!.5!(v1)$);
			\coordinate [label=200:{\scriptsize $W_2$}] (i2) at ($(v0)!.5!(v2)$);
			\coordinate [label=0:{\scriptsize $W_3$}] (i3) at ($(v0)!.5!(v3)$);
			\fill [gray!20] (i1) .. controls +(0:.4) and +(130:.4) .. (i2) .. controls +(80:.2) and +(200:.2) .. (i3) .. controls +(160:.6) and ++(20:.6) .. cycle;
			\draw [black!50] (i3) .. controls +(160:.6) and ++(20:.6) .. (i1);
			\draw (i1) .. controls +(0:.4) and +(130:.4) .. (i2) .. controls +(80:.2) and +(200:.2) .. (i3);
			\draw [dotted] (v1) -- (v3);
			\draw [dotted] (v0) -- (v2) (v0) -- (v1) -- (v2) -- (v3) -- cycle;
			\foreach \i in {0,1,2,3} \fill [black] (v\i) circle [radius=1.5pt];
			\foreach \j in {1,2,3} \fill [black] (i\j) circle [radius=1.5pt];
			\node [anchor=north] at (0,-2.2) {\small (i)};
		\end{scope}
		\begin{scope}[xshift=5cm]
			\coordinate [label=90:{\scriptsize $U_0$}] (v0) at (80:\R);
			\coordinate [label=180:{\scriptsize $U_1$}] (v1) at (190:\R);
			\coordinate [label=270:{\scriptsize $U_2$}] (v2) at (-60:\R);
			\coordinate [label=0:{\scriptsize $U_3$}] (v3) at (-10:\R);
			\coordinate [label=180:{\scriptsize $W_1$}] (i1) at ($(v0)!.6!(v1)$);
			\fill [gray!20] (v0) .. controls +(-100:.5) and +(15:.5) .. (i1) .. controls +(0:.8) and +(-85:.7) .. cycle;
			\draw [black!50] (v0) .. controls +(-100:.5) and +(15:.5) .. (i1);
			\draw (i1) .. controls +(0:.8) and +(-85:.7) .. (v0);
			\draw [dotted] (v1) -- (v3);
			\draw [dotted] (v0) -- (v2) (v0) -- (v1) -- (v2) -- (v3) -- cycle;
			\foreach \i in {0,1,2,3} \fill [black] (v\i) circle [radius=1.5pt];
			\fill [black] (i1) circle [radius=1.5pt];
			\node [anchor=north] at (0,-2.2) {\small (ii)};
		\end{scope}
		\begin{scope}[xshift=10cm]
	        \coordinate [label=90:{\scriptsize $U_0$}] (v0) at (80:\R);
			\coordinate [label=180:{\scriptsize $U_1$}] (v1) at (190:\R);
			\coordinate [label=270:{\scriptsize $U_2$}] (v2) at (-60:\R);
			\coordinate [label=0:{\scriptsize $U_3$}] (v3) at (-10:\R);
			\fill [gray!20] (v0) .. controls +(-100:.8) and +(15:.8) .. (v1) .. controls +(0:1.2) and +(-85:1.2) .. cycle;
			\draw [black!50] (v0) .. controls +(-100:.8) and +(15:.8) .. (v1);
			\draw (v1) .. controls +(0:1.2) and +(-85:1.2) .. (v0);
			\draw [dotted] (v1) -- (v3);
			\draw [dotted] (v0) -- (v2) (v0) -- (v1) -- (v2) -- (v3) -- cycle;
			\foreach \i in {0,1,2,3} \fill [black] (v\i) circle [radius=1.5pt];
			\node [anchor=north] at (0,-2.2) {\small (iii)};
		\end{scope}
	\end{tikzpicture}
    \caption{The longitudinal part of the discontinuity contours for a quadric singularity.}
	\label{fig:DiscontinuityContourShape}
\end{figure}

We can apply the above logic to the case of quadric singularity. As mentioned before, the touching configurations signal the potential logarithmic branch point of the integral. On the real slice, these correspond to either a $0$-face or a $1$-face being embedded in $Q$ in this case. To study the discontinuities, we deform $Q$ slightly away from such configurations. It is then easy to observe that the discontinuities can be classified into three types according to the shape of the longitudinal part of the discontinuity contours. These are illustrated in Figure \ref{fig:DiscontinuityContourShape}, where we use an integral in $\mathbb{CP}^3$ as an example.
\begin{enumerate}
	\item[(i)] In the first type, as in Figure \ref{fig:DiscontinuityContourShape}(i), the contour arises by deforming from touching configurations of $0$-faces. The deformation here can be viewed as inverse operations corresponding to the jumps (a)(c)(d) in the stratification Figure \ref{fig:strata1faceQ}. When we deform $Q$ such that $U_0$ moves off it, $Q$ can cut out a corner from the contour $\triangledown$. Hence $Q\cap\triangledown$ has a simplex-like shape, whose faces are also intersections between $Q$ and faces of $\triangledown$ that are adjacent to $U_0$. This is not an ordinary simplex, since it resides on the quadric. Still, it contains $0$-faces, $W_i=Q\cap\overline{U_0U_i}$, and because we consider small deformations, all $W_i$'s are sufficiently close to $U_0$. \footnote{Later on when we honestly work in complex projective space, $Q\cap\overline{U_0U_i}$ always contains two points. Here by $W_i$ we really mean one of these points that his $U_0$ as the deformation vanishes. This will be crucial in properly selecting the discontinuity contour. We further illustrate this with explicit examples in Section \ref{sec:discJumpc} and Section \ref{sec:twoMassBox}}
	This type is also the unique type of discontinuity that we may encounter if the integrand singularity under study is a hyperplane.
	\item[(ii)] In the second type, as in Figure \ref{fig:DiscontinuityContourShape}(ii), the deformation inverses the jump (b). It starts from a configuration where $\overline{U_0U_1}$ intersects $Q$ at $U_0$ twice. The deformation splits this intersection into two distinguished points, one remaining as the same $U_0$, while the other turning into some $W_1\subset\overline{U_0U_1}$. Because the starting configuration associates to the $1$-face $\underline{U_0U_1}$, we can see that all the faces of $Q\cap\triangledown$ now live on $\overline{U_0U_1}$ and other ambient space of higher faces that border $\underline{U_0U_1}$. As a result, $Q\cap\triangledown$ has the shape of a crescent.
	\item[(iii)] In the third type, as in Figure \ref{fig:DiscontinuityContourShape}(iii), the deformation inverses the jump (e). It starts from the configuration $\overline{U_0U_1}\subset Q$. The discontinuity contour is structurally very similar to the second type, while the only difference is that its two $0$-faces are anchored at $U_0$ and $U_1$ respectively, which is required by the deformation.
\end{enumerate}
The above contours are all what we need when analyzing discontinuities of one-loop integrals, whose integrand singularity is a quadric. Note that these contours are motivated by the touching configurations associated to a specific $k$-face of $\triangledown$ ($k=0,1$). In this sense we can call the discontinuities thus computed as \emph{elementary discontinuities}. In practice we can conveniently label these discontinuities by the corresponding $\mathbb{CP}^k$ ambient space. For example, in Figure \ref{fig:DiscontinuityContourShape}, if the original integral is called $I$, then we can denote the discontinuity in (i) as $\mathrm{Disc}_{\overline{U_0}}I$, and that in (ii) or (iii) as $\mathrm{Disc}_{\overline{U_0U_1}}I$. Note that type (ii) and (iii) cannot both occur in a given integral, and so which $\mathrm{Disc}_{\overline{U_0U_1}}I$ actually refers to should be clear from the context \footnote{From Figure \ref{fig:strata1faceQ}, we see that the only possibility to have several different jumps starting from the same touching configuration is to go from Stratum 1 to Stratum 2, i.e., jump (b) and jump (c). But jump (b) leads to discontinuity of type (ii) while jump (c) leads to discontinuity of type (i). These are easily distinguished using our notations.}. On the other hand, due to the potential two-fold ambiguity tied in jump (a), sometimes we may need extra labels for $\mathrm{Disc}_{\overline{U_0}}I$ in order to distinguish between several different choices of contours of type (i). This will be illustrated in detail in a specific integral in Section \ref{sec:twoMassBox}.

Very often, in a given integral jumps associated to several different faces of $\triangledown$ may be simultaneously triggered by a same condition on the parameters. In this case, the actual discontinuity that one encounters during analytic continuation will be a linear combination of the elementary discontinuities discussed above. Neverthesless, as long as we figure out a proper way to work out these elementary discontinuities, the generic case merely reduces to a linear problem. We will see explicit examples regarding this in Section \ref{sec:symbolConstruction}

Of course, one should pay attention that the picture on the real slice of $\mathbb{CP}^d$ is only a qualitative way to estimate the pattern of the discontinuity contours. Here one may potentially encounter issues such as the empty intersection between $Q$ and $\overline{U_0U_i}$ strictly on the real slice. In order to make the analysis rigorous and quantitative, we need to carefully inspect the structure of the contours in the full $\mathbb{CP}^{d}$. And this is the task of the next two sections.

\section{Bi-projective ``Fibration'' of $\mathbb{CP}^d$ and the simplex contour}\label{sec:fibration}

Before analyzing the discontinuity contours, let us think a bit more about what is done when one actually computes the integrals. The aim is to put this procedure in a more geometric point of view. From the previous section we see that the contour for a discontinuity typically has one less integration to do as compared to the original integral, because the one-dimensional transversal part of the integral contour turns into a circle, thus becoming a residue computation, which is effectively just a matter of Laurent expansion. Nevertheless, we need to learn how to properly extract the transversal part of the contour and how to properly handle the remaining integrals. These are non-trivial when the integrand singularity $S$ is nonlinear. The geometric view point will provide a useful and systematic tool to answer these questions. Its application to the discontinuity computation will be discussed in the next section.

\subsection{Ordinary ``fibration'' of the simplex contour and point projection}\label{sec:projection}

When a specific integral is given, one can typically apply a change of integration variables, which turns the integral into another expression without changing the integral result. This is an indication that an integral should be understood geometrically, and the different expressions correspond to different ways of parametrizing the contour. The virtue of this freedom is that one can usually choose a preferred parametrization that best suits the need of a problem.

It is helpful to have a more detailed look at various parametrizations of multidimensional simplex contours. Let us go back to the integrals for $\mathrm{Li}_2(z)$ discussed in Section \ref{sec:Li2}. In the ordinary Chen's iterated integral a sequence has to be chosen for the integration variables, such as the one presented in \eqref{eq:Li2ChenIntegral}. In this case, we first integrate $x_1$ and then $x_2$. As we integrate $x_1$, $x_2$ is fixed at some value $\tilde{x}_2\in[0,z]$. So the first integral is performed over some line segment, while the second integration variable is a continuous parameter for a collection of line segments, which together cover the entire $2$-simplex contour.

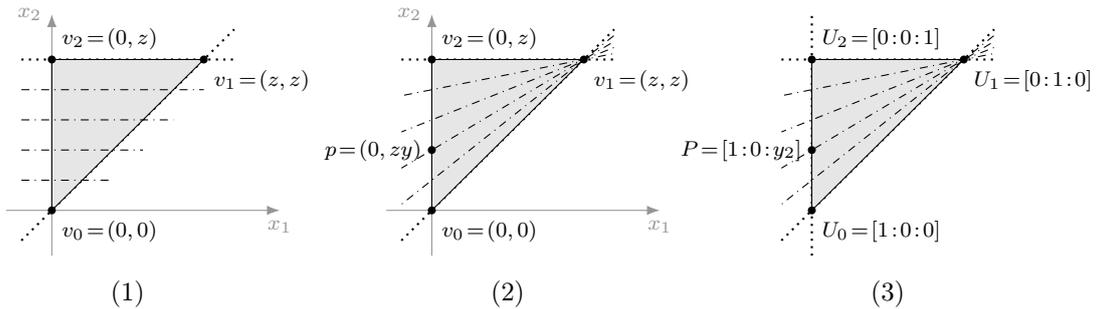
\begin{figure}[ht]
	\centering
	\begin{tikzpicture}
		\begin{scope}
			\draw [-Latex,Gray] (-.6,0) -- (3,0) node [below] {\scriptsize $x_1$};
			\draw [-Latex,Gray] (0,-.6) -- (0,2.6) node [left] {\scriptsize $x_2$};
			\draw [dotted,thick] (-.4,-.4) -- (2.4,2.4) (-.4,2) -- (2.4,2);
			\coordinate [label=-45:{\scriptsize $v_0\!=\!(0,0)$}] (v0) at (0,0);
			\coordinate [label=-60:{\scriptsize $v_1\!=\!(z,z)$}] (v1) at (2,2);
			\coordinate [label=45:{\scriptsize $v_2\!=\!(0,z)$}] (v2) at (0,2);
			\fill [gray!20,draw=black] (v0) -- (v1) -- (v2) -- cycle;
			\foreach \i in {0,1,2} \fill [black] (v\i) circle [radius=1.5pt];
			\foreach \i in {.4,.8,1.2,1.6} \draw [dash dot] (-.4,\i) -- (\i+.4,\i);
			\node [anchor=north] at (1,-.8) {\small (1)};
		\end{scope}
		\begin{scope}[xshift=5cm]
			\draw [-Latex,Gray] (-.6,0) -- (3,0) node [below] {\scriptsize $x_1$};
			\draw [-Latex,Gray] (0,-.6) -- (0,2.6) node [left] {\scriptsize $x_2$};
			\draw [dotted,thick] (-.4,-.4) -- (2.4,2.4) (-.4,2) -- (2.4,2);
			\coordinate [label=-45:{\scriptsize $v_0\!=\!(0,0)$}] (v0) at (0,0);
			\coordinate [label=-60:{\scriptsize $v_1\!=\!(z,z)$}] (v1) at (2,2);
			\coordinate [label=45:{\scriptsize $v_2\!=\!(0,z)$}] (v2) at (0,2);
			\coordinate [label=180:{\scriptsize $p\!=\!(0,zy)$}] (v3) at (0,.8);
			\fill [gray!20,draw=black] (v0) -- (v1) -- (v2) -- cycle;
			\foreach \i in {0,1,2,3} \fill [black] (v\i) circle [radius=1.5pt];
			\begin{scope}
				\clip (-.4,-.4) rectangle (2.4,2.4);
				\foreach \y in {1,2,3,4} \draw [dash dot] ($(v1)!-.4!(0,.4*\y)$) -- ($(v1)!1.2!(0,.4*\y)$);
			\end{scope}
			\node [anchor=north] at (1,-.8) {\small (2)};
		\end{scope}

		\begin{scope}[xshift=10cm]
			\draw [dotted,thick] (0,-.6) -- (0,2.6) (-.4,-.4) -- (2.4,2.4) (-.4,2) -- (2.4,2);
			\coordinate [label=-45:{\scriptsize $U_0\!=\![1\!:\!0\!:\!0]$}] (v0) at (0,0);
			\coordinate [label=-60:{\scriptsize $U_1\!=\![0\!:\!1\!:\!0]$}] (v1) at (2,2);
			\coordinate [label=45:{\scriptsize $U_2\!=\![0\!:\!0\!:\!1]$}] (v2) at (0,2);
			\coordinate [label=180:{\scriptsize $P\!=\![1\!:\!0\!:\!y_2]$}] (v3) at (0,.8);
			\fill [gray!20,draw=black] (v0) -- (v1) -- (v2) -- cycle;
			\foreach \i in {0,1,2,3} \fill [black] (v\i) circle [radius=1.5pt];
			\begin{scope}
				\clip (-.4,-.4) rectangle (2.4,2.4);
				\foreach \y in {1,2,3,4} \draw [dash dot] ($(v1)!-.4!(0,.4*\y)$) -- ($(v1)!1.2!(0,.4*\y)$);
			\end{scope}
			\node [anchor=north] at (1,-.8) {\small (3)};
		\end{scope}
	\end{tikzpicture}
	\caption{Different ``fibrations'' of a simplex contour in $\mathbb{CP}^2$.}
	\label{fig:sliceSimplexCP2}
\end{figure}

To make this description concrete, let us temporarily assume again that the integral is put in $\mathbb{R}^2$. Each value $\tilde{x}_2$ specifies a line in $\mathbb{R}^2$ by the equation $x_2-\tilde{x}_2=0$. Then the above-mentioned line segment is the intersection of this line and the $2$-simplex contour. By considering all $\tilde{x}_2\in\mathbb{R}$ we construct a fibration of $\mathbb{R}^2$ into lines, which further induces a fibration of the $2$-simplex contour (into line segments) by intersection. Correspondingly, the integral is carried out by first integrating along each one-dimensional fibre and then integrating over the moduli space of the fibres. This is illustrated in Figure \ref{fig:sliceSimplexCP2}(1).

We can alternatively ``fibrate'' the contour in other ways. A useful ``fibration'' is to pick one $0$-face of the $2$-simplex contour, e.g., $(z,z)$, and consider all possible lines passing through it \footnote{Rigorously speaking this is not an honest fibration of $\mathbb{R}^2$, because there are non-empty intersection between different lines now. Nevertheless, they only intersect at the reference point, which does not have the full dimension and does no harm to the integral. Hence in a loose sense we still call this a ``fibration''.}. Some subset of the lines have non-empty intersection with the contour, inducing a ``fibration'' of the latter, as illustrated in Figure \ref{fig:sliceSimplexCP2}(2).
Correspondingly, we can introduce two new parameters, one parametrizing points on each line, and the other parametrizing the space of the lines. Note that each ``fibre'' uniquely maps to a point on $\overline{v_0v_2}$ by intersection. Let a point $p\in\overline{v_0v_2}$ be labeled by a parameter $y$ via, e.g., $p=(0,zy)$. Then $p$ and $v_1$ determine the ``fibre'' $\overline{v_1p}$, which can be parametrized by another parameter $w$ via, e.g., $(x_1,x_2)=(z w,z(y+w-wy))$. In this way the $2$-simplex contour is covered by the range $w,y\in[0,1]$, and the integral \eqref{eq:Li2ChenIntegral} becomes
\begin{align}
	\mathrm{Li}_2(z)=\int_0^1\frac{z(1-w)\mathrm{d}w\mathrm{d}y}{(1-zw)(w+y-wy)}.
\end{align}

This second type of ``fibration'' is in fact more natural when viewing in projective space. Based on the canonical representation \eqref{eq:Li2Canonical}, we can pick one of the affine patches by setting $y_0=1$ and perform the integration, so that
\begin{align}
	\mathrm{Li}_2(z)=\int_0^\infty\frac{z\mathrm{d}y_1\mathrm{d}y_2}{(1+y_1+y_2)(y_1+y_2)(1+(1-z)y_1+y_2)}.
\end{align}
In this representation a point on the contour is parametrized by
\begin{align}
	[1:y_1:y_2]\equiv \underbrace{[1:0:y_2]}_{P}+y_1\underbrace{[0:1:0]}_{U_1}.
\end{align}
In particular, by $y_1\to\infty$ we reach the $0$-face $U_1$, in correspondence to $v_1$ in the original affine space. And at $y_1=0$, the resulting coordinates $[1:0:y_2]$ parametrize points $P\in\overline{U_0U_2}$, in correspondence to $p$ in the previous representation.

In fact, by compactifying the affine space into projective space, the fibration discussed in the original Chen integral in Figure \ref{fig:sliceSimplexCP2}(1) belongs to the same type of ''fibration'' using a reference point as described above. The only difference is that this reference point locates at infinity and outside of the contour simplex in the Chen integral, while it is identified with one $0$-face of the contour simplex in the above.

When taking into consideration the full $\mathbb{CP}^2$, the above picture of ``fibration'' becomes slightly subtle. On the one hand, being defined as the linear span of two points with complex coefficients, each ``fibre'' is now structurally a $\mathbb{CP}^1\subset\mathbb{CP}^2$. On the other hand, now the contour can be deformed away from the one on the real slice. It is not always guaranteed that each ``fibre'' always carves out a 1d line segment from the original $2$-simplex contour. For example, we can deform the contour by
\begin{align}\label{eq:badDeformation}
	y_1=t_1,\qquad y_2=t_2+i\frac{t_1t_2}{1+t_1^3+t_2^3},\qquad 0\leq t_1,t_2<\infty.
\end{align}
Here $y_2$ takes complex values at generic points inside the $2$-simplex, but is kept real on all its boundaries. Because the ``fibre'' $y_2$ is determined by $U_1$ and a point $[1:0:t_2]\in\overline{U_0U_2}$, it is described by the function $y_2-t_2=0$. This means that on such ``fibre'' is fixed at a real value while $y_1$ can take arbitrary complex values. Then apparently the ``fibre'' intersects the above $2$-simplex only at two  discrete points $U_1$ and $[1:0:t_2]$ (for whatever $t_2$ values). Therefore, for this specific choice of the contour, the set of $\mathbb{CP}^1$ ``fibres'' do not manage to induce a ``fibration'' of the contour.

Nevertheless, there do exist deformations of the contour which can be properly ``fibrated'' by the above $\mathbb{CP}^1$'s. For instance, instead of \eqref{eq:badDeformation} we can choose
\begin{align}\label{eq:goodDeformation}
	y_1=t_1+i\frac{t_1t_2}{1+t_1^3+t_2^3},\qquad y_2=t_2,\qquad 0\leq t_1,t_2<\infty.
\end{align}
Like the previous one, all faces of this contour again sit on the real slice, and so we get the same equation $y_2-t_2=0$ for the ``fibres''. Because this equation is satisfied by all points on the contour with fixed $t_2$, we immediately see that this new contour is properly ``fibrated'' by the $\mathbb{CP}^1$'s.

As a result, when we single out one integration from a multidimensional integral with a simplex contour, we do not consider arbitrary continuous deformations of the contour, but only those that can be properly ``fibrated''. Let us now generalize the above discussions to integrals in arbitrary $\mathbb{CP}^d$ and make these terms precise.

First of all, we can temporarily ignore the contour and purely consider ``fibrating'' $\mathbb{CP}^d$ into $\mathbb{CP}^1$'s. Here we choose a reference point, calling it $M$, and consider all possible $\mathbb{CP}^1$ subspaces that contain $M$. The space of these $\mathbb{CP}^1$'s can be easily characterized by a hyperplane $H\subset\mathbb{CP}^d$ such that $M\notin H$. This is because every $\mathbb{CP}^1$ intersects $H$ at a unique point, and reversely, together with the reference $M$ every point $N\in H$ uniquely determines a $\mathbb{CP}^1\subset\mathbb{CP}^d$ via their linear span. This means that $H$ serves as a moduli space for $\mathbb{CP}^1$'s containing $M$. 

When using the above ``fibration'' of $\mathbb{CP}^{d}$ to further induce a ``fibration'' of a simplex contour $\triangle$, we typically identify the reference point $M$ as one $0$-face of $\triangle$, say $V_d$. Each of the remaining $0$-face $V_i$ then uniquely determines a $\mathbb{CP}^1$ containing $V_d$, which is in fact the ambient space $\overline{V_iV_d}$. All the remaining $0$-faces together span the codim-1 ambient space $\overline{V_0V_1\cdots V_{d-1}}$, which we can treat as the moduli space $H$ described above. When we specify an actual shape of the contour, we can begin by specifying its $(d-1)$-face $\underline{V_0V_1\cdots V_{d-1}}\subset H$, which by itself is a $(d-1)$-simplex. Then for each point $N\in\underline{V_0V_1\cdots V_{d-1}}$, we can take the linear span $\overline{NV_d}$, and within this $\mathbb{CP}^1$ subspace we further specify one-dimensional real path $\underline{NV_d}$ connecting the two points. When we take the union of these paths for all $N\in\underline{V_0V_1\cdots V_{d-1}}$, if they form a continuous $d$ real dimensional shape, then the resulting shape is a justified choice of the contour $\Delta$, i.e.,
\begin{align}
	\Delta=\bigcup_{N\in\underline{V_0V_1\cdots V_{d-1}}}\underline{NV_d}.
\end{align}
To consider continuous contour deformation, we can deform its face $\underline{V_0V_1\cdots V_{d-1}}$ in a generic way in the sense described in Section \ref{sec:simplexContour}, and further deform each $\underline{NV_d}$ within its own $\mathbb{CP}^1$ ambient space $\overline{NV_d}$, as long as the whole $d$-dimensional shape remains continuous. It is easy to see that the collection of all such deformations form a subset of the most generic deformations allowed for $\Delta$. In this way, we can ensure that the ``fibration'' of $\mathbb{CP}^d$ into $\mathbb{CP}^1$'s always properly induces a corresponding ``fibration'' of the contour $\Delta$, so that one integration is always singled out from the rest.

In the previous section we have observed that one class of singularities arise from jumps in touching configurations that are effectively tied to $0$-faces. These include the unique jump in Figure \ref{fig:strata0faceH} when the integrand singularity $S$ is a hyperplane, as well as the jumps of type (a)(c)(d) in Figure \ref{fig:strata1faceQ} when $S$ is a quadric. Without loss of generality let us assume the $0$-face under study is $V_d$. So the singularity arises when $V_d\subset S$, and the discontinuity around it is denoted as $\mathrm{Disc}_{\overline{V_d}}$ following the notation in Section \ref{sec:discontinuityFirstLook}. To study this discontinuity, we slightly deform $V_d$ away from $S$. The ``fibration'' structure described above provides a convenient way to study such discontinuities. Being a $\mathbb{CP}^1$ subspace, each ``fibre'' $\overline{NV_d}$ exactly intersects $S$ at $g$ points (counting multiplicities), where $g$ is the degree of $S$. These intersection points manifest as poles when viewed within $\overline{NV_n}$. The touching configuration $V_d\subset S$ means that $V_d$ coincides with one of the poles. Hence this effectively reduces the problem of emergence of higher dimensional integrals to that of a one-dimensional integral on each $\overline{NV_d}$. In this way every ``fibre'' defines a transversal direction in the neighborhood of each of the above intersection points with $S$, which was motivated in Section \ref{sec:discontinuityFirstLook}. To compute the discontinuities, it then suffices to compute the residue of the integrand around the corresponding pole, with respect to the integration variable that parameterize the ``fibre''. This operation turns the integral along each ``fibre'' into a Laurent expansion of the integrand, thus reducing the number of integrals by one. By the above discussions on the fibrations, we see that the remaining integral for the discontinuities is exactly an integral in the moduli space of the $\mathbb{CP}^1$ fibres, or equivalently $\overline{V_0V_1\cdots V_{d-1}}$, again with a simplex contour. Therefore, geometrically we can interpret this operation as a projection through the point $V_d$.

For the purpose of computations, we can go to the canonical frame, and express the homogeneous coordinates as
\begin{align}\label{eq:pointFibrationCanonical}
	X=[x_0:x_1:\cdots:x_d]=\underbrace{[x_0:x_1:\cdots:x_{d-1}:0]}_{N}+x_d\underbrace{[0:0:\cdots:0:1]}_{U_d},
\end{align}
where the first term provides the homogeneous coordinates for $N\in\overline{U_0U_1\cdots U_{d-1}}$. $x_d$ is now treated as an inhomogeneous coordinate for points on the ``fibre'' $\overline{NU_d}$. Since $\overline{NU_d}$ is structurally a $\mathbb{CP}^1$ subspace, when necessary we can promote this variable into the full homogeneous coordinates on $\overline{NU_d}$ by the map $x_d\mapsto[x_d:1]\equiv[\frac{t_0}{t_1}:1]\sim[t_0:t_1]$.

\subsection{A generalized class of ``fibrations'' using bi-projection}

In Section \ref{sec:projection} we have discussed ``fibration'' of $\mathbb{CP}^d$ into $\mathbb{CP}^1$'s passing through a common point $M$. As we promise this will be useful in analyzing the discontinuities $\mathrm{Disc}_{\overline{V_i}}$. When it comes to the discontinuities $\mathrm{Disc}_{\overline{V_{i_0}V_{i_1}\cdots V_{i_k}}}$ for some $k>0$ (although we will at most only encounter $\mathrm{Disc}_{\overline{V_iV_j}}$ in one-loop integrals), this notion calls for certain generalization. To see this, let us again consider the massless box \eqref{eq:masslessbox}. Here all $0$-faces are restricted on the quadric $Q$, and we know that a singularity occurs when the ambient space of a $1$-face, say $\overline{U_0U_2}$, is embedded in $Q$, in correspondence to $s=0$. In deforming $Q$ away from such configuration, we still keep the $0$-faces on $Q$, due to the masssless properties of the propagators. This means that for whichever $\mathbb{CP}^1$ passing through a $0$-face, the $0$-face can never be pulled away from the pole induced by $Q$ on this $\mathbb{CP}^1$.

\brief{General discussion on foliations of the simplex contour, beyond point foliation.  Explain how it covers the integration contour.}

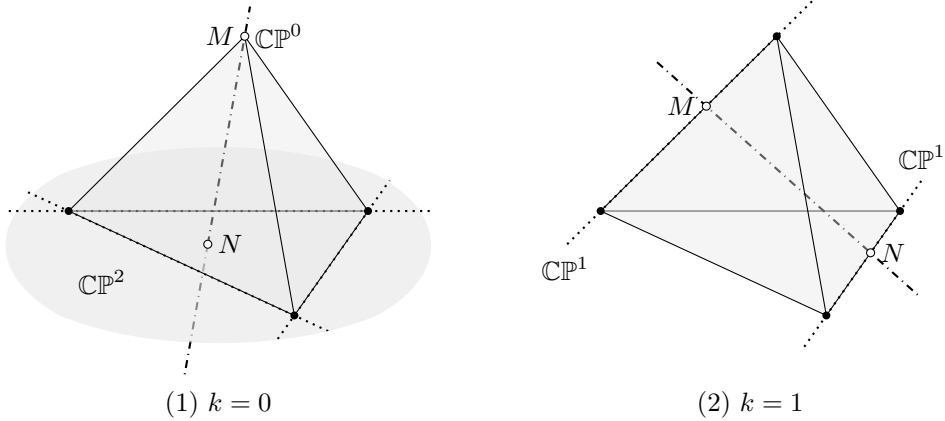
\begin{figure}[ht]
	\centering
	\begin{tikzpicture}
		\begin{scope}
			\coordinate [label=0:{\small $\mathbb{CP}^0$}] (v0) at (80:2);
			\coordinate (v1) at (190:2);
			\coordinate (v2) at (-60:2);
			\coordinate (v3) at (-10:2);
			\draw [dash dot,thick] (-100:.8) -- (-100:2.6);
			\begin{scope}
				\clip (0,-.8) circle [x radius=3,y radius=1.3];
				\fill [gray!20,opacity=.6] (0,-.8) circle [x radius=2.8,y radius=1.4];
				\draw [dotted,thick] ($(v1)!-.3!(v2)$) -- ($(v2)!-.3!(v1)$) ($(v1)!-.3!(v3)$) -- ($(v3)!-.3!(v1)$) ($(v2)!-.4!(v3)$) -- ($(v3)!-.4!(v2)$);
			\end{scope}
			\node [anchor=center] at (-140:2) {\small $\mathbb{CP}^2$};
			\draw (v1) -- (v3);
			\draw [dash dot,thick] (-100:.8) -- (80:2.4);
			\fill [gray!20,opacity=.4] (v0) -- (v1) -- (v2) -- (v3) -- cycle;
			\draw (v0) -- (v2) (v0) -- (v1) -- (v2) -- (v3) -- cycle;
			\foreach \i in {1,2,3} \fill [black] (v\i) circle [radius=1.5pt];
			\draw [black,fill=white] (v0) circle [radius=1.5pt];
			\node [anchor=east] at (v0) {\small $M$};
			\draw [black,fill=white] (-100:.8) circle [radius=1.5pt];
			\node [anchor=west] at (-100:.8) {\small $N$};
			\node [anchor=north] at (0,-2.6) {\small (1) $k=0$};
		\end{scope}
		\begin{scope}[xshift=7cm]
			\coordinate (v0) at (80:2);
			\coordinate (v1) at (190:2);
			\coordinate (v2) at (-60:2);
			\coordinate (v3) at (-10:2);
			\draw (v1) -- (v3);
			\coordinate (m0) at ($(v0)!.4!(v1)$);
			\coordinate (m1) at ($(v2)!.6!(v3)$);
			\draw [dash dot,thick] ($(m0)!-.3!(m1)$) -- ($(m1)!-.3!(m0)$);
			\fill [gray!20,opacity=.4] (v0) -- (v1) -- (v2) -- (v3) -- cycle;
			\draw [dotted,thick] ($(v0)!-.2!(v1)$) -- ($(v1)!-.2!(v0)$) node [below] {\small $\mathbb{CP}^1$} ($(v2)!-.3!(v3)$) -- ($(v3)!-.3!(v2)$) node [above] {\small $\mathbb{CP}^1$};
			\draw (v0) -- (v2) (v0) -- (v1) -- (v2) -- (v3) -- cycle;
			\foreach \i in {0,1,2,3} \fill [black] (v\i) circle [radius=1.5pt];
			\foreach \i in {0,1} \draw [black,fill=white] (m\i) circle [radius=1.5pt];
			\node [anchor=east] at (m0) {\small $M$};
			\node [anchor=west] at (m1) {\small $N$};
			\node [anchor=north] at (0,-2.6) {\small (2) $k=1$};
		\end{scope}
	\end{tikzpicture}
	\caption{Bi-projective ``Fibration'' of $\mathbb{CP}^3$. (1) Use a $\mathbb{CP}^0$ and a $\mathbb{CP}^2$ as the reference subspaces ($k=0$). (2) Use a pair of $\mathbb{CP}^1$'s as the reference subspaces ($k=1$).}
	\label{fig:fiberateCP3}
\end{figure}

The way to solve the above problem is to consider a bigger class of ``fibrations'' of $\mathbb{CP}^d$ that generalizes the ones discussed in Section \ref{sec:projection}. Note that our main aim is to identify a series of $\mathbb{CP}^1$ subspaces and that each $\mathbb{CP}^1$ is spanned by two points $M$ and $N$. In the previous ``fibration'', we always keep $M$ fixed at a $0$-face, e.g., $V_d$, and let $N$ scan over a $\mathbb{CP}^{d-1}$ subspace $\overline{V_0V_1\cdots V_{d-1}}$. In a sense $\overline{V_0V_1\cdots V_{d-1}}$ is complement to $V_d$, since it is spanned by the remaining $0$-faces of $\Delta$ other than $V_d$. The idea for the generalization is to change the dimension of the two subspaces for $M$ and $N$, under the condition that they are complement to each other. So in general we can divide all the $d+1$ $0$-faces of $\Delta$ into two sets, for example $\{V_0,V_1,\ldots,V_{d-k-1}\}$ and $\{V_{d-k},V_{d-k+1},\ldots,V_d\}$, for some integer $0\leq k\leq d-1$. The former set span a $\mathbb{CP}^{d-k-1}$ subspace $\overline{V_0V_1\cdots V_{d-k-1}}$, while the latter set span a $\mathbb{CP}^k$ subspace $\overline{V_{d-k}V_{d-k+1}\cdots V_d}$ (we can treat them as two reference subspaces). For each pair of points $M\in\overline{V_{d-k}V_{d-k+1}\cdots V_d}$ and $N\in\overline{V_0V_1\cdots V_{d-k-1}}$ we can further determine a $\mathbb{CP}^1$ subspace $\overline{MN}$. By scanning over all possible $M$ and $N$, the resulting set of $\overline{MN}$ together can cover the entire $\mathbb{CP}^d$. Two different choices of ``fibrations'' in this manner in the case of $\mathbb{CP}^3$ are presented in Figure \ref{fig:fiberateCP3}, where $k=0,1$ respectively (the case of $k=2$ is equivalent to that of $k=0$).

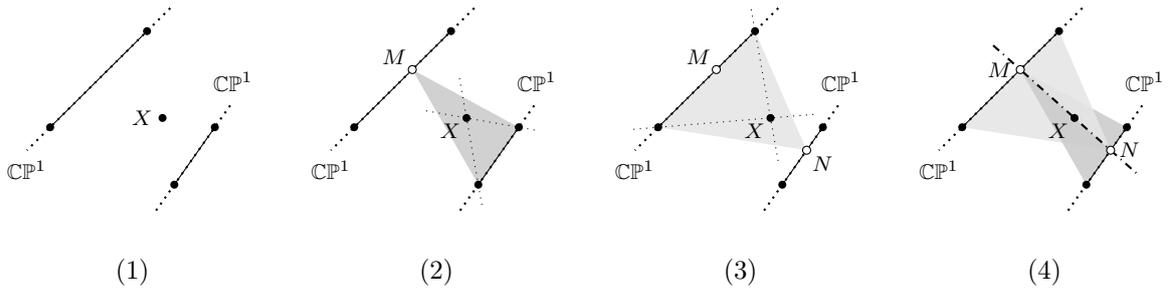
\begin{figure}
	\centering
	\begin{tikzpicture}
		\def\R{1.1}
		\begin{scope}
			\coordinate (v0) at (80:\R);
			\coordinate (v1) at (190:\R);
			\coordinate (v2) at (-60:\R);
			\coordinate (v3) at (-10:\R);
			\coordinate (m0) at ($(v0)!.4!(v1)$);
			\coordinate (m1) at ($(v2)!.6!(v3)$);
			\coordinate [label=180:{\scriptsize $X$}] (p) at ($(m0)!.6!(m1)$);
			\draw [dotted,thick] ($(v0)!-.24!(v1)$) -- ($(v1)!-.24!(v0)$) node [below] {\scriptsize $\mathbb{CP}^1$} ($(v2)!-.45!(v3)$) -- ($(v3)!-.45!(v2)$) node [above] {\scriptsize $\mathbb{CP}^1$};
			\draw (v0) -- (v1) (v2) -- (v3);
			\foreach \i in {0,1,2,3} \fill [black] (v\i) circle [radius=1.5pt];
			\fill (p) circle [radius=1.5pt];
			\node [anchor=north] at (0,-1.8) {\small (1)};
		\end{scope}
		\begin{scope}[xshift=4cm]
			\coordinate (v0) at (80:\R);
			\coordinate (v1) at (190:\R);
			\coordinate (v2) at (-60:\R);
			\coordinate (v3) at (-10:\R);
			\coordinate (m0) at ($(v0)!.4!(v1)$);
			\coordinate (m1) at ($(v2)!.6!(v3)$);
			\coordinate (p) at ($(m0)!.6!(m1)$);
			\fill [gray!40,opacity=.9] (v2) -- (v3) -- (m0) -- cycle;
			\draw [dotted,thick] ($(v0)!-.24!(v1)$) -- ($(v1)!-.24!(v0)$) node [below] {\scriptsize $\mathbb{CP}^1$} ($(v2)!-.45!(v3)$) -- ($(v3)!-.45!(v2)$) node [above] {\scriptsize $\mathbb{CP}^1$};
			\draw [dotted] ($(v2)!-.3!(p)$) -- ($(p)!-.6!(v2)$) ($(v3)!-.3!(p)$) -- ($(p)!-.8!(v3)$);
			\draw (v0) -- (v1) (v2) -- (v3);
			\foreach \i in {0,1,2,3} \fill [black] (v\i) circle [radius=1.5pt];
			\fill (p) circle [radius=1.5pt];
			\draw [black,fill=white] (m0) circle [radius=1.5pt];
			\node [anchor=north east,inner sep=2pt] at (p) {\scriptsize $X$};
			\node [anchor=south east,inner sep=2pt] at (m0) {\scriptsize $M$};
			\node [anchor=north] at (0,-1.8) {\small (2)};
		\end{scope}
		\begin{scope}[xshift=8cm]
			\coordinate (v0) at (80:\R);
			\coordinate (v1) at (190:\R);
			\coordinate (v2) at (-60:\R);
			\coordinate (v3) at (-10:\R);
			\coordinate (m0) at ($(v0)!.4!(v1)$);
			\coordinate (m1) at ($(v2)!.6!(v3)$);
			\coordinate (p) at ($(m0)!.6!(m1)$);
			\fill [gray!20,opacity=.9] (v0) -- (v1) -- (m1) -- cycle;
			\draw [dotted] ($(v0)!-.25!(p)$) -- ($(p)!-.5!(v0)$) ($(v1)!-.22!(p)$) -- ($(p)!-.4!(v1)$);
			\draw [dotted,thick] ($(v0)!-.24!(v1)$) -- ($(v1)!-.24!(v0)$) node [below] {\scriptsize $\mathbb{CP}^1$} ($(v2)!-.45!(v3)$) -- ($(v3)!-.45!(v2)$) node [above] {\scriptsize $\mathbb{CP}^1$};
			\draw (v0) -- (v1) (v2) -- (v3);
			\foreach \i in {0,1,2,3} \fill [black] (v\i) circle [radius=1.5pt];
			\fill (p) circle [radius=1.5pt];
			\draw [black,fill=white] (m0) circle [radius=1.5pt];
			\draw [black,fill=white] (m1) circle [radius=1.5pt];
			\node [anchor=north east,inner sep=2pt] at (p) {\scriptsize $X$};
			\node [anchor=south east,inner sep=2pt] at (m0) {\scriptsize $M$};
			\node [anchor=north west,inner sep=2pt] at (m1) {\scriptsize $N$};
			\node [anchor=north] at (0,-1.8) {\small (3)};
		\end{scope}
		\begin{scope}[xshift=12cm]
			\coordinate (v0) at (80:\R);
			\coordinate (v1) at (190:\R);
			\coordinate (v2) at (-60:\R);
			\coordinate (v3) at (-10:\R);
			\coordinate (m0) at ($(v0)!.4!(v1)$);
			\coordinate (m1) at ($(v2)!.6!(v3)$);
			\coordinate (p) at ($(m0)!.6!(m1)$);
			\fill [gray!40,opacity=.9] (v3) -- (m0) -- (m1) -- cycle;
			\fill [gray!20,opacity=.9] (v0) -- (m1) -- (v1) -- cycle;
			\fill [gray!40,opacity=.9] (v2) -- (m0) -- (m1) -- cycle;
			\draw [dash dot,thick] ($(m0)!-.3!(m1)$) -- ($(m1)!-.3!(m0)$);
			\draw [dotted,thick] ($(v0)!-.24!(v1)$) -- ($(v1)!-.24!(v0)$) node [below] {\scriptsize $\mathbb{CP}^1$} ($(v2)!-.45!(v3)$) -- ($(v3)!-.45!(v2)$) node [above] {\scriptsize $\mathbb{CP}^1$};
			\draw (v0) -- (v1) (v2) -- (v3);
			\foreach \i in {0,1,2,3} \fill [black] (v\i) circle [radius=1.5pt];
			\fill (p) circle [radius=1.5pt];
			\draw [black,fill=white] (m0) circle [radius=1.5pt];
			\draw [black,fill=white] (m1) circle [radius=1.5pt];
			\node [anchor=north east,inner sep=2pt] at (p) {\scriptsize $X$};
			\node [anchor=east,inner sep=3pt] at (m0) {\scriptsize $M$};
			\node [anchor=west,inner sep=3pt] at (m1) {\scriptsize $N$};
			\node [anchor=north] at (0,-1.8) {\small (4)};
		\end{scope}
	\end{tikzpicture}
	\caption{Uniqueness of the bi-projective parametrization in the case of $k=1$ in $\mathbb{CP}^3$.}
	\label{fig:uniqueness}
\end{figure}

The nice thing for this generalization is that, as long as a point $X\in\mathbb{CP}^d$ does not belong to either $\overline{V_0V_1\cdots V_{d-k-1}}$ or $\overline{V_{d-k}V_{d-k+1}\cdots V_d}$ as illustrated in Figure \ref{fig:uniqueness}(1), it then can be identified by a unique geometry of the above type. To see this, note that the span $\overline{XV_0V_1\cdots V_{d-k-1}}$ is a $\mathbb{CP}^{d-k}$ subspace. It intersects $\overline{V_{d-k}V_{d-k+1}\cdots V_d}$ at a unique point, which we name by $M$. This is depicted in Figure \ref{fig:uniqueness}(2). Similarly, the span $\overline{XV_{d-k}V_{d-k+1}\cdots V_d}$ also intersects $\overline{V_0V_1\cdots V_{d-k-1}}$ at a unique point, which we name by $N$ (Figure \ref{fig:uniqueness}(3)). By dimension counting, one can learn that the intersection $\overline{XV_0V_1\cdots V_{d-k-1}}\cap\overline{XV_{d-k}V_{d-k+1}\cdots V_d}$ should be a $\mathbb{CP}^1$ subspace of $\mathbb{CP}^d$. Since $X$, $M$ and $N$ all belong to this intersection, we conclude that $X\in\overline{MN}$, as illustrated in Figure \ref{fig:uniqueness}(4).

Therefore, as long as $X$ does not belong to the two reference subspaces, it uniquely determines $M$ and $N$. This means that such a point can be equally identified by the collection of three sets of homogeneous coordinates, one for $M\in\overline{V_{d-k}V_{d-k+1}\cdots V_d}$, another for $N\in\overline{V_0V_1\cdots V_{d-k-1}}$, and the third one for $X\in\overline{MN}$. Due to this, such construction can also be viewed as a ``fibration'' of $\mathbb{CP}^d$, in the same sense as what was discussed in the previous subsection.

When we work in the canonical frame, this ``fibration'' can be simply achieved by splitting the original homogeneous coordinates
\begin{align}
	X&=[x_0:x_1:\cdots:x_d]\nonumber\\
	&=[x_0:x_1:\cdots:x_{d-k-1}:\underbrace{0:\cdots:0}_{k+1}]+[\underbrace{0:\cdots:0}_{d-k}:x_{d-k}:x_{d-k+1}:\cdots:x_d].
\end{align}
If we now treat each of the above two terms independently as homogeneous coordinates on each of the two reference subspaces, we can manifestly pull out an overall scale from each, without changing the meaning of the coordinates. In other words, we redefine $x_i=t_1n_i$ ($i=0,1,\ldots,d-k-1$) and $x_j=t_0m_{j-d+k}$ ($j=d-k,d-k+1,\ldots,d$). Therefore we can equally write
\begin{align}\label{eq:generalFibrationCanonical}
	X=t_1[n_0:n_1:\cdots:n_{d-k-1}:\underbrace{0:\cdots:0}_{k+1}]+t_0[\underbrace{0:\cdots:0}_{d-k}:m_{0}:m_{1}:\cdots:m_k].
\end{align}
With this treatment, the $m$, $n$ and $t$ variables form homogeneous variables on their own respectively. And so we think about the integral as being performed in $\mathbb{CP}^k\times\mathbb{CP}^{d-k-1}\times\mathbb{CP}^1$. The total number of dimensions is $k+(d-k-1)+1=d$, which matches that of the original $\mathbb{CP}^d$. 
By setting $k=0$, one can think about the previous ``fibration'' \eqref{eq:pointFibrationCanonical} as a special case of this general construction, where one of the reference subspaces $\mathbb{CP}^k$ reduces to a point. 

\brief{Draw connections between choice of foliations and the touching configurations.}

Because the reference subspace $\mathbb{CP}^k$ is a direct generalization of a reference point $\mathbb{CP}^0$, one can naturally expect that the above ``fibration'' at fixed $k$ is best suited for the study of the analytic properties around the touching configuration of a $k$-face ambient space, e.g., $\overline{U_{d-k}U_{d-k+1}\cdots U_d}$, where the reference $\mathbb{CP}^k$ is identified as this ambient space. The $\mathbb{CP}^1$ ``fibres'' again describe the directions transversal to the singularity $S$, so that one can properly compute the discontinuities by taking residues on each ``fibre'' and then integrate over the moduli space of the ``fibres'', $\mathbb{CP}^k\times\mathbb{CP}^{d-k-1}$. Because the ``fibres'' are now determined by two independent reference subspaces, we can view this residue computation as a projection through both of them simultaneously. In contrast to the projection through a single point discussed in Section \ref{sec:projection}, we name such operation a \emph{bi-projection}.

\brief{Discuss the expectation on the $m$ integration.}

Although the moduli space we directly have now is a product space, as we will see in later analysis, at least when we deal with the case $k=1$ in one-loop Feynman integrals, the integrals in $\mathbb{CP}^k$ can be directly performed after taking residues on the ``fibres'', so that effectively the residues are represented in terms of integrals on $\mathbb{CP}^{d-k-1}$, where the contour is induced from the original $(d-k-1)$-face $\underline{U_0U_1\cdots U_{d-k-1}}$.

\brief{Discuss how the operations used in point projection and in spherical contour fit into this new picture.}

\section{Elementary discontinuities from bi-projection}\label{sec:discHex}

\begin{figure}[ht]
	\centering
	\begin{tikzpicture}
		\draw (60:1) -- (120:1) -- (180:1) -- (240:1) -- (300:1) -- (0:1) -- cycle;
		\draw (60:1) -- (60:1.6) node [right] {\small $1$} (120:1) -- (120:1.6) node [left] {\small $2$} (180:1) -- (180:1.6) node [left] {\small $3$} (240:1) -- (240:1.6) node [left] {\small $4$} (300:1) -- (300:1.6) node [right] {\small $5$} (0:1) -- (0:1.6) node [right] {\small $6$};
		\foreach \i in {0,1,...,5} \fill (60*\i:1) circle [radius=1.5pt];
		\node [anchor=center] at (30:1.1) {\small $x_0$};
		\node [anchor=center] at (90:1.1) {\small $x_1$};
		\node [anchor=center] at (150:1.1) {\small $x_2$};
		\node [anchor=center] at (210:1.1) {\small $x_3$};
		\node [anchor=center] at (270:1.1) {\small $x_4$};
		\node [anchor=center] at (330:1.1) {\small $x_5$};
	\end{tikzpicture}
	\caption{A massless hexagon. Feynman parameters for each loop propagator are indicated.}
	\label{fig:hexagon}
\end{figure}
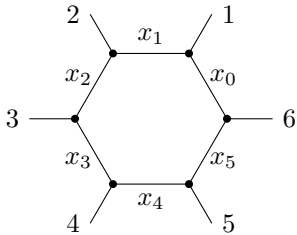

It is good at this point to start looking at some specific integral and show how the method introduced so far applies, especially regarding the detailed computation of the discontinuities. For this purpose we study the following integral in $\mathbb{CP}^5$ in the canonical frame
\begin{align}\label{eq:masslessHexagon}
	I_{\rm hex}=\sqrt{q_0}\int_{\triangledown}\frac{\langle X\mathrm{d}X^5\rangle}{(XQ_0X)^3},\qquad q_0=-\det Q_0
\end{align}
where
\begin{align}
	XQ_0X=&x_0x_2+x_0x_3+x_0x_4+u_1\,x_1x_3+x_1x_4\nonumber\\
	&+x_1x_5+u_2\,x_2x_4+x_2x_5+u_3\,x_3x_5.
\end{align}
This integral comes from a massless hexagon in six spacetime dimensions \cite{Dixon:2011ng,Arkani-Hamed:2017ahv}, as shown in Figure \ref{fig:hexagon}, where the parameters are related to Mandelstam variables by
\begin{align}
	u_1=\frac{s_{23}s_{56}}{s_{234}s_{123}},\qquad
	u_2=\frac{s_{34}s_{61}}{s_{345}s_{234}},\qquad
	u_3=\frac{s_{45}s_{12}}{s_{123}s_{345}}.
\end{align}
The overall normalization factor in \eqref{eq:masslessHexagon} is chosen in order to simplify the final result.

The symbol of this integral has previously been worked out independently by differential equations \cite{Dixon:2011ng} as well as by spherical contours \cite{Arkani-Hamed:2017ahv}. To cleanly express the symbol, we introduce the abbreviation
\begin{align}\label{eq:defxpm}
	x_\pm=\frac{-1+u_1+u_2+u_3\pm\sqrt{q_0}}{2u_1u_2u_3},
\end{align}
together with notations for several ratios
\begin{align}\label{eq:defr}
	r_0=\frac{x_+}{x_-},\qquad r_i=\frac{x_+(1-u_ix_-)}{x_-(1-u_ix_+)},\quad i=1,2,3.
\end{align}
Then
\begin{align}\label{eq:SIhex}
	\mathcal{S}[I_{\rm hex}]=\sum_{\rm cyclic}\left(u_1\otimes(1-u_1)\otimes r_0-u_1\otimes u_2\otimes r_3-u_1\otimes u_3\otimes r_2\right).
\end{align}
Here ``cyclic'' means to sum up cyclic permutations of the labels $\{1,2,3\}$, and note that $x_\pm$ themselves remain invariant under these permutations. This example provides a convenient testing ground for the validity of the method developed in the previous sections.

\subsection{Discontinuities of the massless hexagon $I_{\rm hex}$}\label{sec:discIhex}

Since all the loop propagators are massless, all the $0$-faces are restricted on the quadric singularity $Q_0$. According to the discussions in Section \ref{sec:touchingConfiguration}, to seek for the singularities of $I_{\rm hex}$ we need to identify touching configurations where the ambient space of some $1$-face is embedded in $Q_0$. One easily sees that this can occur only for $\overline{U_1U_3}$, $\overline{U_2U_4}$ and $\overline{U_3U_5}$. Because $Q_0$ has degree two, there is no need to study touching configurations for $k$-faces with $k\geq2$. So the above three cases account for all the singularities of $I_{\rm hex}$. Furthermore, the jumps that lead to these touchings are all of type (e) in Figure \ref{fig:strata1faceQ}. From Section \ref{sec:firstEntries} we learn that their corresponding first entries are $u_1$, $u_2$, and $u_3$, respectively. Since these three parameters are mutually independent, we can already expect that the symbol of $I_{\rm hex}$ has the following structure
\begin{align}\label{eq:SIhexStructure}
	\mathcal{S}[I_{\rm hex}]=u_1\otimes\mathcal{S}[\mathrm{Disc}_{\overline{U_1U_3}}I_{\rm hex}]+u_2\otimes\mathcal{S}[\mathrm{Disc}_{\overline{U_2U_4}}I_{\rm hex}]+u_3\otimes\mathcal{S}[\mathrm{Disc}_{\overline{U_3U_5}}I_{\rm hex}].
\end{align}
Here $\mathrm{Disc}_{\overline{U_iU_j}}$ denotes the discontinuity associated to the touching configuration for $\overline{U_iU_j}$.

Let us first work out $\mathrm{Disc}_{\overline{U_1U_3}}I_{\rm hex}$ in detail. According to the previous section, here we take $\overline{U_1U_3}$ and $\overline{U_0U_2U_4U_5}$ as two reference subspaces and introduce the corresponding ``fibration'' of the contour. Since we are already in the canonical frame, following \eqref{eq:generalFibrationCanonical} this is realized by
\begin{align}
	X=[c_1n_0:c_0m_0:c_1n_1:c_0m_1:c_1n_2:c_1n_3].
\end{align}
To set conventions for the analysis from now on, we always choose $M$ (in correspondence to $[c_0:c_1]=[1:0]$) to sit on the reference subspace that is identical to the ambient space that labels the discontinuity, and choose $N$ (in correspondence to $[c_0:c_1]=[0:1]$) to sit on the other reference subspace. 

When we perform the change of coordinates in an actual integral, of course we need to fix a gauge for each set of homogeneous coordinates. This can be done most conveniently by picking up an affine patch. Without loss of generality, let us fix
\begin{align}
	c_0=m_0=n_0=1,
\end{align}
and correspondingly $x_1=1$ in the original coordinates, then we have the relation
\begin{align}\label{eq:Xcnmgaugefixed}
	X=[x_0:1:x_2:x_3:x_4:x_5]=[c_1:1:c_1n_1:m_1:c_1n_2:c_1n_3].
\end{align}
This is a non-linear transformation, and the Jacobian is
\begin{align}
	\left|\frac{\partial(x_0,x_2,x_3,x_4,x_5)}{\partial(c_1,m_1,n_1,n_2,n_3)}\right|=c_1^3.
\end{align}
In the integral we integrate each new variable over the range $[0,+\infty)$ in the integral. The sign in the Jacobian is determined by requiring that it is positive when the variables are valued in this range.

\begin{figure}[ht]
	\centering
	\begin{tikzpicture}
		\def\R{2.5}
		\coordinate [label=120:{\small $U_1$}] (v0) at (80:\R);
		\coordinate [label=180:{\small $U_3$}] (v1) at (190:\R);
		\coordinate (v2) at (-60:\R);
		\coordinate (v3) at (-10:\R);
		\coordinate [label=180:{\small $M$}] (m) at ($(v0)!.3!(v1)$);
		\coordinate [label=0:{\small $N$}] (n) at ($(v2)!.2!(v3)$);
		\coordinate (c) at ($(m)!.15!(n)$);
		\draw [dash dot,thick] ($(m)!-.2!(n)$) -- (c);
		\fill [gray!20,opacity=.8] (v0) .. controls +(-100:.9) and +(15:.9) .. (v1) .. controls +(0:1.6) and +(-85:1.8) .. cycle;
		\draw [black!50] (v1) .. controls +(0:1.6) and +(-85:1.8) .. (v0);
		\draw (v0) .. controls +(-100:.9) and +(15:.9) .. (v1);
		\draw [dash dot,thick] (c) -- ($(n)!-.2!(m)$);
		\draw [dotted,thick] ($(v0)!-.2!(v1)$) -- ($(v1)!-.2!(v0)$) ($(v2)!-.3!(v3)$) -- ($(v3)!-.3!(v2)$) node [anchor=west] {\small $\overline{U_0U_1U_4U_5}$};
		\draw [dotted] (v0) -- (v2) -- (v1) -- (v3) -- cycle;
		\foreach \i in {0,1,2,3} \fill [black] (v\i) circle [radius=1.5pt];
		\node [anchor=north east,inner sep=1pt] at ($(c)+(-60:3pt)$) {\small $C^\pm$};
		\fill (m) circle [radius=1.5pt];
		\fill (n) circle [radius=1.5pt];
		\fill (c) circle [radius=1.5pt];
	\end{tikzpicture}
	\caption{``Fibration'' in $\mathbb{CP}^5$: Massless hexagon example. In order to make the structure manifest, we represent the $0145$ directions by a single line.}
	\label{fig:discHexagon}
\end{figure}
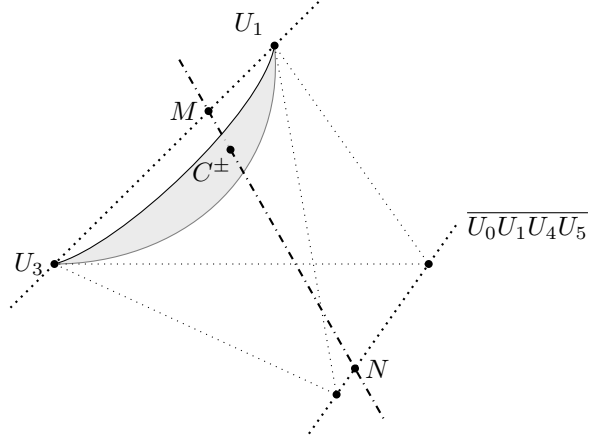

With the new coordinates, the quadric reads
\begin{align}
	XQ_0X=u_1m_1+\underbrace{(m_1+n_2+n_3+u_3m_1n_3)}_{A_1}c_1+\underbrace{(n_1+n_2+u_2n_1n_2+n_1n_3)}_{A_2}c_1^2.
\end{align}
On each ``fibre'', the quadric carves out two points $C^{\pm}$, as illustrated in Figure \ref{fig:discHexagon}, whose corresponding $c_1$ values are the two roots of the above polynomial
\begin{align}\label{eq:cA1A2}
	c_1=\frac{-A_1\pm\sqrt{A_1^2-4u_1m_1A_2}}{2A_2}\equiv c_1^\pm.
\end{align}
These are manifest as poles in $c_1$ in the integrand. To compute the discontinuity we should take residue of the integrand at one of these poles, and the choice has to be in accordance with the intuitive picture in Figure \ref{fig:discHexagon}. This means that, for any fixed values of $n$'s and treating $c_1$ as a function of $m_1$ via \eqref{eq:cA1A2}, we should choose from $C^\pm$ such that it collides with $U_1$ when $m_1\to0$ and with $U_3$ when $m_1\to\infty$.

Let us assume that $n$'s are all strictly non-negative real numbers. By setting $m_1=0$, we have $c_1^+=0$, $c_1^-=-\frac{n_2+n_3}{A_2}$. In this case we always have $C^+=[0:1:0:0:0:0]=U_1$, and so at least in the neighborhood of $m_1=0$ we should compute residue at $c_1=c_1^+$.

When $m_1\to\infty$, by \eqref{eq:Xcnmgaugefixed} the homogeneous coordinate $x_4\to\infty$ as well. In order to see the structure clearly, we'd better rescale the homogeneous coordinates by $1/m_1$ so that they remain finite. By expanding $c_1^\pm/m_1$ with respect to $1/m_1$ we get
\begin{align}
	\frac{c_1^\pm}{m_1}&=\frac{-(1+u_3n_3)\pm\frac{1}{m_1}\sqrt{(1+u_3n_3)^2m_1^2}}{2A_2}+\mathcal{O}(m_1^{-1}).
\end{align}
Now the situation divides into two cases. When $\sqrt{(1+u_3n_3)^2m_1^2}=(1+u_3n_3)m_1$, we have $C^+\to U_3$ at $m_1\to\infty$.  On the contrary, when $\sqrt{(1+u_3n_3)^2m_1^2}=-(1+u_3n_3)m_1$, we have $C^-\to U_3$ instead.
At first sight this seems to cause a problem. While in the former case we can safely take residues at $c_1=c_1^+$ and perform the rest of the integrals, in the latter case it appears that we need to divide the $m$ and $n$ integrals into regions such that the residue is computed either at $c_1=c_1^+$ or $c_1=c_1^-$ accordingly.

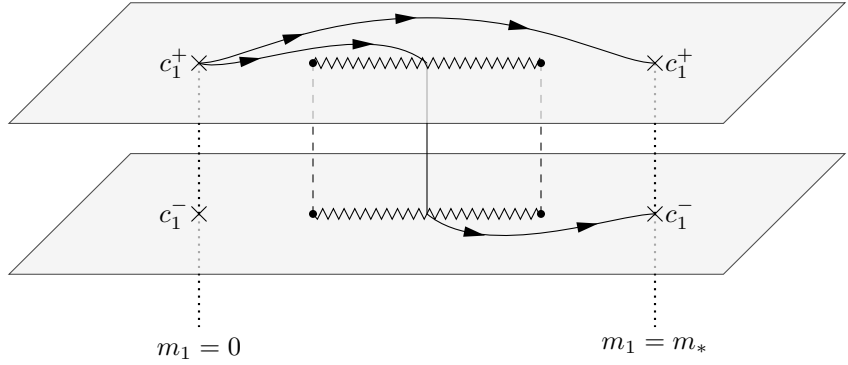
\begin{figure}[ht]
	\centering
	\begin{tikzpicture}
		\draw [dotted,thick] (-1.5,-2) -- ++(0,-1.5) node [below] {\small $m_1=0$} (4.5,-2) -- ++(0,-1.5) node [below] {\small $m_1=m_*$};
		\fill [gray!10,opacity=0.7,draw=black] (-4,-2.8) -- (-2.4,-1.2) -- (7,-1.2) -- (5.4,-2.8) -- cycle;
		\draw [dashed] (0,0) -- (0,-2) (3,0) -- (3,-2);
		\draw [dotted,thick] (-1.5,-2) -- +(0,2) (4.5,-2) -- +(0,2);
		\begin{scope}[decoration={markings,mark={between positions 0.2 and 0.8 step 1.5cm with {\fill (-2pt,2pt) -- (6pt,0) -- (-2pt,-2pt) -- cycle;}}}]
			\draw (1.5,0) -- (1.5,-2);
			\draw [postaction={decorate}] (1.5,-2) .. controls +(-40:1) and +(180:.5) .. (4.5,-2);
		\end{scope}
		\fill [gray!10,opacity=0.7,draw=black] (-4,-.8) -- (-2.4,.8) -- (7,.8) -- (5.4,-.8) -- cycle;
		\draw [decorate,decoration={zigzag,segment length=4pt,amplitude=2pt}] (0,0) -- (3,0) (0,-2) -- (3,-2);
		\fill (0,0) circle [radius=1.5pt];
		\fill (3,0) circle [radius=1.5pt];
		\fill (0,-2) circle [radius=1.5pt];
		\fill (3,-2) circle [radius=1.5pt];
		\draw ($(-1.5,0)+(45:4pt)$) -- +(-135:8pt) ($(-1.5,0)+(-45:4pt)$) -- +(135:8pt);
		\draw ($(-1.5,-2)+(45:4pt)$) -- +(-135:8pt) ($(-1.5,-2)+(-45:4pt)$) -- +(135:8pt);
		\draw ($(4.5,0)+(45:4pt)$) -- +(-135:8pt) ($(4.5,0)+(-45:4pt)$) -- +(135:8pt);
		\draw ($(4.5,-2)+(45:4pt)$) -- +(-135:8pt) ($(4.5,-2)+(-45:4pt)$) -- +(135:8pt);
		\begin{scope}[decoration={markings,mark={between positions 0.2 and 0.8 step 1.5cm with {\fill (-2pt,2pt) -- (6pt,0) -- (-2pt,-2pt) -- cycle;}}}]
			\draw [postaction={decorate}] (-1.5,0) .. controls +(0:.5) and +(180:1.6) .. (1.5,.6) .. controls +(0:1.6) and +(180:.5) .. (4.5,0);
			\draw [postaction={decorate}] (-1.5,0) .. controls +(-20:.5) and +(140:1) .. (1.5,0);
		\end{scope}
		\node [anchor=east] at (-1.5,0) {\small $c_1^+$};
		\node [anchor=east] at (-1.5,-2) {\small $c_1^-$};
		\node [anchor=west] at (4.5,0) {\small $c_1^+$};
		\node [anchor=west] at (4.5,-2) {\small $c_1^-$};
	\end{tikzpicture}
	\caption{Analytic continuation of the end point in the $m_1$ integral. $c_1^+$ may pass the branch cut and turn into $c_1^-$ as $m_1$ continuously deforms.}
	\label{fig:cContinuation}
\end{figure}

It should be emphasized that the above naive understanding is not correct. There is in fact not an essential difference between $c_1^+$ and $c_1^-$. If we temporarily fix $n$'s and treat $c_1^\pm$ as functions of $m_1$, they actually live on a common double-cover of $\mathbb{CP}^1$. The two Riemann sheets in this double-cover are glued by a branch cut anchored at $c_1=-\frac{A_1}{2A_2}\big|_{m_1=m_1^\pm}$, where $m_1^\pm$ are the two roots of $A_1^2-2u_1m_1A_2$, as illustrated in Figure \ref{fig:cContinuation}. For either $c_1^+\big|_{m_1\to\infty}$ or $c_1^-\big|_{m_1\to\infty}$, we can always find a smooth path that connects it with $c_1^+\big|_{m_1\to0}$, and the path is parametrized by $m_1$ (note that $m_1$ does not have to always stay within the real axis during integration). The only difference is that in the former case the path entirely belongs to the first Riemann sheet, while in the latter case it has to pass the branch cut and enter the second sheet. Therefore two cases are merely related by analytic continuations.

To clarify the analysis, we can begin by thinking that the $m_1$ integral is performed along a contour that goes from $0$ to some sufficiently small $m_*$ instead. In this situation we can safely take residues at $c_1=c_1^+$ and then directly carry out the $m_1$ integral. Let us first identify the residue contour for $c_1$. Because we have associated $u_1$ as the first entry in correspondence to this discontinuity, we expand $c_1^+$ around $u_1=0$, yielding
\begin{align}
	c_1^+=\frac{-A_1+\sqrt{A_1^2}}{2A_2}-\frac{m_1\sqrt{A_1^2}}{A_1^2}u_1+\mathcal{O}(u_1^2)=-\frac{m_1}{A_1}u_1+\mathcal{O}(u_1^2).
\end{align}
Here the second equality is justified by  $|m_1|$ being sufficiently small. Regardless of the coefficient in the linear term, since the discontinuity is defined by analytically continuing $u_1$ around $u_1=0$ counter-clockwisely, during this process, $c_1^+$ also continues around $c_1=0$ in the same direction. Therefore the residue contour on the ``fibre'' for the discontinuity is in the clockwise direction. This means that, to compute the discontinuity $\mathrm{Disc}_{\overline{U_1U_3}}I_{\rm hex}$, we should modify the integral on the ``fibre'' to the negative of the residue of the integrand at $c_1=c_1^+$
\begin{align}\label{eq:c1residue}
	-\underset{c_1=c_1^+}{\mathrm{Res}}\frac{c_1^3\sqrt{q_0}}{(XQ_0X)^3}
	=\frac{3\sqrt{q_0}u_1m_1A_1}{(A_1^2-4u_1m_1A_2)^{5/2}}=\frac{\partial R(m_1)}{\partial m_1}.
\end{align}
Before we move on, it is worth to comment that if we began by gauge-fixing $c_1=1$ instead and integrating $c_0$ from $0$ to $+\infty$ in the original integral, then the singularity occurs when the pole hits $c_0=\infty$. By the same logic as above one can find that in this case we simply compute the residue without adding an extra minus sign. This guarantees to yield the same expression as in \eqref{eq:c1residue}.

Note that $A_1$ is linear in $m_1$ while $A_2$ is independent of $m_1$. Regardless of their detailed expressions, it can be verified that the above result is always a derivative of some function $R(m_1)$, which has the schematic form
\begin{align}
	R(m_1)=\frac{P(m_1)}{(A_1^2-4u_1m_1A_2)^{3/2}},
\end{align}
where $P(m_1)$ is a degree-$3$ polynomial in $m_1$. Therefore the integral of $m_1$ boils down to the evaluation of $R(m_1)$ at the two end points of the $m_1$ contour
\begin{align}
	-\int_0^{m_*}\mathrm{d}m_1\,\underset{c_1=c_1^+}{\mathrm{Res}}\frac{c_1^3\sqrt{q_0}}{(XQ_0X)^3}=R(m_*)-R(0).
\end{align}

Now to compute the actual discontinuity $\mathrm{Disc}_{\overline{U_1U_3}}I_{\rm hex}$ we should analytically continue $m_*$ to $\infty$. Pay attention that as a function of $m_1$, $R(m_1)$ contains exactly the same branch cut as $c_1^+(m_1)$ in \eqref{eq:cA1A2}. If it occurs that, during this analytic continuation, $c_1^+(m_*)$ passes the branch cut and turns into $c_1^-(m_*)$, then $R(m_*)$ should simultaneously pass the branch cut and turn into $-R(m_*)$. Therefore, depending on which of $C^\pm(m_*)\to U_3$, the $m_1$ integral in the discontinuity should be computed as
\begin{align}
	-\int_0^{\infty}\mathrm{d}m_1\,\underset{c_1=c_1^+}{\mathrm{Res}}\frac{c_1^3\sqrt{q_0}}{(XQ_0X)^3}=
	\begin{cases}
		R(\infty)-R(0),&C^+(\infty)=U_3,\\
		-R(\infty)-R(0),&C^-(\infty)= U_3.
	\end{cases}
\end{align}
Recall the evaluation of the square root differs by a sign in the two cases. This means that even though the $m_1$ integrals are carried out in different ways in the two cases, they actually lead to exactly the same result. Therefore, the discontinuity is insensitive to the region of the parameters during intermediate analysis, and in practice we can choose whatever region that we feel convenient to carry out the computation. This is reminiscent of the fact that discontinuities themselves can be analytically continued as independent functions, as already commented in the $\mathrm{Li}_2(z)$ example in Figure \ref{fig:Li2Disc}(5).

Taking into consideration the remaining $n$ integrals, the explicit expression for the discontinuity is thus
\begin{align}
	\mathrm{Disc}_{\overline{U_1U_3}}I_{\rm hex}
	&=\int_0^\infty\mathrm{d}n_1\mathrm{d}n_2\mathrm{d}n_3\,\frac{\sqrt{q_0}\,u_1}{2(N\widetilde{Q}_1N)^2},\label{eq:mIntegration1}
\end{align}
where $N=[1:n_1:n_2:n_3]$ and
\begin{align}\label{eq:Q1matrix}
	\widetilde{Q}_1=\left(\begin{matrix}0&\frac{u_1}{2}&-\frac{1-u_1}{2}&-\frac{1}{2}\\\frac{u_1}{2}&0&\frac{u_1u_2}{2}&\frac{u_1}{2}\\-\frac{1-u_1}{2}&\frac{u_1u_2}{2}&0&-\frac{u_3}{2}\\-\frac{1}{2}&\frac{u_1}{2}&-\frac{u_3}{2}&-u_3\end{matrix}\right).
\end{align}

In the next subsection we will study the singularities and discontinuities of this new function. For the convenience of notation, let us complete the $n$ variables into homogeneous coordinates of $\mathbb{CP}^3$ and denote them again by $X$, i.e., $N\mapsto X=[x_0:x_1:x_2:x_3]$. This turns the discontinuity into an integral in the canonical frame
\begin{align}\label{eq:discU13canonical}
	I_{\rm hex}^{(1)}\equiv\mathrm{Disc}_{\overline{U_1U_3}}I_{\rm hex}=\int_{\triangledown}\frac{\sqrt{q_0}\,\langle X\mathrm{d}X^3\rangle}{2u_1(XQ_1X)^2},\qquad Q_1=\widetilde{Q}_1/u_1.
\end{align}
Here we also choose to rescale parameters in the quadric. Such operation turns more coefficients in the quadric to constants, which will be convenient for later analysis.

To emphasize once again, even though we obtained $I_{\rm hex}^{(1)}$ by studying the analytic continuation of $I_{\rm hex}$ in the neighborhood of $u_1=0$, once we have this function we can continue $u_1$ (as well as the other parameters $u_2$ and $u_3$) to arbitrary values when investigating its own analytic properties. In other words, $\mathrm{Disc}_{\overline{U_1U_3}}I_{\rm hex}$ can be treated as an independent function, regardless of where it originates.

\subsection{Discontinuities of the discontinuity $I_{\rm hex}^{(1)}$}\label{sec:discIhex1}

We now analyze the analytic structure of the discontinuity $I_{\rm hex}^{(1)}\equiv\mathrm{Disc}_{\overline{U_1U_3}}I_{\rm hex}$, whose explicit integral expression is presented in \eqref{eq:discU13canonical}. From the expression of the quadric $Q_1$ in \eqref{eq:Q1matrix}, we see that three of the $0$-faces $U_0,U_1,U_2\in Q_1$, while $U_3\notin Q_1$. From the perspective of the stratification of touching configurations in Figure \ref{fig:strata1faceQ}, this integral possesses three types of jumps
\begin{itemize}
	\item[(b)] Such jump exists when considering the relation between $\overline{U_iU_3}$ ($i=0,1,2$) and $Q_1$. When the jump occurs $\overline{U_iU_3}$ becomes tangent to $Q_1$ at $U_i$. From the discussion in Section \ref{sec:firstEntries}, in each of the three cases we identify the first entry by the corresponding off-diagonal element of $Q_1$, i.e.,
	\begin{align}
		\overline{U_0U_3}:\; \frac{1}{u_1},\qquad
		\overline{U_2U_3}:\; \frac{u_3}{u_1}.
	\end{align}
	Note here we omit all constant factors, since they play no role in the symbol. In particular, because $(Q_1)_{13}$ is purely a constant, we do not consider any first entry for $\overline{U_1U_3}$ (i.e., it never becomes tangent to $Q_1$ at $U_1$).
	\item[(c)] Such jump exists when considering $\overline{U_iU_3}$ ($i=0,1,2$) as well. In all three cases $U_3$ falls onto $Q_1$ when the jump occurs. Because all the other three $0$-faces are already on $Q_1$, in all these cases we commonly associate $(Q_1)_{33}$ to the first entry, and so they are equivalent to merely focusing on the stratification with respect to the $0$-face $U_3$. Therefore we identify the first entry
	\begin{align}
		\overline{U_3}:\; \frac{u_3}{u_1}.
	\end{align}
	\item[(e)] Such jump exists in the stratifications for $\overline{U_0U_2}$ and $\overline{U_1U_2}$, respectively. When it happens, the corresponding ambient space becomes fully embedded in $Q_1$. Accordingly the first entries are
	\begin{align}
		\overline{U_0U_2}:\; \frac{1-u_1}{u_1},\qquad
		\overline{U_1U_2}:\; u_2.
	\end{align}
\end{itemize}
From the discussion above, we may expect the symbol of this discontinuity to have the following form
\begin{align}\label{eq:I1hexStructureRaw}
	\mathcal{S}[I_{\rm hex}^{(1)}]&=\underbrace{\frac{1}{u_1}\otimes\mathcal{S}[\mathrm{Disc}_{\overline{U_0U_3}}I_{\rm hex}^{(1)}]+\frac{u_3}{u_1}\otimes\mathcal{S}[\mathrm{Disc}_{\overline{U_2U_3}}I_{\rm hex}^{(1)}]}_{\text{(b)}}+\underbrace{\frac{u_3}{u_1}\otimes\mathcal{S}[\mathrm{Disc}_{\overline{U_3}}I_{\rm hex}^{(1)}]}_{\text{(c)}}\nonumber\\
	&\quad+\underbrace{\frac{1-u_1}{u_1}\otimes\mathcal{S}[\mathrm{Disc}_{\overline{U_0U_2}}I_{\rm hex}^{(1)}]+u_2\otimes\mathcal{S}[\mathrm{Disc}_{\overline{U_1U_2}}I_{\rm hex}^{(1)}]}_{\text{(e)}}.
\end{align}
Note that the first entries for this integral are not all independent, and different symbol terms may contain a common factor in their first entries, e.g., $u_3$. This just means that, when we analytically continue $u_3$ around $u_3=0$, the complete discontinuity will receive several different contributions. 

According to the discussion in Section \ref{sec:discontinuityFirstLook}, there are three types of discontinuity contour depicted in Figure \ref{fig:DiscontinuityContourShape}, which need to be treated differently. They come from different jump types. We will deal with them one by one in the following.

\subsubsection{Elementary discontinuities of type (iii)}\label{sec:discJumpe}

Let us begin by studying elementary discontinuities of type (iii). The jump type (e) leads to this contour, and we have already encountered such type when studying $I_{\rm hex}$. Hence the analysis follows that in Section \ref{sec:discIhex}.

Take $\mathrm{Disc}_{\overline{U_0U_2}}I_{\rm hex}^{(1)}$ as an example. Here we reparametrize the coordinates by
\begin{align}
	X=[c_0m_0:c_1n_0:c_0m_1:c_1n_1],
\end{align}
and gauge-fix $c_0=m_0=n_0=1$. The transformation gives rise to the Jacobian $c_1$. Under this parametrization the quadric becomes
\begin{align}
	XQ_1X\propto \underbrace{(u_1-1)m_1}_{A_0}+\underbrace{(u_1-n_1+(u_1u_2-u_3n_1)m_1)}_{A_1}c_1+\underbrace{(u_1-u_3n_1)n_1}_{A_2}c_1^2.
\end{align}
Therefore on each ``fibre'' the two poles are located at $c_1^\pm=\frac{-A_1\pm\sqrt{A_1^2-4A_0A_2}}{2A_2}$. To find the correct contour, we again inspect these solutions in two limits of $m_1$
\begin{subequations}
	\begin{align}
		m_1\to0:&\quad c_1^\pm=\frac{n_1-u_1\pm\sqrt{(u_1-n_1)^2}}{2n_1(u_1-u_3n_1)}+\mathcal{O}(m_1),\\
		m_1\to\infty:&\quad \frac{c_1^\pm}{m_1}=\frac{-u_1u_2+u_3n_1\pm\sqrt{(u_1u_2-u_3n_1)^2}}{2n_1(u_1-u_3n_1)}+\mathcal{O}(m_1^{-1}).
	\end{align}
\end{subequations}
Apparently, depending on the region of the remaining variables, we can have either $C^+$ or $C^-$ to approach the desired $0$-face. So altogether there are four possible situations. For the time being let us assume we are in the region such that $C^+$ approaches $U_0$ and $U_2$ in the two limits respectively, then we have the conditions
\begin{align}\label{eq:I1hex02condition}
	\sqrt{(u_1-n_1)^2}=u_1-n_1,\qquad
	\sqrt{(u_1u_2-u_3n_1)^2}=u_1u_2-u_3n_1.
\end{align}
By taking residue of the integrand and integrating $m_1$, it can again be shown that
\begin{align}
	-\underset{c_1=c_1^+}{\mathrm{Res}}\frac{c_1\sqrt{q_1}}{2u_1(XQ_1X)^2}=\frac{\partial}{\partial m_1}\underbrace{\frac{P(m_1)}{\sqrt{A_1^2-4A_0A_2}}}_{R(m_1)},
\end{align}
where $P(m_1)$ is now a degree-$1$ polynomial in $m_1$. Therefore we have
\begin{align}\label{eq:I1hexDisc02nInt}
	\mathrm{Disc}_{\overline{U_0U_2}}I_{\rm hex}^{(1)}&=-\int_0^\infty\mathrm{d}n_1\mathrm{d}m_1\underset{c_1=c_1^+}{\mathrm{Res}}\frac{c_1\sqrt{q_1}}{2u_1(XQ_1X)^2}=\int_0^\infty\mathrm{d}n_1(R(\infty)-R(0))\nonumber\\
	&=\int_0^\infty\mathrm{d}n_1\,\frac{\sqrt{q_0}}{-u_1u_2+(u_1+u_2+u_3-1)n_1-u_3n_1^2}.
\end{align}

For the other three types of parameter regions, we have to choose $C^-$ to approach the desired $0$-face at $m_1\to0$ and/or $m_1\to\infty$. On the one hand, either or both of the conditions in \eqref{eq:I1hex02condition} should flip sign accordingly. On the other hand, before we evaluate $R(m_1)$ at $0$ and/or $\infty$ we need to analytically continue it to the other Riemann sheet so that it acquires an extra sign as well. The explicit computations in all these cases are
\begin{subequations}
	\begin{align}
		C^+(0)=U_0,\;C^-(\infty)=U_2:&\quad\mathrm{Disc}_{\overline{U_0U_2}}I_{\rm hex}^{(1)}=\int_0^\infty\mathrm{d}n_1(-R(\infty)-R(0)),\\
		C^-(0)=U_0,\;C^+(\infty)=U_2:&\quad\mathrm{Disc}_{\overline{U_0U_2}}I_{\rm hex}^{(1)}=\int_0^\infty\mathrm{d}n_1(R(\infty)+R(0)),\\
		C^-(0)=U_0,\;C^-(\infty)=U_2:&\quad\mathrm{Disc}_{\overline{U_0U_2}}I_{\rm hex}^{(1)}=\int_0^\infty\mathrm{d}n_1(-R(\infty)+R(0)).
	\end{align}
\end{subequations}
In particular, in the last case we can alternatively think that the residue is computed at $c_1=c_1^-$ for every value of $m_1$ along its integration. Nevertheless, we are allowed to first compute the residue at $c_1=c_1^+$ and analytically continue it to the second sheet afterwards. By the reasoning in Section \ref{sec:discIhex}, the computation in all these cases yields the same result \eqref{eq:I1hexDisc02nInt}, despite they are carried out in different ways. The remaining $n_1$ integration is simple, which finally gives
\begin{align}
	\mathrm{Disc}_{\overline{U_0U_2}}I_{\rm hex}^{(1)}=\log r_0,
\end{align}
where $r_0$ is the ratio defined in \eqref{eq:defr}.

The exact same workflow also determines the other discontinuity of this type. We omit the details of the computation and just print the result
\begin{align}
	\mathrm{Disc}_{\overline{U_1U_2}}I_{\rm hex}^{(1)}=\log\frac{1}{r_3},
\end{align}
where $r_3$ was defined in \eqref{eq:defr} as well.

\subsubsection{Elementary discontinuities of type (ii)}\label{sec:discJumpb}

We now move to elementary discontinuities of type (ii), which comes from jump type (b). The analysis of type (ii) elementary discontinuities is in a way very similar to that of type (iii) discussed previously. The reason is that, as shown in Figure \ref{fig:DiscontinuityContourShape}, the shapes of discontinuity contours in both types share the same crescent topology. The main difference is that, while type (iii) both tips of the crescent coincide with some $0$-face of $\triangledown$, in type (ii) only one tip is fixed in this way, and the other tip (let us call it $W$) can freely move along a $1$-face (its position depends on the shape of the quadric).

Let us use $\mathrm{Disc}_{\overline{U_0U_3}}I_{\rm hex}^{(1)}$ to illustrate the analysis in this type. This again begins by separating the four $0$-faces into two sets $\{U_0,U_3\}$ and $\{U_1,U_2\}$, and introducing the reparametrization
\begin{align}
	X=[c_0m_0:c_1n_0:c_1n_1:c_0m_1],
\end{align}
with the gauge-fixing $c_0=m_0=n_0=1$. The Jacobian for the corresponding transformation is $c_1$. With this reparametrization the quadric reads
\begin{align}
	XQ_1X\propto\underbrace{-m_1(1+u_3m_1)}_{A_0}+\underbrace{(u_1-n_1+u_1n_1+(u_1-u_3n_1)m_1)}_{A_1}c_1+\underbrace{u_1u_2n_1}_{A_2}c_1^2.
\end{align}
Again we name the two roots as $c_1^\pm=\frac{-A_1\pm\sqrt{A_1^2-4A_0A_2}}{2A_2}$. Because the quadric $Q_1$ intersects $\overline{U_0U_3}$ at $U_0$ and $W=[1:0:0:-1/u_3]$. Therefore the two tips of the crescent correspond to $m_1\to0$ and $m_1\to-1/u_3$. In these limits the roots behave like
\begin{subequations}
	\begin{align}
		c_1^\pm\big|_{m_1=0}&=\frac{-u_1+n_1-u_1n_1\pm\sqrt{(u_1-n_1+u_1n_1)^2}}{2u_1u_2n_1},\\
		c_1^\pm\big|_{m_1=-\frac{1}{u_3}}&=\frac{-u_1(1-\frac{1}{u_3}+n_1)\pm\sqrt{\frac{u_1^2(1-u_3-u_3n_1)^2}{u_3^2}}}{2u_1u_2n_1}.
	\end{align}
\end{subequations}
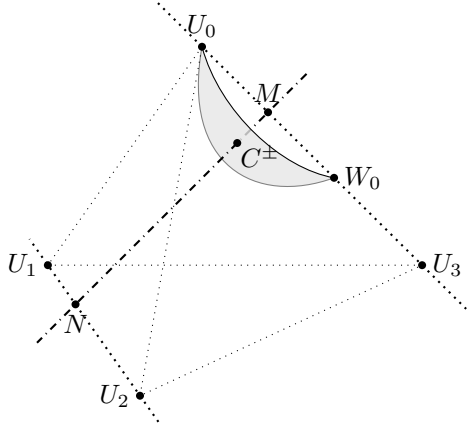
\begin{figure}[ht]
	\centering
	\begin{tikzpicture}
		\def\R{2.5}
		\coordinate [label=90:{\small $U_0$}] (v0) at (100:\R);
		\coordinate [label=180:{\small $U_1$}] (v1) at (190:\R);
		\coordinate [label=180:{\small $U_2$}] (v2) at (-120:\R);
		\coordinate [label=0:{\small $U_3$}] (v3) at (-10:\R);
		\coordinate [label=0:{\small $W_0$}] (w) at ($(v0)!.6!(v3)$);
		\coordinate [label=90:{\small $M$}] (m) at ($(v0)!.3!(v3)$);
		\coordinate [label=-90:{\small $N$}] (n) at ($(v2)!.7!(v1)$);
		\coordinate (c) at ($(m)!.16!(n)$);
		\draw [dash dot,thick] ($(m)!-.2!(n)$) -- (c);
		\fill [gray!20,opacity=.8] (v0) .. controls +(-75:.9) and +(170:.7) .. (w) .. controls +(200:1.1) and +(-100:1.6) .. cycle;
		\draw [black!50] (w) .. controls +(200:1.1) and +(-100:1.6) .. (v0);
		\draw (v0) .. controls +(-75:.9) and +(170:.7) .. (w);
		\draw [dash dot,thick] (c) -- ($(n)!-.2!(m)$);
		\draw [dotted,thick] ($(v0)!-.2!(v3)$) -- ($(v3)!-.2!(v0)$) ($(v2)!-.2!(v1)$) -- ($(v1)!-.2!(v2)$);
		\draw [dotted] (v0) -- (v1) -- (v3) -- (v2) -- cycle;
		\foreach \i in {0,1,2,3} \fill [black] (v\i) circle [radius=1.5pt];
		\node [anchor=north west,inner sep=1pt] at (c) {\small $C^\pm$};
		\fill (w) circle [radius=1.5pt];
		\fill (m) circle [radius=1.5pt];
		\fill (n) circle [radius=1.5pt];
		\fill (c) circle [radius=1.5pt];
	\end{tikzpicture}
	\caption{``Fibration'' in $\mathbb{CP}^3$: Elementary discontinuities of type (ii).}
	\label{fig:discTangent}
\end{figure}

According to the discussion previously, we can choose whatever parameter region that we feel convenient (even if it may not be the actual region that we encounter during the $n$ integration). So without loss of generality we take the region such that both tips associate to $c_1^+$, in correspondence to the conditions
\begin{align}
	\sqrt{(u_1-n_1+u_1n_1)^2}=(u_1-n_1+u_1n_1),\qquad
	\sqrt{\frac{u_1^2(1-u_3-u_3n_1)^2}{u_3^2}}=\frac{u_1(1-u_3-u_3n_1)}{u_3}.
\end{align}
The result of the $m_1$ integral obtained in this region should be valid for any other region as well. Consequently, we can compute the discontinuity by
\begin{align}
	\mathrm{Disc}_{\overline{U_0U_3}}I_{\rm hex}^{(1)}&=-\int_0^\infty\mathrm{d}n_1\int_0^{-\frac{1}{u_3}}\mathrm{d}m_1\underset{c_1=c_1^+}{\mathrm{Res}}\frac{c_1\sqrt{q_1}}{2u_1(XQ_1X)^2}\nonumber\\
	&=\log\frac{1}{r_1r_3}.\label{eq:discU0U3}
\end{align}

A similar computation also yields
\begin{align}
	\mathrm{Disc}_{\overline{U_2U_3}}I_{\rm hex}^{(1)}=\log\frac{1}{r_1r_2}.
\end{align}

\subsubsection{Elementary discontinuities of type (i)}\label{sec:discJumpc}

Finally let us analyze elementary discontinuities of type (i), which comes from the jump type (c), or in other words, the discontinuity $\mathrm{Disc}_{\overline{U_3}}I_{\rm hex}^{(1)}$ in \eqref{eq:I1hexStructureRaw}. In this case $M$ is fixed at $U_3$, so the problem exactly reduces to the point projection discussed in \cite{Gong:2022erh}. Nevertheless, in the following we will analyze this case in detail for completeness of the discussion, and towards the end we will show that this point projection gives rise to an alternative way to compute the previous elementary discontinuities of type (ii).

Because $M$ is fixed, we do not introduce any $m$ variables, and the reparametrization for the computation of $\mathrm{Disc}_{\overline{U_3}}I_{\rm hex}^{(1)}$ reads
\begin{align}
	X=[c_1n_0:c_1n_1:c_1n_2:c_0].
\end{align}
This time we use the gauge-fixing $c_1=n_0=1$, and the corresponding Jacobian is simply $1$. The quadric now reads
\begin{align}
	XQ_1X\propto\underbrace{-u_1n_1+(1-u_1)n_2-u_1u_2n_1n_2}_{A_0}+\underbrace{(1-u_1n_1+u_3n_2)}_{A_1}c_0+u_3c_0^2.
\end{align}
Like before, we denote the two roots as $c_0^\pm=\frac{-A_1\pm\sqrt{A_1^2-4u_3A_0}}{2u_3}$.

From Figure \ref{fig:DiscontinuityContourShape} we know that the contour for the discontinuity is anchored at three vertices, which arise from the intersections $\overline{U_iU_3}\cap Q_1$ ($i=0,1,2$). Recall that every point on the contour is obtained by computing residues at either $c_0=c_0^+$ or $c_0=c_0^-$. To fully specify the contour we again need to learn which choice leads to the desired vertex (while the other one leads to $U_i$). To approach these limits we need to tune values of $[n_0:n_1:n_2]$ to $[1:0:0]$, $[0:1:0]$ and $[0:0:1]$, respectively. Since we have already gauge-fixed $n_0=1$, these limits are taken by
\begin{subequations}
	\begin{align}
		[1:0:0]:&\quad c_0^+\big|_{n_1,n_2\to0}=0,\quad c_0^-\big|_{n_1,n_2\to0}=-\frac{1}{u_3},\\
		[0:1:0]:&\quad \frac{c_0^{\pm}}{n_1}\big|_{n_1\to\infty}=\frac{u_1\pm\sqrt{u_1^2}}{2u_3},\\
		[0:0:1]:&\quad \frac{c_0^\pm}{n_2}\big|_{n_2\to\infty}=\frac{-u_3\pm\sqrt{u_3^2}}{2u_3}.
	\end{align}
\end{subequations}
It is easy to see that, in the limit $[n_0:n_1:n_2]\to[1:0:0]$, $C^+\sim U_0$. Therefore in this limit the contour is anchored at $C^-=[u_3:0:0:-1]$. In the other two limits we again observe the issue of parameter regions. For simplicity, let us assume that
\begin{align}\label{eq:discU3Assumption1}
	\sqrt{u_1^2}=-u_1,\qquad \sqrt{u_2^3}=u_3.
\end{align}
In this way $C^+$ always approaches $0$-faces of $\triangledown$ in these limits, and so we can simply think that all points on the contour are tied to the root $c_0^-$ in the corresponding ``fibre''. Hence the discontinuity is computed by
\begin{align}
	\mathrm{Disc}_{\overline{U_3}}I_{\rm hex}^{(1)}&=\int_0^\infty\mathrm{d}n_1\mathrm{d}n_2\,\underset{c_0=c_0^-}{\mathrm{Res}}\frac{\sqrt{q_0}}{2u_1(XQ_1X)^2}\nonumber\\
	&=\int_0^\infty\mathrm{d}n_1\mathrm{d}n_2\,\frac{\sqrt{q_0}u_1u_3}{(A_1^2-4u_3A_0)^{3/2}}.
\end{align}
Note that with the assumption \eqref{eq:discU3Assumption1} this integrand never passes the branch cut during the $n$ integration, as the three vertices of the contour all stay on the same first Riemann sheet.
\begin{figure}[ht]
	\centering
	\begin{tikzpicture}
		\def\R{2.5}
		\coordinate (v0) at (100:\R);
		\coordinate (v1) at (190:\R);
		\coordinate (v2) at (-120:\R);
		\coordinate [label=90:{\small $U_3$}] (v3) at (-10:\R);
		\coordinate [label=90:{\small $W_0$}] (w0) at ($(v0)!.6!(v3)$);
		\coordinate [label=-90:{\small $W_1$}] (w1) at ($(v1)!.4!(v3)$);
		\coordinate [label=-90:{\small $W_2$}] (w2) at ($(v2)!.5!(v3)$);
		\coordinate (n) at ($(v0)+(-115:\R)$);
		\coordinate (c) at ($(v3)!.45!(n)$);
		\draw [dash dot,thick] ($(v3)!-.2!(n)$) -- (c);
		\draw [dotted] (v3) -- (v0) (v3) -- (v1) (v3) -- (v2);
		\fill [gray!20,opacity=.8] (w0) .. controls +(-130:.6) and +(15:.6) .. (w1) .. controls +(-10:.4) and +(120:.4) .. (w2) .. controls +(40:.5) and +(-80:.5) .. cycle;
		\draw [black!50] (w0) .. controls +(-130:.6) and +(15:.6) .. (w1);
		\draw (w1) .. controls +(-10:.4) and +(120:.4) .. (w2) .. controls +(40:.5) and +(-80:.5) .. (w0);
		\draw [dash dot,thick] (c) node [below] {\small $C^\pm$} -- (n);
		\fill [gray!20,opacity=.6] (178:{.46*\R}) circle [x radius={.52*\R},y radius={1.15*\R},rotate=-20];
		\draw [dash dot,thick] (n) -- ($(n)!-.4!(v3)$);
		\draw [dotted,thick] ($(v0)!-.2!(v1)$) -- ($(v1)!-.2!(v0)$) ($(v1)!-.3!(v2)$) -- ($(v2)!-.3!(v1)$) ($(v2)!-.1!(v0)$) -- ($(v0)!-.1!(v2)$);
		\foreach \i in {0,1,2,3} \fill [black] (v\i) circle [radius=1.5pt];
		\foreach \i in {0,1,2} \fill [black] (w\i) circle [radius=1.5pt];
		\fill (n) circle [radius=1.5pt];
		\fill (c) circle [radius=1.5pt];
		\node [anchor=north] at (n) {\small $N$};
		\node [anchor=east] at (v0) {\small $U_0$};
		\node [anchor=east] at (v1) {\small $U_1$};
		\node [anchor=east] at (v2) {\small $U_2$};
	\end{tikzpicture}
	\caption{``Fibration'' in $\mathbb{CP}^3$: Elementary discontinuities of type (i).}
	\label{fig:discPoint}
\end{figure}
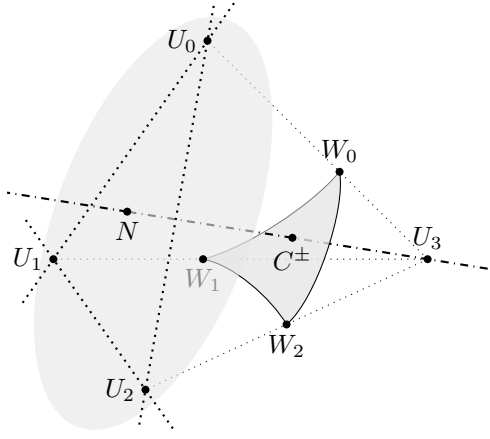

Despite the appearance of the square root, the above two-fold integral can be done analytically and yield a logarithm in the end. Being obtained from a residue computation, this integral can also be viewed as living on the quadric, and the square root in its integrand can be rationalized by introducing stereographic coordinates to the quadric. To do this explicitly, we first recover the homogeneous coordinates of $\mathbb{CP}^2$ by putting back $n_0$
\begin{align}\label{eq:I1discU3ninth}
	\mathrm{Disc}_{\overline{U_3}}I_{\rm hex}^{(1)}&=\int_\triangledown\,\frac{\sqrt{q_0}u_1u_3\,\langle N\mathrm{d}N^2\rangle}{(A_1^2-4u_3A_0)^{3/2}},
\end{align}
where now $A_0=-u_1n_0n_1+(1-u_1)n_0n_2-u_1u_2n_1n_2$ and $A_1=n_0-u_1n_1+u_3n_2$. To resolve the ambiguity in the square root we introduce a new variable $n_3$ and require
\begin{align}\label{eq:discU3quadric}
	A_1^2-4u_3A_0-n_3^2=0.
\end{align}
Let us denote $\widetilde{N}=[n_0:n_1:n_2:n_3]$. Then in the $\widetilde{N}$ space the three vertices of the contour correspond to
\begin{align}
	W_0=[1:0:0:1],\quad
	W_1=[0:1:0:-u_1],\quad
	W_2=[0:0:1:u_3].
\end{align}
Moreover, the three original $0$-faces read
\begin{align}
	U_0=[1:0:0:-1],\quad
	U_1=[0:1:0:u_1],\quad
	U_2=[0:0:1:-u_3].
\end{align}
Hence $U_i$ and $M_i$ look the same in the $\mathbb{CP}^2$ coordinates $[n_0:n_1:n_2]$, but they actually live on different Riemann sheets. To introduce stereographic coordinates we need to choose a reference point on the quadric. A useful choice can be, e.g., $U_0$. Correspondingly, we reparametrize $\widetilde{N}$ by the linear sum
\begin{align}\label{eq:tNy}
	\widetilde{N}=y_0W_0+y_1W_1+y_2W_2+tU_0.
\end{align}
Plugging this into the equation \eqref{eq:discU3quadric} results in a linear equation in $t$ (it is not quadratic because $U_0$ is a solution to \eqref{eq:discU3quadric}), which we can solve to give
\begin{align}
	t=-\frac{u_3(-y_0y_2+u_1u_2y_1y_2+u_1y_0(y_1+y_2))}{y_0-u_1(1-u_3)y_1+u_1u_3y_2}.
\end{align}
This together with \eqref{eq:tNy} provides a one-to-one map from $Y=[y_0:y_1:y_2]$ to the quadric \eqref{eq:discU3quadric} (which is a double-cover of $[n_0:n_1:n_2]$). With this we can change the integral variables in the integral \eqref{eq:I1discU3ninth} to $Y$ and get
\begin{align}\label{eq:pointProjectionRationalize}
	\mathrm{Disc}_{\overline{U_3}}I_{\rm hex}^{(1)}&=-\frac{\sqrt{q_0}u_1u_3}{2}\int_{\triangledown'}\langle Y\mathrm{d}Y^2\rangle\,\frac{\partial}{\partial y_0}\big(\frac{1}{YQ'_1Y}\big),
\end{align}
where
\begin{align}
	YQ'_1Y=&y_0^2+u_1^2(1-u_3)y_1^2+u_1u_3^2y_2^2-2u_1(1-u_3)y_0y_1\nonumber\\
	&+2u_1u_3y_0y_2-u_1(1+u_1-u_2-u_3)u_3y_1y_2.
\end{align}
The new contour $\triangledown'$ in $Y$ space is almost a simplex. By construction of the map \eqref{eq:tNy} one can see its vertices are again at the canonical positions $[1:0:0]$, $[0:1:0]$ and $[0:0:1]$. The ambient spaces of its $1$-faces descend from their counterparts in the original $N$ space, i.e., the lines defined by $n_0=0$, $n_1=0$ and $n_2=0$, respectively. Via the map \eqref{eq:tNy} these ambient spaces are \footnote{In the first boundary one may ask why we do not choose the other sign for the square root, which also satisfies the map. The reason is that, after localizing $y_0$ to this value, it should further vanish when taking $[y_1:y_2]$ to $[1:0]$ or $[0:1]$, under the assumed conditions \eqref{eq:discU3Assumption1}.}
\begin{align}
	y_0=\frac{u_1y_1-u_3y_2+\sqrt{\Delta}}{2}\equiv y_*,\quad y_1=0,\quad y_2=0,
\end{align}
where $\Delta=4u_1u_2u_3y_1y_2+(u_1y_1-u_3y_2)^2$. This means we can first directly integrate $y_0$ away and get
\begin{align}
	\mathrm{Disc}_{\overline{U_3}}I_{\rm hex}^{(1)}&=\frac{\sqrt{q_0}u_1u_3}{2}\int_0^\infty\mathrm{d}y_2\,\frac{1}{YQ'_1Y}\big|_{y_0=\frac{u_1y_1-u_3y_2+\sqrt{\Delta}}{2},y_1=1}\nonumber\\
	&=\int_0^\infty\mathrm{d}y_2\,\frac{\sqrt{q_0}u_1u_3}{\Delta_1+(u_1(2u_3-1)+(2u_1-1)u_3y_2)\sqrt{\Delta_1}},
\end{align}
where $\Delta_1=\Delta\big|_{y_1=1}$. The final one-dimensional integral can again be carried out via rationalization, which finally yields
\begin{align}
	\mathrm{Disc}_{\overline{U_3}}I_{\rm hex}^{(1)}&=\log r_1.
\end{align}
For readers' convenience, in the ancillary file we present a Mathematica function that performs the rationalization procedure in our analysis. Note that the rationalization always begins with introducing one extra variable that identifies the square root, turning it into a quadric in an enlarged space. Then with the input of vertices of the original contour and a reference point, the codes automatically generate the corresponding rationalization map. We also present a function that computes the change of integral measure under this map.

\subsubsection{An alternative approach to elementary discontinuities of type (ii)}

The point projection discussed in the previous subsection can in fact provide an alternative way to compute the elementary discontinuities of type (ii).
Take $\mathrm{Disc}_{\overline{U_0U_3}}I_{\rm hex}^{(1)}$ as an example, because $U_3\notin Q_1$, lines through $U_3$ can properly intersect the crescent area, so that the residue can be computed in a well-defined way. The interesting thing here is that, for every relevant "fibre", both points in its intersection with $Q_1$ contribute to the discontinuity. These points overlap from the view of $N$ space but lie on different Riemann sheets. The only exception is when the "fibre" is tangent to the crescent, such that the two points collide and hit the branch cut. Therefore, the entire crescent contour is always folded in $N$ space, with its two end points $
W_0$ and $U_0$ sharing the same $n$ values.
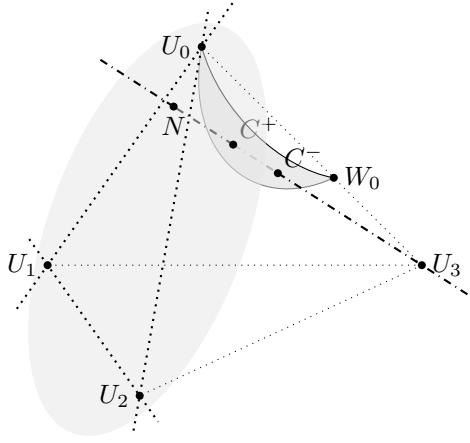
\begin{figure}[ht]
	\centering
	\begin{tikzpicture}
		\def\R{2.5}
		\coordinate (v0) at (100:\R);
		\coordinate (v1) at (190:\R);
		\coordinate (v2) at (-120:\R);
		\coordinate [label=0:{\small $U_3$}] (v3) at (-10:\R);
		\coordinate [label=0:{\small $W_0$}] (w) at ($(v0)!.6!(v3)$);
		\coordinate (n) at ($(v0)+(-115:{.35*\R})$);
		\coordinate (cm) at ($(n)!.42!(v3)$);
		\coordinate (cp) at ($(n)!.24!(v3)$);
		\draw [dash dot,thick] (cp) -- (cm);
		\fill [gray!20,opacity=.8] (v0) .. controls +(-75:.9) and +(170:.7) .. (w) .. controls +(205:1.2) and +(-100:1.4) .. cycle;
		\draw [black!50] (w) .. controls +(205:1.2) and +(-100:1.4) .. (v0);
		\draw (v0) .. controls +(-75:.9) and +(170:.7) .. (w);
		\draw [dash dot,thick] (cm) node [anchor=south west,inner sep=2pt] {\small $C^-$} -- ($(v3)!-.2!(n)$);
		\draw [dash dot,thick] (cp) node [anchor=south west,inner sep=2pt] {\small $C^+$} -- (n);
		\fill [gray!20,opacity=.5] (178:{.46*\R}) circle [x radius={.52*\R},y radius={1.15*\R},rotate=-20];
		\draw [dash dot,thick] (n) -- ($(n)!-.3!(v3)$);
		\draw [dotted,thick] ($(v0)!-.2!(v1)$) -- ($(v1)!-.2!(v0)$) ($(v1)!-.2!(v2)$) -- ($(v2)!-.2!(v1)$) ($(v2)!-.1!(v0)$) -- ($(v0)!-.1!(v2)$);
		\draw [dotted] (v3) -- (v0) (v3) -- (v1) (v3) -- (v2);
		\foreach \i in {0,1,2,3} \fill [black] (v\i) circle [radius=1.5pt];
		\fill (w) circle [radius=1.5pt];
		\fill (n) circle [radius=1.5pt];
		\fill (cp) circle [radius=1.5pt];
		\fill (cm) circle [radius=1.5pt];
		\node [anchor=north] at (n) {\small $N$};
		\node [anchor=east] at (v0) {\small $U_0$};
		\node [anchor=east] at (v1) {\small $U_1$};
		\node [anchor=east] at (v2) {\small $U_2$};
	\end{tikzpicture}
	\caption{An alternative ``fibration'' of elementary discontinuities of type (ii).}
	\label{fig:discTangentAlt}
\end{figure}

To compute $\mathrm{Disc}_{\overline{U_0U_3}}I_{\rm hex}^{(1)}$ following the above picture, we can follow exactly the same steps presented in the previous subsection up to \eqref{eq:pointProjectionRationalize}, i.e., deriving the residue and rationalizing the resulting integral. Note here we still choose to compute the residue at $c_0=c_0^-$, which means we treat the Riemann sheet on which $W_0$ is located as the principal sheet. Although the residue at $c^0=c_0^+$ should also matter, it is rather regarded as a different contribution as we move to the second sheet, and will be automatically included while performing the $n$ integration. Now the contour in \eqref{eq:pointProjectionRationalize} is no longer a "simplex". For each pair of $[n_1:n_2]$, the variable $n_0$ is integrated from $\infty$, where $N$ coincides with $W_0$, to a value $\tilde{n}_0$ such that the "fibre" $\overline{U_3N}$ is tangent to the quadric $Q_1$, and then further back to $\infty$ where it coincides with $U_0$. With the rationalization map \eqref{eq:tNy}, $[y_1:y_2]$ is integrated from $[0:1]$ to $[1:0]$, while for each pair $[y_1:y_2]$, $y_0$ is integrated from a value $\tilde{y}_0$ to $\infty$. The value can be solved by setting $n_1=n_2=0$, which yields
\begin{align}
    \tilde{y}_0=(1-u_3)u_1y_1-u_1u_3y_3.
\end{align}
Hence we have
\begin{align}
	\mathrm{Disc}_{\overline{U_0U_3}}I_{\rm hex}^{(1)}&=\frac{\sqrt{q_0}u_1u_3}{2}\int_0^\infty\mathrm{d}y_2\,\frac{1}{YQ'_1Y}\big|_{y_0=\tilde{y}_0,y_1=1}\nonumber\\
	&=\int_0^\infty\mathrm{d}y_2\,\frac{\frac{1}{2}\sqrt{q_0}}{u_1(u_3-1)+(1-u_1-u_2-u_3+2u_1u_3y_2+(u_1-1)u_3y_2^2)}\nonumber\\
	&=\log\frac{1}{r_1r_3}.
\end{align}
As expected, this yields the same result as \eqref{eq:discU0U3}.

\section{Symbol construction}\label{sec:symbolConstruction}

Given that various types of elementary discontinuities have been computed following the previous section, we can now move on to construct the symbol.  We first finish the discussion on the massless hexagon \eqref{eq:masslessHexagon}, and draw some comments on the robustness of the method. Then we quickly visit a case of box integral with two massive and two massless propagators, to illustrate some new features in the analysis. 

\subsection{Complete symbol of the massless hexagon}

In \eqref{eq:SIhexStructure} we have observed that the symbol of the hexagon integral $I_{\rm hex}$ \eqref{eq:masslessHexagon} depends on three of its discontinuities, one of which being $I_{\rm hex}^{(1)}\equiv\mathrm{Disc}_{\overline{U_1U_3}}I_{\rm hex}$, whose integral representation is given in \eqref{eq:discU13canonical}. Furthermore, the symbol of $I_{\rm hex}^{(1)}$ itself further depends on its own discontinuities according to \eqref{eq:I1hexStructureRaw}, and in Section \ref{sec:discIhex1} we have determined all these elementary discontinuities by explicit computation. These latter discontinuities turn out to all be pure logarithms, which means that $I^{(1)}_{\rm hex}$ is a function of weight 2. By inserting the symbol of the logarithms into \eqref{eq:I1hexStructureRaw} we thus get
\begin{align}\label{eq:I1hexStructure}
	\mathcal{S}[I_{\rm hex}^{(1)}]\equiv\mathcal{S}[\mathrm{Disc}_{\overline{U_1U_3}}I_{\rm hex}]&=\underbrace{\frac{1}{u_1}\otimes\frac{1}{r_1r_3}+\frac{u_3}{u_1}\otimes\frac{1}{r_1r_2}}_{\text{(b)}}+\underbrace{\frac{u_3}{u_1}\otimes r_1}_{\text{(c)}}+\underbrace{\frac{1-u_1}{u_1}\otimes r_0+u_2\otimes\frac{1}{r_3}}_{\text{(e)}}\nonumber\\
	&=(1-u_1)\otimes r_0-u_2\otimes r_3-u_3\otimes r_2.
\end{align}
In the second equality we have applied the identity that
\begin{align}\label{eq:r123r0}
    r_1r_2r_3=r_0.
\end{align}

Similar computations can be performed on the other two discontinuities of $I_{\rm hex}$. We collect the details in the ancillary file, and simply print the results of their symbols in the following
\begin{subequations}
    \begin{align}
        \mathcal{S}[\mathrm{Disc}_{\overline{U_2U_4}}I_{\rm hex}]&=(1-u_2)\otimes r_0-r_3\otimes r_1-r_1\otimes r_3,\\
        \mathcal{S}[\mathrm{Disc}_{\overline{U_3U_5}}I_{\rm hex}]&=(1-u_3)\otimes r_0-r_1\otimes r_2-r_2\otimes r_1.
    \end{align}
\end{subequations}
Plugging these data further into \eqref{eq:SIhexStructure} we thus obtain the entire symbol of the original hexagon integral $I_{\rm hex}$, as presented in \eqref{eq:SIhex}.

It is worth drawing a further inspection on the symbol structure of the integral $I_{\rm hex}^{(1)}$. Note that in \eqref{eq:discU13canonical} we have intentionally rescaled the quadric before analyzing singularities of this integral. One may ask whether we may still land on the same result if we do not perform this rescaling first. This is important for the consistency of the method.

If we directly use the original $\widetilde{Q}_1$ as presented in \eqref{eq:Q1matrix}, more entries in the corresponding matrix depend on the parameters $u_i$. Following our procedure for deriving the first symbol entries, we may conclude that the symbol of $I_{\rm hex}^{(1)}$ should have the following structure instead
\begin{align}\label{eq:I1hexStructureRawAlt}
	\mathcal{S}[I_{\rm hex}^{(1)}]&=\underbrace{u_1\otimes\mathcal{S}[\mathrm{Disc}_{\overline{U_1U_3}}I_{\rm hex}^{(1)}]+u_3\otimes\mathcal{S}[\mathrm{Disc}_{\overline{U_2U_3}}I_{\rm hex}^{(1)}]}_{\text{(b)}}+\underbrace{u_3\otimes\mathcal{S}[\mathrm{Disc}_{\overline{U_3}}I_{\rm hex}^{(1)}]}_{\text{(c)}}\nonumber\\
	&\quad+\underbrace{u_1\otimes\mathcal{S}[\mathrm{Disc}_{\overline{U_0U_1}}I_{\rm hex}^{(1)}]+(1-u_1)\otimes\mathcal{S}[\mathrm{Disc}_{\overline{U_0U_2}}I_{\rm hex}^{(1)}]+(u_1u_2)\otimes\mathcal{S}[\mathrm{Disc}_{\overline{U_1U_2}}I_{\rm hex}^{(1)}]}_{\text{(e)}}.
\end{align}
Since the integral does not change under the rescaling, the four elementary discontinuities that already showed up in \eqref{eq:I1hexStructureRaw} of our previous analysis remain the same as well, so that we can directly apply the results from the previous section. Apart from these, \eqref{eq:I1hexStructureRawAlt} receives two new contributions from $\mathrm{Disc}_{\overline{U_0U_1}}I_{\rm hex}^{(1)}$ and $\mathrm{Disc}_{\overline{U_1U_3}}I_{\rm hex}^{(1)}$. Applying the analysis for elementary discontinuities of type (iii), in Subsection \ref{sec:discJumpe}, one can learn that for the jump type (e) term
\begin{align}
    \mathrm{Disc}_{\overline{U_0U_1}}I_{\rm hex}^{(1)}=\log\frac{1}{r_2},
\end{align}
while applying the analysis for elementary discontinuities of type (ii) in Section \ref{sec:discJumpb}, the jump type (b) case yields
\begin{align}
    \mathrm{Disc}_{\overline{U_1U_3}}I_{\rm hex}^{(1)}=\log(r_2r_3).
\end{align}
With these we conclude that
\begin{align}\label{eq:I1hexStructureAlt}
	\mathcal{S}[I_{\rm hex}^{(1)}]&=\underbrace{u_1\otimes(r_2r_3)+u_3\otimes\frac{1}{r_1r_2}}_{\text{(b)}}+\underbrace{u_3\otimes r_1}_{\text{(c)}}+\underbrace{u_1\otimes\frac{1}{r_2}+(1-u_1)\otimes r_0+(u_1u_2)\otimes\frac{1}{r_3}}_{\text{(e)}}\nonumber\\
	&=(1-u_1)\otimes r_0-u_2\otimes r_3-u_3\otimes r_2,
\end{align}
which is exactly the same as the result \eqref{eq:I1hexStructure} from the previous analysis.

Generalizing this, if we consider applying a generic rescaling to the quadric, with some function $f(u_1,u_2,u_3)$, that will mean that we should include this function as a multiplicative factor in the first entries corresponding to all possible elementary discontinuities of $I_{\rm hex}^{(1)}$. There are altogether seven such discontinuities, so that the potential extra contribution to $\mathcal{S}[I_{\rm hex}^{(1)}]$ is
\begin{align}
    f(u_1,u_2,u_3)\otimes\big(&\mathcal{S}[\mathrm{Disc}_{\overline{U_0U_1}}I_{\rm hex}^{(1)}]+\mathcal{S}[\mathrm{Disc}_{\overline{U_0U_2}}I_{\rm hex}^{(1)}]+\mathcal{S}[\mathrm{Disc}_{\overline{U_0U_3}}I_{\rm hex}^{(1)}]\nonumber\\
    &+\mathcal{S}[\mathrm{Disc}_{\overline{U_1U_2}}I_{\rm hex}^{(1)}]+\mathcal{S}[\mathrm{Disc}_{\overline{U_1U_3}}I_{\rm hex}^{(1)}]+\mathcal{S}[\mathrm{Disc}_{\overline{U_2U_3}}I_{\rm hex}^{(1)}]\nonumber\\
    &+\mathcal{S}[\mathrm{Disc}_{\overline{U_3}}I_{\rm hex}^{(1)}]\big).
\end{align}
However, with the identity \eqref{eq:r123r0} one can verify that the summation of these discontinuities simply vanishes. This guarantees that the symbol obtained from our analysis is indeed invariant under any rescaling.

A better way to make the above invariance transparent is to organize the symbol structure according to letters in the first entries even before we solve the discontinuities in detail. For brevity let us denote $\mathcal{S}_{ij}\equiv\mathcal{S}[\mathrm{Disc}_{\overline{U_iU_j}}I_{\rm hex}^{(1)}]$ and $\mathcal{S}_i\equiv\mathcal{S}[\mathrm{Disc}_{\overline{U_i}}I_{\rm hex}^{(1)}]$. Following this organizing principle, in the analysis with rescaling, the structure \eqref{eq:I1hexStructureRaw} reads
\begin{align}\label{eq:symbolByFirstEntry}
    \mathcal{S}[I_{\rm hex}^{(1)}]=\frac{1}{u_1}\otimes(\mathcal{S}_{02}\mathcal{S}_{03}\mathcal{S}_{23}\mathcal{S}_3)+(1-u_1)\otimes\mathcal{S}_{02}+u_2\otimes\mathcal{S}_{12}+u_3\otimes(\mathcal{S}_{23}\mathcal{S}_3),
\end{align}
while in the analysis without rescaling, the structure \eqref{eq:I1hexStructureRawAlt} reads
\begin{align}\label{eq:symbolByFirstEntryAlt}
    \mathcal{S}[I_{\rm hex}^{(1)}]=u_1\otimes(\mathcal{S}_{01}\mathcal{S}_{12}\mathcal{S}_{13})+(1-u_1)\otimes\mathcal{S}_{02}+u_2\otimes\mathcal{S}_{12}+u_3\otimes(\mathcal{S}_{23}\mathcal{S}_3).
\end{align}
Again, the equivalence of the two expressions is based on the identity \eqref{eq:r123r0}. Nevertheless, the two expressions view the first letters in relation to the discontinuities in two different ways. In \eqref{eq:symbolByFirstEntryAlt} we learn the first term from the neighborhood of $u_1=0$, since analytic continuation around $u_1=0$ simultaneously induces a combination of three discontinuities $\mathrm{Disc}_{\overline{U_0U_1}}I_{\rm hex}^{(1)}$, $\mathrm{Disc}_{\overline{U_1U_2}}I_{\rm hex}^{(1)}$ and $\mathrm{Disc}_{\overline{U_1U_3}}I_{\rm hex}^{(1)}$. In comparison, in \eqref{eq:symbolByFirstEntry} we learn the first term from the neighborhood of $u_1=\infty$. Hence we see that, even though a first entry $f$ indicates logarithmic singularities both at $f=0$ and $f=\infty$ for the function, when determining the symbol it suffices to just focus on one of them. This is exactly what we did in the previous analyses.

\subsection{Two-mass off-shell box}\label{sec:twoMassBox}

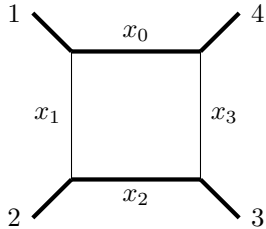
\begin{figure}[ht]
    \centering
    \begin{tikzpicture}
        \def\R{1.2};
        \coordinate (v1) at (135:\R);
        \coordinate (v2) at (-135:\R);
        \coordinate (v3) at (-45:\R);
        \coordinate (v4) at (45:\R);
        \draw (v1) -- (v2) (v3) -- (v4);
        \draw [ultra thick] (v1) -- (v4) (v2) -- (v3);
        \draw [ultra thick] (v1) -- ++(135:{.6*\R}) node [left] {\small $1$};
        \draw [ultra thick] (v2) -- ++(-135:{.6*\R}) node [left] {\small $2$};
        \draw [ultra thick] (v3) -- ++(-45:{.6*\R}) node [right] {\small $3$};
        \draw [ultra thick] (v4) -- ++(45:{.6*\R}) node [right] {\small $4$};
        \node [anchor=south] at ($(v1)!.5!(v4)$) {\small $x_0$};
        \node [anchor=east] at ($(v1)!.5!(v2)$) {\small $x_1$};
        \node [anchor=north] at ($(v2)!.5!(v3)$) {\small $x_2$};
        \node [anchor=west] at ($(v3)!.5!(v4)$) {\small $x_3$};
    \end{tikzpicture}
    \caption{A scalar box in 4 dimensions. The thick lines denote massive particles of mass $m$, while the thin line denotes massless particles.}
    \label{fig:offshellbox}
\end{figure}

To gain a better understanding of the symbol construction in the presence of square root ambiguity in the first entries, let us consider another example of an integral in $\mathbb{CP}^3$
\begin{align}
    I_{\rm box}=\int_\triangledown\frac{4\sqrt{q_2}\langle X\mathrm{d}X^3\rangle}{(XQ_2X)^2},
\end{align}
where $X=[x_0:x_1:x_2:x_3]$ and $q_2=\det Q_2$, with the matrix
\begin{align}\label{eq:Q2matrix}
    Q_2=\frac{1}{2}\left(\begin{matrix}
    -2u & 1 & s-2u & 1 \\
    1 & 0 & 1 & t \\
    s-2u & 1 & -2u & 1 \\
    1 & t & 1 & 0
    \end{matrix}\right).
\end{align}
This integral comes from a box diagram in 4 dimensions, constructed by exchanging two massless particles between two massive particles of mass $m$, as illustrated in Figure \ref{fig:offshellbox}. We set all the massive external legs off-shell, with momentum squared being $p^2$. Then the three parameters in $I_{\rm box}$ are tied to the kinematic variables by
\begin{align}
    s=\frac{s_{12}}{p^2-m^2},\quad
    t=\frac{s_{14}}{p^2-m^2},\quad
    u=\frac{m^2}{p^2-m^2}.
\end{align}
To obtain $I_{\rm box}$ from the Feynman parameter representation of this box diagram, we have also rescaled the Feynman parameters.

From \eqref{eq:Q2matrix} it is easy to observe there are three jump types that effectively lead to singularities of $I_{\rm box}$. These types as well as the one-dimensional ambient spaces in which they are present are listed as follows
\begin{center}
    \begin{tabular}{c|c}
         \toprule
         jump type & $\mathbb{CP}^1$ ambient spaces \\
         \midrule
         (a) & $\overline{U_0U_2}$ \\
         (c) & $\overline{U_0U_1}$, $\overline{U_0U_3}$, $\overline{U_1U_2}$, $\overline{U_2U_3}$ \\
         (e) & $\overline{U_1U_3}$ \\
         \bottomrule
    \end{tabular}
\end{center}
For the unique ambient space associated to jump type (e), from $Q_2$ we can read off the symbol letter $t$. For each of the four ambient spaces associated to jump type (c), we further read off a common letter $u$. For $\overline{U_0U_2}$ associated to jump type (a), note that it intersects $Q_2$ at two points $W_{02}^\pm=[1:0:\frac{(\sqrt{s}\pm\sqrt{s-4u})^2}{4u}:0]$. Therefore we read off two letters from this type, so that the symbol $\mathcal{S}[I_{\rm box}]$ is expected to have the following structure
\begin{align}\label{eq:SIboxStructure}
    \mathcal{S}[I_{\rm box}]=t\otimes D_1+u\otimes D_2+\frac{(\sqrt{s}+\sqrt{s-4u})^2}{4u}\otimes D_3^++\frac{(\sqrt{s}-\sqrt{s-4u})^2}{4u}\otimes D_3^-.
\end{align}
The $D_1$, $D_2$ and $D_3^\pm$ here are the remaining parts of the symbol to be determined by discontinuities.

\begin{figure}[ht]
	\centering
	\begin{tikzpicture}
		\def\R{2.5}
		\coordinate [label=180:{\small $U_0$}] (v0) at (100:\R);
		\coordinate [label=180:{\small $U_3$}] (v3) at (190:\R);
		\coordinate [label=180:{\small $U_1$}] (v1) at (-120:\R);
		\coordinate [label=0:{\small $U_2$}] (v2) at (-10:\R);
		\coordinate [label=0:{\small $W_{02}^+$}] (w02p) at ($(v0)!.7!(v2)$);
		\coordinate [label=0:{\small $W_{02}^-$}] (w02m) at ($(v0)!.3!(v2)$);
		\coordinate [label=180:{\small $W_{01}$}] (w01) at ($(v0)!.5!(v1)$);
		\coordinate [label=180:{\small $W_{03}$}] (w03) at ($(v0)!.5!(v3)$);
		\fill [gray!20,opacity=.8] (w03) .. controls +(-30:.3) and +(105:.3) .. (w01) .. controls +(25:.7) and +(170:.9) .. (w02p) .. controls +(150:1.2) and +(10:1) .. cycle;
		\draw [black!50] (w02p) .. controls +(150:1.2) and +(10:1) .. (w03);
		\draw (w03) .. controls +(-30:.3) and +(105:.3) .. (w01) .. controls +(25:.7) and +(170:.9) .. (w02p);
		\draw [dotted] (v0)-- (v2) (v1) -- (v3) (v0) -- (v1) -- (v2) -- (v3) -- cycle;
		\fill [gray!20,opacity=.8] (w03) .. controls +(-30:.3) and +(105:.3) .. (w01) .. controls +(60:.7) and +(-145:.7) .. (w02m) .. controls +(170:.6) and +(45:.6) .. cycle;
		\draw [black!50] (w02m) .. controls +(170:.6) and +(45:.6) .. (w03);
		\draw (w03) .. controls +(-30:.3) and +(105:.3) .. (w01) .. controls +(60:.7) and +(-145:.7) .. (w02m);
		\draw [dotted] (v0)-- (v2) (v1) -- (v3) (v0) -- (v1) -- (v2) -- (v3) -- cycle;
		\foreach \i in {0,1,2,3} \fill [black] (v\i) circle [radius=1.5pt];
		\fill [black] (w02p) circle [radius=1.5pt];
		\fill [black] (w02m) circle [radius=1.5pt];
		\fill [black] (w01) circle [radius=1.5pt];
		\fill [black] (w03) circle [radius=1.5pt];
	\end{tikzpicture}
	\caption{Two different discontinuity contours associated to $\overline{U_0}$.}
	\label{fig:discContourSqrt}
\end{figure}
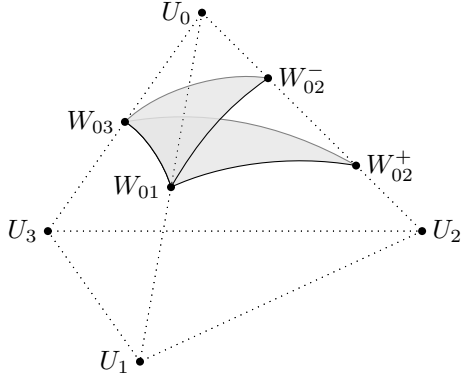

Then we enumerate the elementary discontinuities. Firstly, the jump type (e) in $\overline{U_1U_3}$ leads to the discontinuity $\mathrm{Disc}_{\overline{U_1U_3}}I_{\rm box}$. This can be computed following the procedure described in Section \ref{sec:discJumpe}, and we obtain
\begin{align}
    \mathrm{Disc}_{\overline{U_1U_3}}I_{\rm box}=\log\left(\frac{\sqrt{st}+\sqrt{-4+st-4tu}}{\sqrt{st}-\sqrt{-4+st-4tu}}\right)^2.
\end{align}
Apart from this, jumps of both type (a) and type (c) are effectively equivalent to jumps in the $0$-faces $\overline{U_0}$ and $\overline{U_2}$. So the corresponding discontinuities arise from analytic continuation around such touching configurations. The new ingredient in this example is that, taking $\overline{U_0}$ for instance, there exist multiple choices of discontinuity contours due to the two-fold ambiguity in the intersection points $Q_2\cap\overline{U_0U_2}$. Note that the discontinuity contour should be anchored at the intersection points between $Q_2$ and $\overline{U_0U_i}$ ($i=1,2,3$), and that for $i=1,3$ this is unique since the other intersection point coincides with $0$-faces of the original contour. Therefore we have two different discontinuities $\mathrm{Disc}_{\overline{U_0}}^\pm I_{\rm box}$, as illustrated in Figure \ref{fig:discContourSqrt}, where $\mathrm{Disc}_{\overline{U_0}}^+ I_{\rm box}$ has its contour anchored at $W_{02}^+$, while $\mathrm{Disc}_{\overline{U_0}}^- I_{\rm box}$ at $W_{02}^-$. Similarly, we have two other different discontinuities $\mathrm{Disc}_{\overline{U_3}}^\pm I_{\rm box}$ associated to the $0$-face $\overline{U_3}$. All these discontinuities can be computed following the procedure described in Section \ref{sec:discJumpc}, and which yields
\begin{align}
    \mathrm{Disc}_{\overline{U_0}}^+ I_{\rm box}&=\mathrm{Disc}_{\overline{U_3}}^- I_{\rm box}=\log\frac{(\sqrt{t(s-4u)}+\sqrt{-4+st-4tu})^2u+(\sqrt{s}-\sqrt{s-4u})^2}{(\sqrt{t(s-4u)}-\sqrt{-4+st-4tu})^2u+(\sqrt{s}-\sqrt{s-4u})^2},\\
    \mathrm{Disc}_{\overline{U_0}}^- I_{\rm box}&=\mathrm{Disc}_{\overline{U_3}}^+ I_{\rm box}=\log\frac{(\sqrt{t(s-4u)}-\sqrt{-4+st-4tu})^2u+(\sqrt{s}+\sqrt{s-4u})^2}{(\sqrt{t(s-4u)}+\sqrt{-4+st-4tu})^2u+(\sqrt{s}+\sqrt{s-4u})^2}.
\end{align}

Now we match the above elementary discontinuities with the ones assumed in the structure \eqref{eq:SIboxStructure}. Firstly, considering analytic continuation in $t$ around $t=0$, by definition we directly have that 
\begin{align}
    D_1=\mathcal{S}[\mathrm{Disc}_{\overline{U_1U_3}}I_{\rm box}].
\end{align}
The remaining three terms need extra care. When we analytically continue $u$ around $u=0$, we can observe that
\begin{align}\label{eq:W02u0}
    \frac{(\sqrt{s}-\sqrt{s-4u})^2}{4u}=\frac{u}{s}+\mathcal{O}(u^2),\qquad
    \frac{(\sqrt{s}+\sqrt{s-4u})^2}{4u}=\frac{s}{u}+\mathcal{O}(u^0).
\end{align}
This means that the corresponding discontinuity receives contributions from all the three remaining terms in \eqref{eq:SIboxStructure}, among which the the third term contributes with a minus sign (since the direction of analytic continuation in its first entry is reversed). On the other hand, \eqref{eq:W02u0} also means that as $u\to0$ we simultaneously have $W_{02}^-\to U_0$ and $W_{02}^+\to U_2$. This gives rise to the relation
\begin{align}\label{eq:IboxDrelationM}
    \frac{D_2D_3^-}{D_3^+}=\mathcal{S}[\mathrm{Disc}_{\overline{U_0}}^-I_{\rm box}+\mathrm{Disc}_{\overline{U_3}}^+I_{\rm box}].
\end{align}
Furthermore, if we first consider analytic continuation such that the role of $W_{02}^\pm$ is switched, and then perform the same analytic continuation around $u=0$, the above reasoning will also imply
\begin{align}\label{eq:IboxDrelationP}
    \frac{D_2D_3^+}{D_3^-}=\mathcal{S}[\mathrm{Disc}_{\overline{U_0}}^+I_{\rm box}+\mathrm{Disc}_{\overline{U_3}}^-I_{\rm box}].
\end{align}
With the results on the elementary discontinuities, \eqref{eq:IboxDrelationM} and \eqref{eq:IboxDrelationP} together yield
\begin{align}
    D_2=\left(\frac{\sqrt{st}+\sqrt{-4+st-4tu}}{\sqrt{st}-\sqrt{-4+st-4tu}}\right)^2,\quad
    \frac{D_3^-}{D_3^+}=\left(\frac{\sqrt{t(s-4u)}-\sqrt{-4+st-4tu}}{\sqrt{t(s-4u)}+\sqrt{-4+st-4tu}}\right)^2.
\end{align}

Note that the two roots in first entries of \eqref{eq:SIboxStructure} satisfy
\begin{align}
    \frac{(\sqrt{s}-\sqrt{s-4u})^2}{4u}=\frac{4u}{(\sqrt{s}+\sqrt{s-4u})^2}=\sqrt{\left(\frac{\sqrt{s}-\sqrt{s-4u}}{\sqrt{s}+\sqrt{s-4u}}\right)^2}.
\end{align}
Therefore we can fully solve the symbol as
\begin{align}\label{eq:SIboxStructureResult}
    \mathcal{S}[I_{\rm box}]&=t\otimes D_1+u\otimes D_2+\frac{\sqrt{s}-\sqrt{s-4u}}{\sqrt{s}+\sqrt{s-4u}}\otimes\frac{D_3^-}{D_3^+}\nonumber\\
    &=2\Big((tu)\otimes\frac{\sqrt{st}+\sqrt{-4+st-4tu}}{\sqrt{st}-\sqrt{-4+st-4tu}}+\frac{\sqrt{s}-\sqrt{s-4u}}{\sqrt{s}+\sqrt{s-4u}}\otimes\frac{\sqrt{t(s-4u)}-\sqrt{-4+st-4tu}}{\sqrt{t(s-4u)}+\sqrt{-4+st-4tu}}\Big).
\end{align}
As a consistency check, one can verify that this result satisfies symbol integrability in the sense that
\begin{align}
    \mathcal{S}=\sum_{i,j}w_i\otimes w_j\quad\Rightarrow\quad\mathrm{d}\log(w_i)\wedge\mathrm{d}\log(w_j)=0.
\end{align}

\section{Discussions and Outlook}\label{sec:outlook}

In this work we initiate a systematic study of analytic properties of Feynman parameter integrals in view of discontinuities. There are many related problems that are interesting to explore in future, which we briefly comment on in the following.

\brief{Mixed Weight, Degenerate $Q$}

First of all, there are several problems for integrals with quadric singularity that need to be understood further. One of them is about the mixed weight integrals whose numerators are some tensors. A useful idea in this regard involves the identification of part of the integrand that is already an exact form, so that it effectively has less integrals as compared to the remaining part. Another problem associates to the situation when $Q$ is degenerate, or $\rm{det}(Q)=0$. This typically gives rise to non-trivial coefficients in front of symbol terms, but the classification of geometry configurations should in principle remain the same. It should be interesting to work out some explicit cases and compare with, e.g., the result obtained by spherical contours \cite{Arkani-Hamed:2017ahv}.

\brief{Higher degree hypersurfaces, Elliptic integrals, Reducible Component}

The analysis in this paper has a great potential to generalize to cases with higher-degree integrand singularities. This may occur in Feynman parameter representations of two or higher-loop Feynman integrals as well as four or higher-point energy correlators in the collinear limit. Take the cubic singularity as an example. Different from quadrics where only $0$ and $1$-faces of the contour should be considered in the stratification of touching configurations, in the cubic case $2$-faces also becomes relevant. The geometry of cubics is also more involved, as it can be irrational. Nevertheless, the classification of logarithmic singularities on the principal branch should still work, and we expect that at least when the cubic is rational the first entries of symbol can still be read off in a convenient way. The operation of bi-projection should still be well-define for generic cubics, yet it is very interesting to check what such operation may yield. In the case when the cubic is irrational so that the integral becomes elliptic, we hope that this geometric treatment may inspire some useful tools in handling the related analytic behaviors.

It is also worth to study situations when the integrand singularity include multiple irreducible components of various degrees. For instance, a cubic that consists of a quadric and a hyperplane. In fact, similar problems have already been studied in the Aomoto polylogarithms, where the integrand singularity is a higher degree hypersurface with linear irreducible components. It is known that integrals with multiple quadric singularity components may lead to elliptic integrals, even though each component itself is rational. It is important to understand better how the elliptic behavior emerges in the current context.

\brief{Symbol Alphabet instead of the full symbol}

Our current analysis deals with the construction of the full symbol. Yet it is interesting to check the existence of even more efficient method if one only targets on the symbol alphabets. A proper answer to this question may help boost the bootstrap study of Feynman integrals.

\acknowledgments

The authors would like to thank Ruth Britto, Einan Gardi, Wei Gu, Song He, Johannes Henn, Lecheng Ren and Qinglin Yang for useful discussions. EYY would also like to thank the hospitality of \emph{The Third Summer School on Frontiers of Quantum Field Theories} organized at Northwest University, during which this work was finishing and part of the results was presented. The authors are supported by National Natural Science Foundation of China under Grant No.~12175197 and Grand No.~12347103. EYY is also supported by National Natural Science Foundation of China under Grant No.~11935013 and Grant No.~12535003, and by the Fundamental Research Funds for the Chinese Central Universities under Grant No.~226-2022-00216. JYG is also supported by the National Natural Science Foundation of China under Grant No.~12357077.

\bibliographystyle{JHEP}
\bibliography{projection}

\end{document}